\documentclass[12pt]{article}
\usepackage[latin9]{inputenc}
\usepackage{microtype}
\usepackage{float} 
\usepackage[colorlinks=true, linkcolor=black, citecolor=black, urlcolor=black]{hyperref}
\usepackage{cite}
\usepackage{babel}
\usepackage{mathdots}
\usepackage{amsmath}
\usepackage{amssymb}
\usepackage{amsfonts}
\usepackage{mathrsfs}
\usepackage{esint}
\usepackage{graphicx}
\usepackage{youngtab}
\usepackage{multicol}
\usepackage{multirow}
\usepackage{slashed}
\usepackage{simplewick}
\usepackage{braket}
\usepackage{bm}
\usepackage{relsize}
\usepackage[dvipsnames,table,xcdraw]{xcolor}
\usepackage{subfigure}
\allowdisplaybreaks
\usepackage{geometry}
\newlength{\figline}\newlength{\figsize}
\newlength{\figlineV}\newlength{\figsizeV}
\def\non{\nonumber\\}

\usepackage{color} 

\newcommand{\bl}[1]{#1}

\usepackage[dvipsnames]{xcolor}
\newcommand{\VF}[1]{#1}

\usepackage{hyperref}
\usepackage{cleveref}
\crefname{equation}{Eq.}{Eqs.}
\crefname{figure}{Fig.}{Figs.}
\crefname{table}{Table}{Tables}
\crefname{section}{Section}{Sections}

\everymath{\displaystyle}

\begin{document}
\author{Wan-Zhe Feng$^a$\footnote{\href{mailto:vicf@tju.edu.cn}{vicf@tju.edu.cn}},~
Jinzheng Li$^b$\footnote{\href{mailto:li.jinzh@northeastern.edu}{li.jinzh@northeastern.edu}},~
Pran Nath$^b$\footnote{\href{mailto:p.nath@northeastern.edu}{p.nath@northeastern.edu}}~~and
Zong-Huan Ye$^a$\footnote{\href{mailto:y2953083702@tju.edu.cn}{y2953083702@tju.edu.cn}}\\
$^{a}$\textit{\normalsize Center for Joint Quantum Studies and Department of Physics, School of Science,}\\\textit{\normalsize
Tianjin University, Tianjin 300350, PR. China}\\
$^{b}$\textit{\normalsize Department of Physics, Northeastern University, Boston, MA 02115-5000, USA} }

\title{\vspace{-2cm}\begin{flushright}
\end{flushright}
\vspace{1cm}
\Large \bf
Gauge-independent Gravitational Waves from Cogenesis in a \texorpdfstring{$B-L$}{B-L} Conserving Universe}
\vspace{0.0cm}
\date{}
\maketitle
\begin{abstract}
An analysis of baryogenesis and stochastic gravitational wave production is presented for an extension of the standard model where the dark sector consists of dark matter particles charged under a $U(1)_x$ gauge symmetry, while a subset of dark fields also carry lepton number but no $U(1)_x$ charge. We demonstrate that with CP violation induced by Yukawa couplings, equal and opposite lepton asymmetries are generated in the visible and hidden sectors. Subsequent evolution preserves lepton number separately in each
sector, and sphaleron interactions partially convert the lepton asymmetry into baryon asymmetry near the temperature of the first-order phase transition.
Further, we discuss  stochastic gravitational wave background  production for the first-order phase transition using a gauge-independent bubble nucleation dynamics which yields spectra also valid in the
 supercooled low-temperature regime with {$T_p/m_{A_x} \ll 1$} where $T_p$ is the percolation temperature and $m_{A_x}$ is the dark photon mass.  A parameter-space scan identifies regions that simultaneously account for cogenesis of baryon asymmetry and dark matter
  and predict stochastic gravitational wave signals within reach of current (NANOGrav, EPTA, PPTA) and future detectors at higher frequencies, providing a unified framework for cogenesis and associated gravitational
  wave production.
\end{abstract}
\numberwithin{equation}{section}
\newpage

{  \hrule height 0.4mm \hypersetup{colorlinks=black,linktocpage=true}
\tableofcontents
\vspace{0.5cm}
 \hrule height 0.4mm}

\thispagestyle{empty}
\section{Introduction \label{sec:1}}

After the discovery of gravitational waves by the LIGO Collaboration in 2016
\cite{LIGOScientific:2016aoc}, the study of gravitational waves has established
itself as a new tool for the exploration of fundamental physics. Thus, while the
observation by LIGO was confined to the observation of gravitational waves arising from
black hole mergers, the broader study now is related to
stochastic backgrounds of gravitational waves.
Recently, several pulsar timing array (PTA) collaborations, including NANOGrav~\cite{NANOGrav:2023gor}, EPTA~\cite{EPTA:2023fyk}, and PPTA~\cite{Reardon:2023gzh}, have reported strong evidence for a stochastic gravitational wave background (SGWB) at nanohertz frequencies.
 These observations have sparked significant interest in cosmological interpretations involving new physics beyond the standard model~\cite{Agarwal:2025cag,Zheng:2025tcm}.
Thus such backgrounds can arise from a variety of sources such as from primordial black holes and neutron stars, the end of inflation decay of the  inflaton~\cite{Khlebnikov:1997di,Easther:2006vd,GarciaBellido:2007af},
and cosmic phase transitions in the early universe~\cite{Kirzhnits:1972iw,Kirzhnits:1972ut,Witten:1980ez,Guth:1981uk,Steinhardt:1981ct}.
An interesting possibility that arises in this case is that such cosmic phase transitions near the electroweak
scale~\cite{Adams:1993zs,Parwani:1991gq,Arnold:1992rz,Espinosa:1992kf,Quiros:1992ez,Curtin:2016urg,Espinosa:2007qk,Espinosa:2008kw,Azevedo:2018fmj,Mohamadnejad:2021tke,Biermann:2022meg,Wang:2022akn,Addazi:2016fbj,Zhu:2025pht,Aoki:2017aws,Pasechnik:2023hwv,Schwaller:2015tja,Addazi:2017gpt,Fairbairn:2019xog,Co:2021rhi,Borah:2021ocu,Abe:2023zja,Imtiaz:2018dfn,Paul:2020wbz,Breitbach:2018ddu,Ertas:2021xeh,Freese:2022qrl,Bringmann:2023iuz,Banik:2024zwj,Jaeckel:2016jlh,Li:2023bxy,Ghosh:2023aum,DiBari:2021dri,Chen:2023rrl,Gouttenoire:2023naa,Kanemura:2023jiw,Kierkla:2025vwp,Ghosh:2022fzp,Roy:2022gop,Balan:2025uke,Gonstal:2025qky,Carena:2025flp,Dasgupta:2025uzi,Biekotter:2025fjx,Chala:2025oul}
may lead to the production of gravitational waves.
Another important  phenomena at this scale is related to the possibility for baryogenesis with sufficient strength consistent with observation~\cite{Roy:2025zvo,Cohen:1990py,Carena:2000id,Cline:2006ts,White:2016nbo,Cline:2018fuq}.
The standard model alone fails to provide adequate solutions to either phenomena. Thus the electroweak baryogenesis within the standard model cannot generate a sufficient CP violation to explain the observed baryon asymmetry in the universe, and the first-order phase transitions in the standard model do not produce gravitational waves at detectable levels~\cite{Kajantie:1996mn,Gurtler:1997hr,Csikor:1998eu}.
These inadequacies motivate extensions of physics beyond the standard model involving hidden sectors, which appear naturally in supergravity, string theory, D-brane constructions, and quiver theories~\cite{Hill:2000mu}. In this case the analysis of first order phase transitions at the electroweak scale is significantly modified and a possibility exists for achieving
both observable gravitational waves as well as the desired matter-antimatter
asymmetry in the universe in such a set-up~\cite{Girmohanta:2025wcq}.
Further, it is entirely possible that the hidden sector
and the visible sector have interactions beyond the gravitational interactions.
However, the success of the standard model in explaining a large amount of
electroweak data indicates that such interactions must remain small.
Such small interactions between the hidden sector and the visible sector can come about via a variety of
portals: the Higgs portal~\cite{Patt:2006fw}, kinetic mixing~\cite{Holdom:1985ag}, Stueckelberg mass mixing~\cite{Kors:2004dx}, mixed kinetic and Stueckelberg mechanisms~\cite{Feldman:2007wj}.\footnote{The Stueckelberg mass mixing mechanism produces milli-charged dark matter~\cite{Kors:2004dx,Cheung:2007ut,Feldman:2007wj,Feng:2023ubl} in the hidden sector, and a small amount ($\sim 0.3\%$) might explain the EDGES anomaly~\cite{Aboubrahim:2021ohe}, Stueckelberg-Higgs mixing~\cite{Du:2022fqv}, extra-weak interactions~\cite{Feldman:2006wd}, or higher dimensional operators.}

In this work, we present an analysis of cogenesis, i.e., the simultaneous generation of baryon asymmetry and dark matter, within a $U(1)_x$ extension of the Standard Model where $B-L$ symmetry is conserved~\cite{Feng:2013zda}.
In this framework, both the observed matter-antimatter asymmetry and the dark matter abundance arise from out-of-equilibrium decays of heavy Majorana fermions that generate equal and opposite $B-L$ asymmetries in the visible and hidden sectors.
Cogenesis mechanisms naturally yields asymmetric dark matter with a mass of order the proton mass, thereby explaining the observed similarity between the cosmic abundances of baryons and dark matter,
$\Omega_{\rm DM}/\Omega_B \simeq 5.5$~\cite{Kaplan:2009ag,Cohen:2010kn,Graesser:2011wi,Ibe:2011hq,Feng:2012jn,Feng:2013wn,Feng:2016juw}.
In the above framework we further investigate production of
gravitational waves from first-order phase transitions,
which applies broadly to dark matter models featuring asymmetric dark matter with masses around $\mathcal{O}(\mathrm{GeV})$.
The analysis utilizes gauge invariant computation of nucleation and gravitational wave production~\cite{Metaxas:1995ab,Garny:2012cg, Arunasalam:2021zrs,Lofgren:2021ogg, Hirvonen:2021zej,Kierkla:2023von},
particularly in the low-temperature regime ($T/m \lesssim 1$).
In the analyses we use the Metaxas-Weinberg approach,
and obtain robust predictions for bubble nucleation dynamics and the resulting gravitational wave power spectrum.
Our parameter space analysis indicates regions where the model can account for both the observed matter-antimatter asymmetry and the gravitational wave signals that may be within reach of the current and future experiments.
We observe that parameter choices of particular phenomenological interest often involve supercooled or near-supercooled phase transitions with $T_p/m_{A_x} \ll 1$, where $T_p$ is the percolation temperature.

The paper is organized as follows: Section~\ref{Sec:Cogen} presents the $U(1)_x$ extended model and demonstrates the cogenesis mechanism, including asymmetry generation, neutrino mass generation, and dark matter dynamics. Section~\ref{Sec:CW} develops the gauge-invariant analysis of first-order phase transitions and presents parameter space scans showing regions of observability. Section~\ref{Sec:Con} concludes with a summary of our findings and their implications for future gravitational wave observations.
In Appendix~\ref{App:GenAsy}, we derive the full Boltzmann equations for the decays of heavy Majorana fermions, including washout effects.
In Appendix~\ref{App:NuPara}, we present a benchmark model demonstrating that an extended neutrino sector can naturally account for the observed neutrino masses and mixings. \bl{
Appendix~\ref{App:KM} discusses the kinetic mixing between the $U(1)_x$ dark photon and the Standard Model hypercharge gauge boson, including experimental constraints on the kinetic mixing parameter.}
Appendix~\ref{App:GIEA} provides additional discussion of the gauge-invariant effective action for first-order phase transitions. Appendix~\ref{App:Details} presents further details on phase transition physics, hydrodynamics, and gravitational wave production.

\section{Extension of the standard model to include a hidden sector}\label{Sec:Cogen}

\subsection{The model with a \texorpdfstring{$B-L$}{B-L} conserving  hidden sector}

The observed matter-antimatter asymmetry, along with the fact that
dark matter and baryonic matter have abundances of the same order
of magnitude ${\Omega_{{\rm DM}}h^{2}}/{\Omega_{{\rm B}}h^{2}}\approx5.5$~\cite{Planck:2013oqw},
suggests a potential common origin in the early universe. As proposed
in 2013~\cite{Feng:2013zda}, both the baryon asymmetry and dark matter relic abundance
can arise in a framework where $B-L$ symmetry is conserved. In this
paper, we extend this line of investigation by exploring the associated
gravitational wave signatures and other phenomenological implications
of such a scenario.
We consider a $U(1)_{X}$ gauge extension of the standard model, which consists
of at least two Majorana fermions $N_{i}$, a massive Dirac field
$\psi$, a complex scalar dark matter $\phi$, a dark $U(1)_{x}$
Higgs field $\Phi_{x}$ and the $U(1)_{x}$ dark photon $A_{x}$,
two dark matter candidates $X,X^{\prime}$ which are Dirac fermions
charged under $U(1)_{x}$ and also carry non-vanishing lepton number.
The quantum numbers of these fields are summarized in the following
table:
\begin{center}
\begin{tabular}{|c|c|c|c|c|c|c|c|c|}
\hline
 & $L$ & $H$ & $N$ & $\psi$ & $\phi$ & $X$ & $X^{\prime}$ & $\Phi_{x}$\tabularnewline
\hline
\hline
$L$ No. & $+1$ & $0$ & $0$ & $+1$ & $-1$ & $+\tfrac{1}{2}$ & $+\tfrac{1}{2}$ & $0$\tabularnewline
\hline
$B-L$ & $-1$ & $0$ & $0$ & $-1$ & $+1$ & $-\tfrac{1}{2}$ & $-\tfrac{1}{2}$ & $0$\tabularnewline
\hline
$U(1)_{x}$ & $0$ & $0$ & $0$ & $0$ & $0$ & $+1$ & $-1$ & $+1$\tabularnewline
\hline
\end{tabular}
\par\end{center}

The complete Lagrangian of the model is given by
\begin{align}
\mathcal{L} & =\mathcal{L}_{{\rm SM}}+\lambda_{i}\overline{N_{i}}\psi\phi+y_{i}^{\psi}\overline{\psi}L_{i}H+y_{x}\phi\overline{X^{c}}X^{\prime}+h.c.
+ \tfrac{1}{2} \overline{N_i} ({\rm i} \slashed \partial - M_i) N_i \nonumber \\
 & +\overline{\psi}{\rm i}\slashed\partial\psi-m_{\psi}\overline{\psi}\psi+|\partial_{\mu}\phi|^{2}-m_{\phi}^{2}\phi^{*}\phi
 +\overline{X} ({\rm i} \slashed D - m_{\rm DM}) X +\overline{X^\prime} ({\rm i} \slashed D - m_{\rm DM}) X^\prime
 \nonumber \\
 & -\tfrac{1}{4}F_{x\,\mu\nu}F_{x}^{\mu\nu}-\tfrac{\epsilon}{2}F_{x\,\mu\nu}F^{\mu\nu}+|D_{\mu}\Phi_{x}|^{2}
  -V_{{\rm eff}}(\Phi_{x})\,,\label{eq:Lagrangian}
\end{align}
where $F_{x}^{\mu\nu}$ is the $U(1)_{x}$ field strength for the field $A^\mu_x$
and the corresponding covariant derivative of $U(1)_{x}$ is given
by $D_{\mu}=\partial_{\mu}-ig_{x}A_{x\,\mu}$.
Both $X$ and $X^\prime$ are dark matter candidates which have the same mass $m_{\rm DM}$,
have similar lepton number and  $B-L$ number but are oppositely charged under $U(1)_x$.
 Their masses will be  determined in Section~\ref{Sec:DM}.

In a universe where $B-L$ symmetry  (not necessarily
to be a gauge symmetry)  is exactly conserved,  the cogenesis proceeds as follows: at high
energies, heavy Majorana fermions $N_{i}$ decay out of equilibrium,
generating an asymmetry between the fields $\psi,\phi$ and their
antiparticles. Due to the chargeless nature of $N_{i}$, and the fact
that both $\psi,\phi$ carry non-zero $B-L$ charges, these decays
produce equal and opposite $B-L$ asymmetries. The asymmetry is subsequently
transferred to the visible sector via the decay $\psi\to L_{i}H$
and to the dark sector via $\phi\to XX^{\prime}$, resulting in equal
and opposite $B-L$ asymmetries stored in each sector. In the visible
sector, the lepton asymmetry is partially converted into a baryon
asymmetry through sphaleron processes. The similarity in the cosmic
abundances of dark matter and baryons is naturally explained if the
dark matter mass is of the same order as that of the proton.
The symmetric components of $\mathcal{O}({\rm GeV})$ dark matter
primarily annihilate into two $U(1)_{x}$ dark photons $A_{x}$,
leaving the relic density dominated by the asymmetric component.

\subsection{Baryogenesis  via cogenesis}
The key step for the cogenesis mechanism is relying on the heavy Majorana
right-handed neutrinos decay into $\psi,\phi$ at high energies. The
Yukawa coupling constant $\lambda_{i}$ being complex ensures the
CP violating decay of such a process produces the same amount of
asymmetry in the $\psi$ and in $\phi$. The asymmetry of $\psi$
then transforms to the Standard Model via the interaction term
$\overline{\psi}L_{i}H$ and eventually disappear and there would be no vestige left of them
in the current universe. In the dark sector, all $\phi$ particles
eventually decay into dark matter. At later stages of the universe,
the symmetric component of dark matter $X,X^{\prime}$ efficiently
annihilates, leaving the residual asymmetric component as the dominant
constituent of dark matter.
In Eq.~(\ref{eq:Lagrangian}) the couplings $\lambda_{i}$ are assumed to be complex
and the couplings $y_{i}^{\psi},y_{x}$ are assumed to be real. Thus
the only couplings that enter in generating the asymmetry are $\lambda_{i}$.
The decay of chargeless Majorana fermion $N_{i}$ will generate an equal and opposite net $B-L$ asymmetry in the field $\psi$ and $\phi$.
The asymmetry generated via the decaying processes
$N_{i}\to\psi+\phi$ is through the interference of the tree amplitude
and loop diagrams, and is given by\cite{Covi:1996wh,Feng:2013wn}
\begin{align}
\epsilon_{L} & =\frac{\Gamma(N_{1}\to\psi\phi)-\Gamma(N_{1}\to\overline{\psi}\phi^{*})}{\Gamma(N_{1}\to\psi\phi)+\Gamma(N_{1}\to\overline{\psi}\phi^{*})}\simeq-\frac{1}{8\pi}\frac{{\rm Im}(\lambda_{1}^{2}\lambda_{2}^{*2})}{|\lambda_{1}|^{2}}\frac{M_{1}}{M_{2}}\,,\label{eq:ASM}
\end{align}
where we assume there are two Majorana fields $N_{1}$ and $N_{2}$
with $N_{2}$ mass $M_{2}$ being much larger than the $N_{1}$ mass
$M_{1}$, i.e., $M_{2}\gg M_{1}$. When the temperature drops down
below the mass of $N_{i}$, the out-of-equilibrium decay of $N_{i}$
will occur, which will generate the asymmetry in $\psi$ and $\phi$.
The evolution of $N_1$ and the created net lepton number ${L}$, which is equal to the net $\Psi$ number generated ($Y_{L}=Y_{\Psi}$),
are governed by the following set of Boltzmann equations
\begin{align}
\frac{{\rm d}Y_{N_1}}{{\rm d}T} & =\frac{1}{T\bar{H}}\left[\left\langle \Gamma_{N_1}\right\rangle (Y_{N_1}-Y_{N_1}^{{\rm EQ}})-\epsilon_L \left\langle \Gamma_{N_1}\right\rangle Y_{N_1}^{\rm EQ}\times\frac{4\pi^{2}g_{*S}Y_{L}}{15}\right]\nonumber \\
 & +\frac{2s}{T\bar{H}}\Big(\left\langle \sigma v\right\rangle _{N_1N_1\rightarrow\phi\overline{\phi}}+\left\langle \sigma v\right\rangle _{N_1N_1\rightarrow\psi\overline{\psi}}\Big)\Big(Y_{N_1}Y_{N_1}-Y_{N_1}^{{\rm EQ}}Y_{N_1}^{{\rm EQ}}\Big)\,,\label{Eq:DecayN}\\
\frac{{\rm d}Y_{L}}{{\rm d}T} & =-\frac{1}{T\bar{H}}\left(\epsilon_L\left\langle \Gamma_{N_1}\right\rangle (Y_{N_1}-Y_{N_1}^{{\rm EQ}})
-\frac{4\pi^{2}g_{*S}Y_{L}}{15} \left\langle \Gamma_{N_1}\right\rangle Y_{N_1}^{{\rm EQ}} \right)
\nonumber \\
 & +\frac{s}{T\bar{H}}\frac{4\pi^{2}g_{*S}Y_{L}}{15}\left(8\left\langle \sigma v\right\rangle^{\rm RIS} _{\psi\psi\rightarrow\overline{\phi}\overline{\phi}}n_{\psi}^{{\rm EQ}}n_{\psi}^{{\rm EQ}}+4\left\langle \sigma v\right\rangle^{\rm RIS}  _{\psi\phi\rightarrow\overline{\psi}\overline{\phi}}n_{\psi}^{{\rm EQ}}n_{\phi}^{{\rm EQ}}\right)\,.\label{Eq:GenAsy}
\end{align}
where $s=2\pi^{2}h_{{\rm eff}}T^{3}/45$ and
\begin{align}
\bar{H}=\frac{H}{1+\frac{1}{3}\frac{T}{h_{{\rm eff}}}\frac{{\rm d}h_{{\rm eff}}}{{\rm d}T}}=\sqrt{\frac{\pi^{2}g_{{\rm eff}}}{90}}\frac{T^{2}/M_{{\rm Pl}}}{1+\frac{1}{3}\frac{T}{h_{{\rm eff}}}\frac{{\rm d}h_{{\rm eff}}}{{\rm d}T}}\,.
\end{align}
The detailed derivations of Eqs.~(\ref{Eq:DecayN}) and (\ref{Eq:GenAsy}),
including all washout effects, are provided in Appendix~\ref{App:GenAsy}.
The cross-sections of processes ${\psi\psi\rightarrow\overline{\phi}\overline{\phi}}$ and
${\psi\phi\rightarrow\overline{\psi}\overline{\phi}}$ indicated by the superscript ``RIS'',
have been RIS-subtracted to eliminate the double counting associated with the on-shell contribution of the intermediate $N_1$ propagator.
After the sphaleron processes go out of equilibrium, at a temperature $\sim100$~GeV by which the top quark has already decoupled from
the thermal bath, the baryon number and lepton number become separately
conserved, corresponding to the values observed in the present universe.
By analyzing the chemical potentials of the relevant particle species~\cite{Harvey:1990qw,Feng:2012jn},
one can determine the current value of $B_{{\rm f}}$ in term of $(B-L)_{{\rm v}}$
is given by
${B_{{\rm f}}}/{(B-L)_{{\rm v}}}={30}/{97}\,,$
where $B_{{\rm f}}$ denotes the final (the currently observed) value
of the baryon number density.\footnote{The commonly used conversion factor $\tfrac{28}{79}$ is derived under
the assumption that all Standard Model fermions, including the top
quark, remain in thermal equilibrium while sphaleron processes are
active. However, it is known that the top quark decouples prior to
the electroweak crossover, whereas sphaleron processes typically remain
efficient down to temperatures around 100--130~GeV. Consequently,
the top quark should be excluded from the chemical potential relations
used to derive the baryon-to-lepton conversion factor, leading to
a revised coefficient of $\tfrac{30}{97}$~\cite{Harvey:1990qw,Feng:2012jn}.}
Hence, the converted baryon asymmetry is given by
\begin{align}
Y_{B}=\frac{30}{97}Y_{B-L}=\frac{30}{97}Y_{L}\,, \label{eq:Bfinal}
\end{align}
which should match the experimental value $B/n_{\gamma}\sim6\times10^{-10}$~\cite{WMAP:2010qai}.
The current entropy $s_0=7.04n^0_{\gamma}$ leads to $(B/s)_{\rm ex}\simeq 8.6\times10^{-11}$,
and thus $Y_L \simeq 3 \times 10^{-10} $.
For $m_\psi,m_\phi \sim \mathcal{O}({\rm TeV})$,
if $M_i \gg 10^6~{\rm GeV}$ and $|\lambda_i| \sim 10^{-5}$,
the washout effect is less than $1\%$ and is thus negligible.
From above we estimate the size of $\epsilon_{L}\sim 10^{-8}$,
which can be easily manufactured using Eq.~(\ref{eq:ASM}).

\subsection{The active neutrino masses}

The observed tiny neutrino mass is explained by introducing three
right-handed Dirac neutrinos with the Yukawa interactions
\begin{align}
\mathcal{L}_{\nu}=-y_{i}^{\nu}\overline{{\nu}_{i\,R}} L_{i}H+h.c.\,.
\end{align}
We notice that the Yukawa coupling $\overline{\psi}L_{i}H$ also contributes
to the neutrino mass term, and the right-handed neutrino can also
mix with the left-handed component of $\psi$. Hence, the full mass
terms in the neutrino sector can be written as
\begin{align}
\mathcal{L}_{\nu}^{{\rm mass}}=-y_{ij}^{\nu}\overline{ {\nu}_{j\,R}} L_{i}H-y_{i}^{\psi}\overline{{\psi}_{R}} L_{i}H-\mu_{i}\overline{{\nu}_{i\,R}} \psi_{L}+h.c.-m\overline{\psi}\psi\,,
\end{align}
which generate the neutrino mass terms
\begin{align}
\mathcal{L}_{\nu}^{{\rm mass}} & =-m_{ij}^{\nu}(\overline{{\nu}_{i\,L}}\nu_{j\,R}+\overline{{\nu}_{j\,R}}\nu_{i\,L})
-m_{i}^{\psi}(\overline{{\psi}_{R}}\nu_{i\,L}+\overline{{\nu}_{i\,L}}\psi_{R})\nonumber \\
 & \quad-\mu_{i}(\overline{{\psi}_{L}}\nu_{i\,R}+\overline{{\nu}_{i\,R}}\psi_{L})
 -m_{\psi}(\overline{{\psi}_{R}}\psi_{L}+\overline{{\psi}_{L}}\psi_{R})\,,
\end{align}
where $m_{ij}^{\nu}=\tfrac{1}{\sqrt{2}}y_{ij}^{\nu}v_{{\rm SM}}$,
$m_{i}^{\psi}=\tfrac{1}{\sqrt{2}}y_{i}^{\psi}v_{{\rm SM}}$, and can
be rewritten in the matrix form
\begin{align}
\mathcal{L}_{\nu}^{{\rm mass}}=-\left(\begin{array}{cccc}
\overline{{\nu}_{e\,L}} & \overline{{\nu}_{\mu\,L}} & \overline{{\nu}_{\tau\,L}} & \overline{{\psi}_{L}}\end{array}\right)\left(\begin{array}{cccc}
m_{ee}^{\nu} & m_{e\mu}^{\nu} & m_{e\tau}^{\nu} & m_{e}^{\psi}\\
m_{\mu e}^{\nu} & m_{\mu\mu}^{\nu} & m_{\mu\tau}^{\nu} & m_{\mu}^{\psi}\\
m_{\tau e}^{\nu} & m_{\tau\mu}^{\nu} & m_{\tau\tau}^{\nu} & m_{\tau}^{\psi}\\
\mu_{e} & \mu_{\mu} & \mu_{\tau} & m_{\psi}
\end{array}\right)\left(\begin{array}{c}
\nu_{e\,R}\\
\nu_{\mu\,R}\\
\nu_{\tau\,R}\\
\psi_{R}
\end{array}\right)+h.c.\,.\label{eq:nuMM}
\end{align}
A $4\times 4$ unitary matrix $U$ transforms the neutrino flavor eigenbasis to mass eigenbasis,
and the top-left $3\times3$ block of the rotation matrix
$U$ should match the PMNS matrix determined from neutrino oscillation data:
\begin{align}
\big|U_{{\rm PMNS}}^{{\rm ex}}\big|=\left(\begin{array}{ccc}
0.803\sim0.845 & 0.514\sim0.578 & 0.142\sim0.155\\
0.233\sim0.505 & 0.460\sim0.693 & 0.630\sim0.799\\
0.262\sim0.525 & 0.473\sim0.702 & 0.610\sim0.762
\end{array}\right)\,.\label{eq:exPMNS}
\end{align}
The masses of the three light neutrinos should also satisfy: (1) the
upper bound of neutrino mass sum, i.e.,
\begin{align}
\sum_{i}m_{i}^{\nu} \lesssim 0.12~{\rm eV}
\end{align}
(2) the experimental fitting values of the mass-square differences
\begin{align}
\triangle m_{12}^{2} & =m_{\nu2}^{2}-m_{\nu1}^{2}=7.54\times10^{-5}~{\rm eV}^{2}\,,\\
\triangle m_{13}^{2} & =m_{\nu3}^{2}-m_{\nu1}^{2}=2.47\times10^{-3}~{\rm eV}^{2}\,.
\end{align}

We apply the Particle Swarm Optimization (PSO) algorithm to find the
values of all the input parameters. The optimization is done by requiring
the calculated values from seesaw mixing matrix matching all above
mentioned experiment values. To this end, we define 6 fitting functions,
the first 4 of which capture the deviation of the calculated $|U_{{\rm PMNS}}|$,
neutrino mass sum, two mass-square differences to the respective experimental values.
The 5th fitting function ensures that the non-unitary constraint is satisfied,
which sets limits on the size of the mixing between the Standard Model neutrinos and the massive fermion $\psi$.
Such requirement is needed for considerable large mixings between
the Standard Model neutrinos and extra massive fermions.
We further define a 6th fitting function implementing the measured Dirac phase,
resulting in complex-valued couplings $y_i^\psi$.

A set of parameters that reproduce all experimentally measured neutrino data is provided in Appendix~\ref{App:NuPara}.
In particular, we require the $4\times 4$ neutrino mass matrix given in Eq.~(\ref{eq:nuMM})
to take the following form
\begin{align}
\left(\begin{array}{cccc}
m_{ee}^{\nu} & 0 & 0 & \mu_{e}\\
0 & m_{\mu\mu}^{\nu} & 0 & \mu_{\mu}\\
0 & 0 & m_{\tau\tau}^{\nu} & \mu_{\tau}\\
m_{e}^{\psi} & m_{\mu}^{\psi} & m_{\tau}^{\psi} & m_{\psi}
\end{array}\right)\,,\label{eq:nuMMS}
\end{align}
which is remarkable in that the three Standard Model neutrino flavors initially exhibit no mixing in the upper-left
$3\times 3$ block of the mass matrix,
and all observed mixing arises solely from the presence of the new particle $\psi$.
Although the couplings  $y_i^\psi$ are complex,
they do not generate additional asymmetry in the Standard Model leptonic sector via the decay $\psi \to LH$,
due to the extremely strong washout effects arising from
the mass of $\psi$ that lies not far from the electroweak scale.

\subsection{Dark matter budget}\label{Sec:DM}

In the model there are two dark matter candidates $X,X^{\prime}$
which are produced equally from the decay of the heavy $\phi$ particle.
The asymmetric component of $X,X^{\prime}$ constitutes the dominant
contribution to the total dark matter relic abundance,
$\Omega_X h^2 = \Omega_{X^\prime} h^2 \approx 0.06$.
Using ${\Omega_{{\rm DM}}h^{2}}/{\Omega_{{\rm B}}h^{2}}\approx5.5$ and Eq.~(\ref{eq:Bfinal}),
one finds $m_{{\rm DM}}=5.5\times({30}/{97})\times ({1}/{2})\approx0.85\,{\rm GeV}$,
where the $1/2$ factor arises because both $X,X^{\prime}$ carry
a $B-L$ number $-1/2$.
The symmetric component of $X,X^{\prime}$ can be efficiently reduced from the annihilation processes
\begin{align}
\overline{X}X\to A_{x}A_{x}\,,\qquad\overline{X^{\prime}}X^{\prime}\to A_{x}A_{x}\,,
\end{align}
into sub-GeV $U(1)_x$ dark photons $A_x$.
Such dark photon will decay into Standard Model fermion and antifermion pairs ($\overline{f}f$)
because of the kinetic mixing between the Standard Model,
detailed in Appendix~\ref{App:KM}.
The Boltzmann equations govern the evolution of the dark matter particle
$X$ is given by ($X^\prime$ constitutes the other half of the dark matter relic density and the computation of $X^{\prime}$ evolution is similar)
\begin{align}
\frac{{\rm d}Y_{X}}{{\rm d}T} & =\frac{s}{T\bar{H}}\left(Y_{X}Y_{\overline{X}}\langle\sigma v\rangle_{\overline{X}X\to A_{x}A_{x}}-Y_{A_{x}}^{2}\langle\sigma v\rangle_{A_{x}A_{x}\to\overline{X}X}\right)\\
\frac{{\rm d}Y_{\overline{X}}}{{\rm d}T} & =\frac{s}{T\bar{H}}\left(Y_{X}Y_{\overline{X}}\langle\sigma v\rangle_{\overline{X}X\to A_{x}A_{x}}-Y_{A_{x}}^{2}\langle\sigma v\rangle_{A_{x}A_{x}\to\overline{X}X}\right)\\
\frac{{\rm d}Y_{A_{x}}}{{\rm d}T} & =\frac{s}{T\bar{H}}\Big(2Y_{A_{x}}^{2}\langle\sigma v\rangle_{A_{x}A_{x}\to\overline{X}X}-2Y_{X}Y_{\overline{X}}\langle\sigma v\rangle_{\overline{X}X\to A_{x}A_{x}} \non
 & \qquad\qquad+\frac{1}{s}\langle\Gamma\rangle_{A_{x}\to\overline{f}f}+Y_{f}Y_{\bar{f}}\langle\sigma v\rangle_{\overline{f}f\to A_{x}}\Big)
\end{align}
where, $Y_{X}-Y_{\overline{X}}=Y_{L}\simeq 3\times 10^{-10}$,
determined by the cogenesis and was computed in
Eq.~(\ref{Eq:GenAsy}).

\begin{table}[h]
    \begin{center}
    \begin{tabular}{c|ccccccc}
    \hline
    Model & $v_{x}$[GeV] & $g_{x}$ &$m_{A_x}$ & $\epsilon$ & $\tau_{A_x}$[sec] & $\Omega_{sym^{\prime}}h^{2}$ & $\Omega_{sym}h^{2}$\tabularnewline
    \hline
    (A)&1& $0.2$ & 0.2  & $10^{-4}$ & $7\times 10^{-16}$ &$8.3\times10^{-7}$ & $\ll10^{-10}$\\
    (B)&1& $0.1$ & 0.1  & $10^{-4}$ & $2\times 10^{-15}$ &$8.5\times10^{-7}$ & $\ll10^{-10}$\\
    (C)&1& $0.2$ & 0.2  & $10^{-9}$ & $7\times 10^{-6}$ &$8.6\times10^{-7}$ & $\ll10^{-10}$\\
    (D)&1& $0.1$ & 0.1  & $10^{-9}$ & $2\times 10^{-5}$ &$8.9\times10^{-7}$ & $\ll10^{-10}$
    \end{tabular}
    \par\end{center}
    \caption{Benchmark models illustrating annihilation of the symmetric $X$ component into pairs of sub-GeV dark photons.
    We list $\Omega_{\mathrm{sym}} h^{2}$ and $\Omega_{\mathrm{sym}'} h^{2}$,
    the relic abundances of the $X$ symmetric component computed with and without the dark matter asymmetry.
    \VF{We consider two representative values for the kinetic mixing parameter,
    $\epsilon = 10^{-4}$ and $\epsilon = 10^{-9}$,
    which are shown as black lines within the allowed regions in Fig.~\ref{fig:DPKM}.
    The kinetic mixing leads to a dark photon lifetime much shorter than 1~second, so the dark photon decays before BBN.}}
    \label{Tab:DMEvo}
    \end{table}

\begin{figure}
    \centering
     \includegraphics[width=0.46\linewidth]{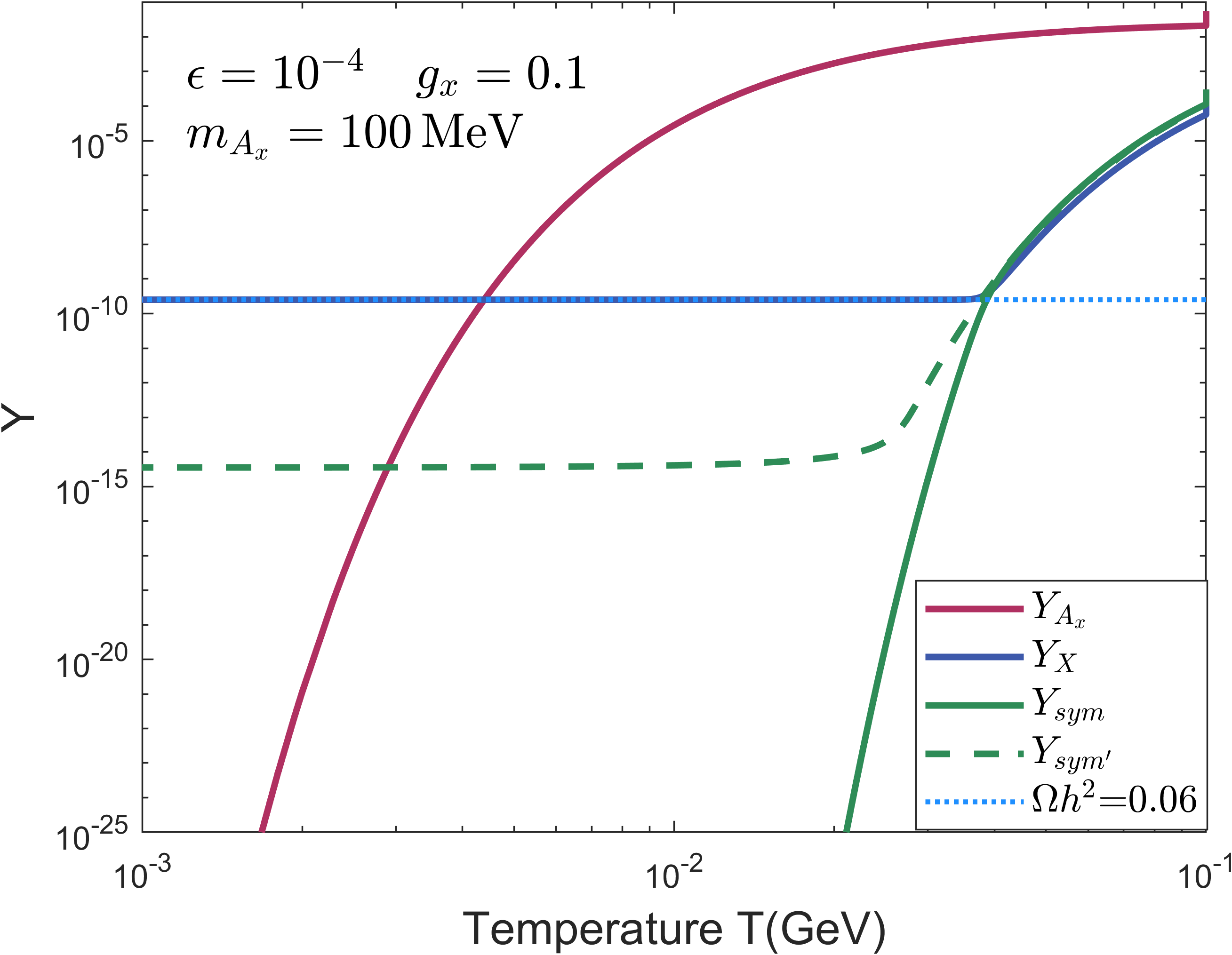}\quad
     \includegraphics[width=0.46\linewidth]{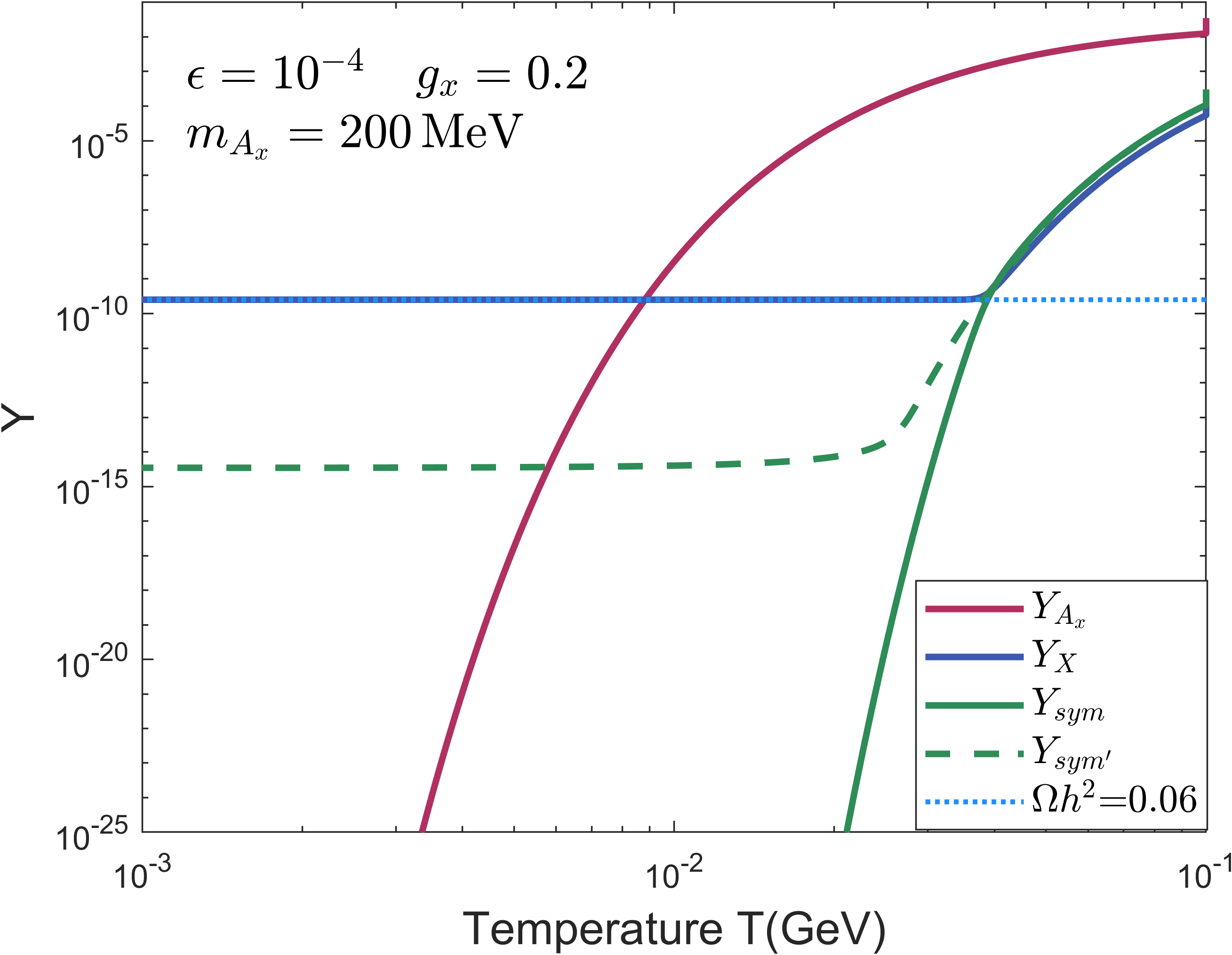}\\~\\~\\
     \includegraphics[width=0.46\linewidth]{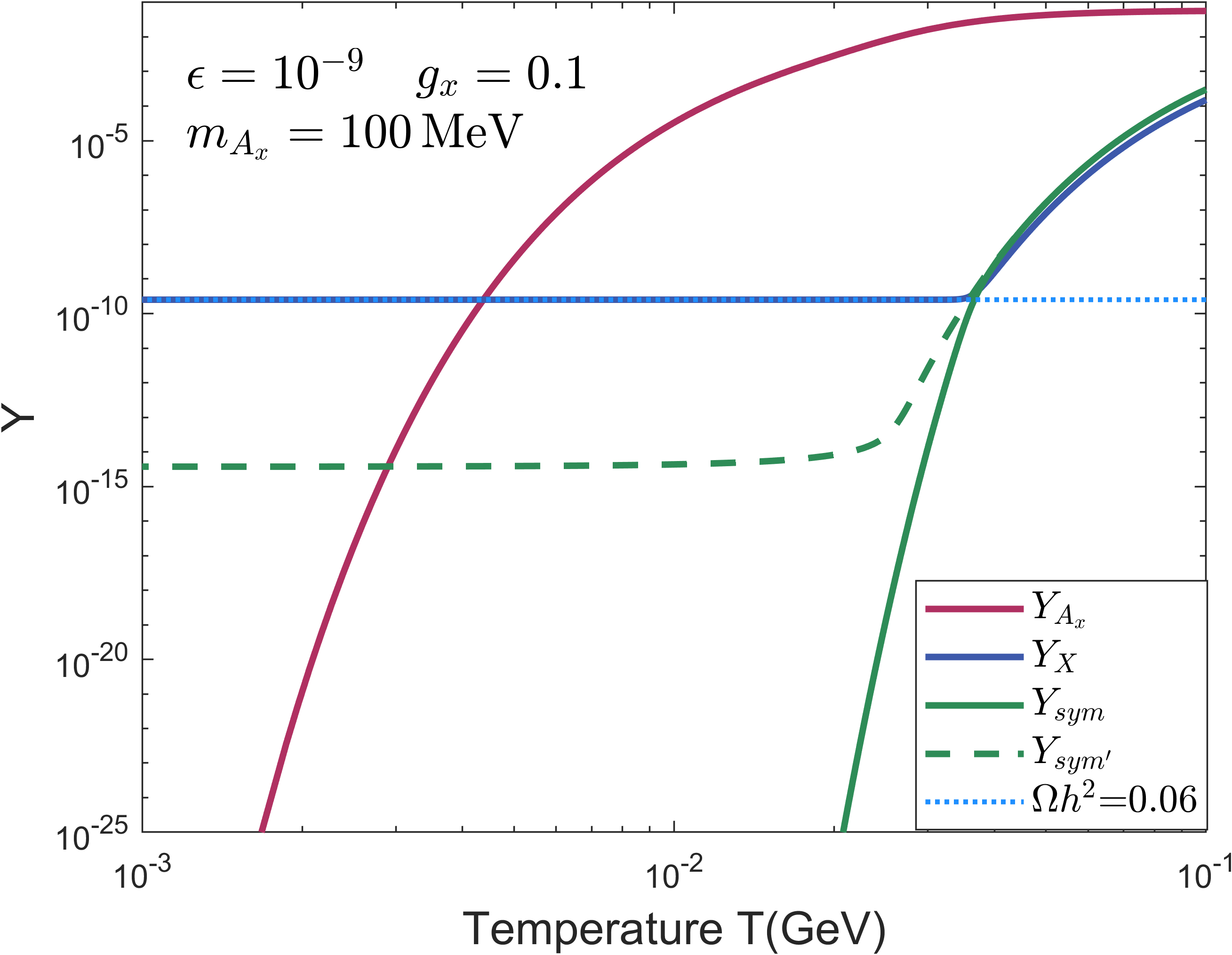}\quad
     \includegraphics[width=0.46\linewidth]{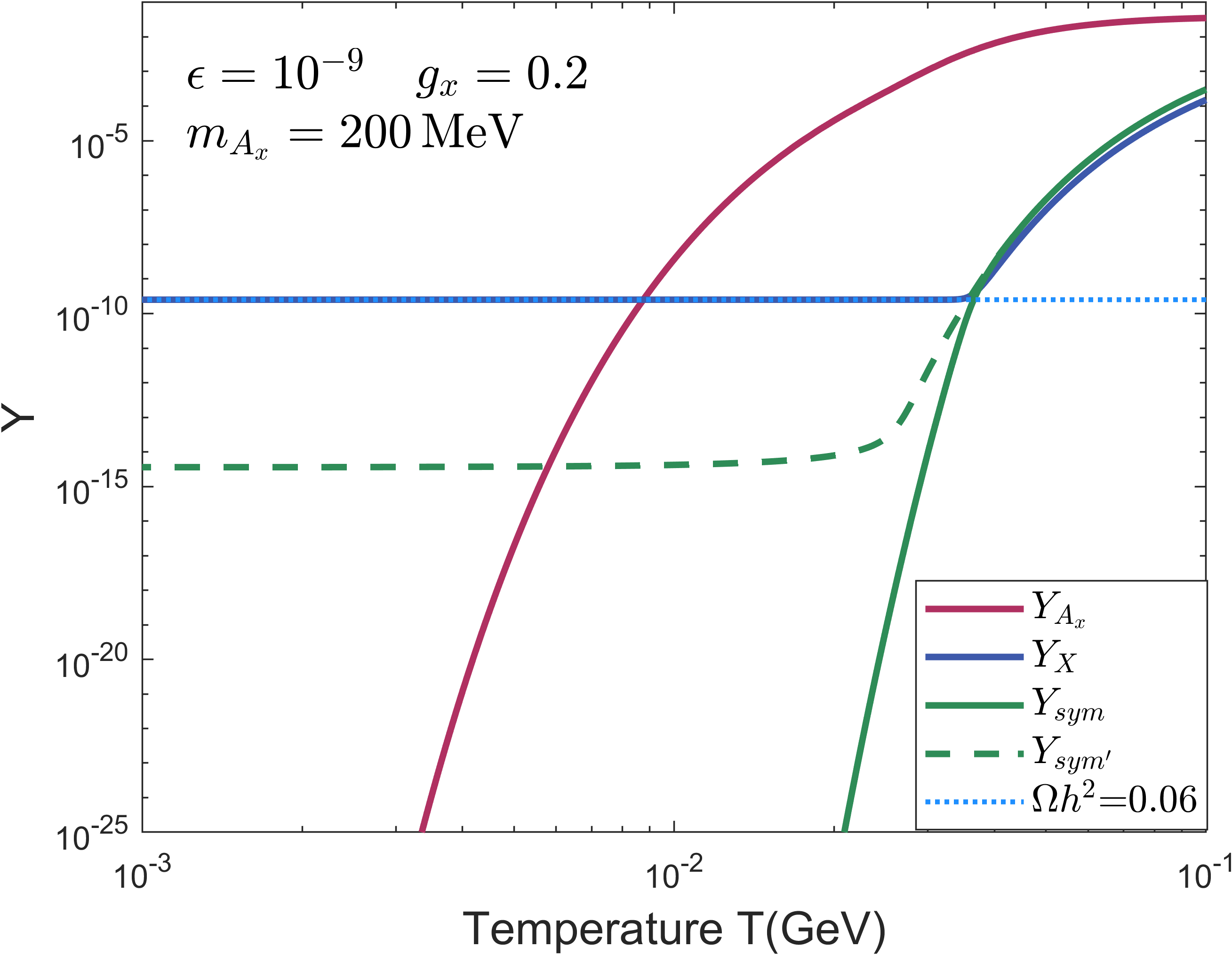}
    \caption{Thermal evolution of $Y_{A_x}, Y_X, Y_{sym}, Y_{sym^\prime}$ for
    four benchmark models labeled by $\epsilon, g_x, m_{A_x}$ as given in each of the four panels.
     The plots show the comoving number densities of asymmetric dark matter $X$, its symmetric component ($Y_{sym}=2Y_{\overline{X}}$), and the dark photon $A_x$ as functions of the universe temperature.
    $Y_{sym^\prime}$ denotes the symmetric abundance of the dark matter in the absence of dark matter asymmetry,
    i.e., when $Y_X - Y_{\overline{X}} = 0$.
    The final abundance of  $Y_X$ corresponds to a relic density $\Omega_X h^2 = 0.06$,
    as shown by the light blue dotted line, while $X^\prime$ contributes the other half.
    In the presence of  dark matter asymmetry, the symmetric component is more efficiently annihilated through the process
    $\overline{X}X\to A_x A_x$.
    The $U(1)_x$ dark photon $A_x$  which mixes with the  hypercharge gauge field via a kinetic term with kinetic parameter $\epsilon$
    will eventually decay into Standard Model fermion pairs before BBN.}
    \label{fig:DMEvo}
\end{figure}

We present the dark matter evolution for the four representative benchmarks in Table~\ref{Tab:DMEvo},
and the evolution of $X,\overline{X}$ and the dark photon $A_x$ for the four models are shown in Fig.~\ref{fig:DMEvo}.
The symmetric component of dark matter can be efficiently annihilated through the secluded process  $\overline{X}X\to A_x A_x$,
leaving the asymmetric component as the dominant contributor to the total dark matter relic density.
Compared to the scenario without a dark matter asymmetry ($Y_X - Y_{\overline{X}} = 0$),
the symmetric component is more efficiently depleted in the presence of an asymmetry.

\begin{figure}[h]
    \centering
     \includegraphics[width=0.43\linewidth]{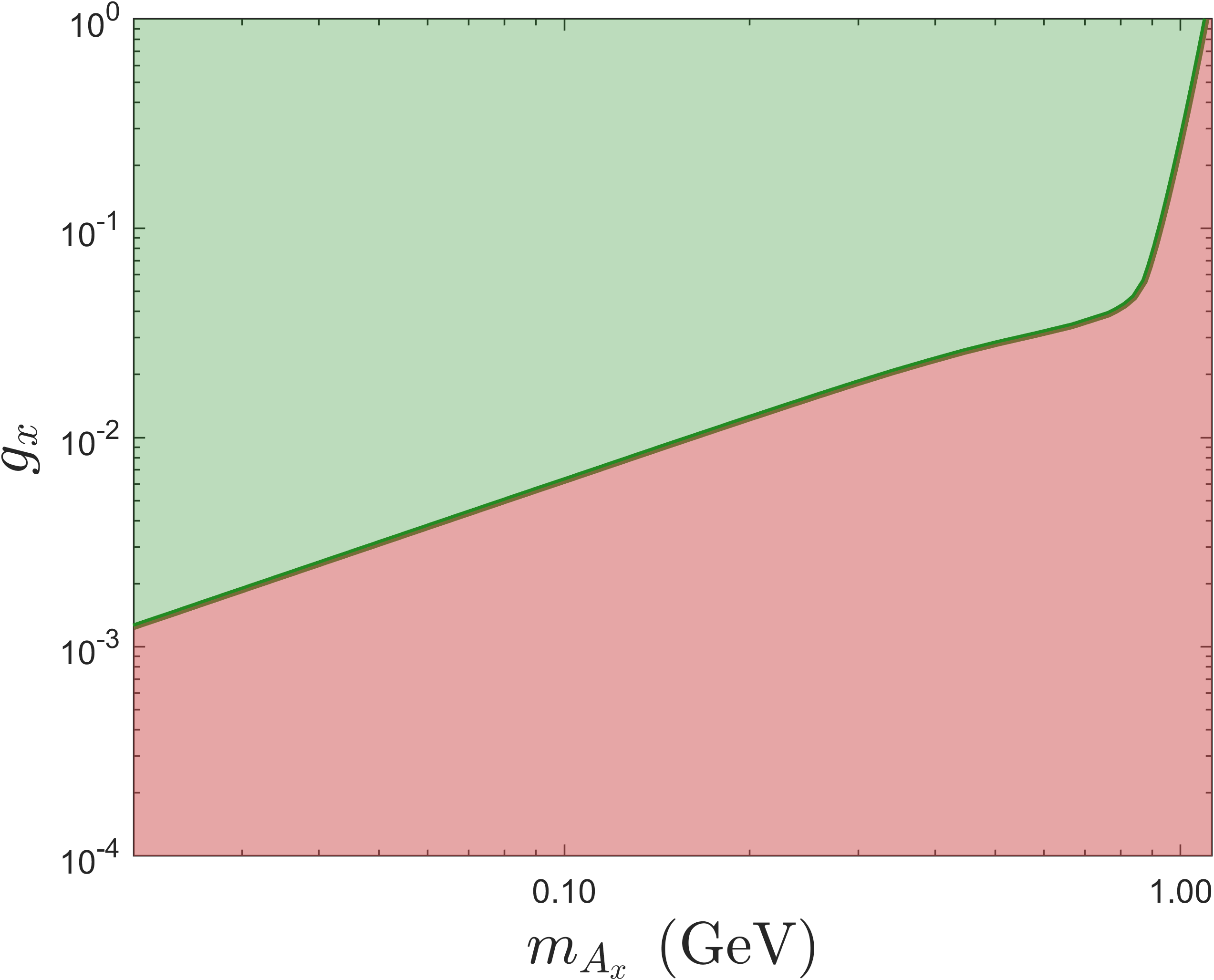}\qquad
     \includegraphics[width=0.44\linewidth]{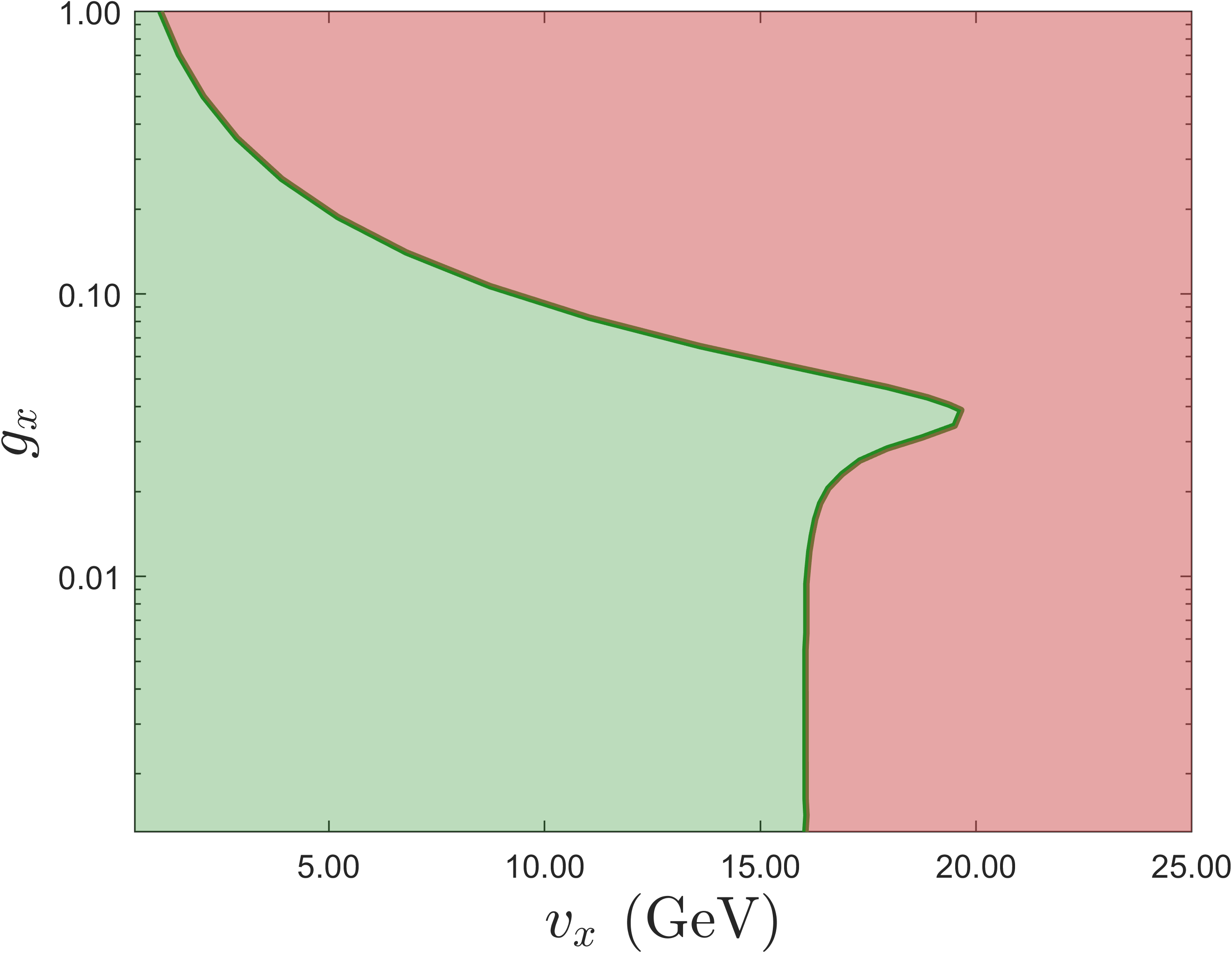}
     \caption{Left panel: $g_x$ versus the dark photon mass $m_{A_x}$.  Right panel: $g_x$ versus the vacuum expectation value  of the $U(1)_x$ Higgs field, $v_x$.
The shaded light-green regions in both panels indicate the parameter space of $g_x$ that yields sufficient secluded annihilation of the symmetric dark matter component, for dark photon masses in the range $20~{\rm MeV}-1~{\rm GeV}$.
The kinetic mixing between $U(1)_x$ and the hypercharge gauge field is fixed at $\epsilon = 10^{-4}$,
consistent with current experimental constraints.}
\label{fig:gxbd}
\end{figure}
Further, we note that kinetic mixing between the hidden sector and the visible sector is tightly constrained for dark photon masses below
20~MeV~\cite{Fabbrichesi:2020wbt,Aboubrahim:2022qln}.
In the analysis here we compute the minimal $U(1)_x$ gauge coupling that yields a sufficient annihilation of the symmetric dark matter component,
so that the constraint
 $\Omega_{sym}h^2\lesssim 1\% \times 0.12$, for dark photon masses in the range $20~{\rm MeV}-1~{\rm GeV}$ is satisfied.
The result of our analysis is shown in Fig.~\ref{fig:gxbd}, where the light green regions indicate the allowed values of $g_x$ for the case when
the kinetic mixing parameter
$\epsilon = 10^{-4}$, a value consistent with experimental bounds for dark photon masses in the range $20~{\rm MeV}-1~{\rm GeV}$.

\newpage
\section{Gauge invariant analysis of first order phase transition from a \texorpdfstring{$U(1)_{x}$}{U(1)x} gauge invariant hidden sector~\label{Sec:CW}}
While the model of Eq.~(\ref{eq:Lagrangian}) provides a possible cogenesis mechanism that explains the matter-antimatter asymmetry,
a first-order phase transition could also occur, potentially generating observable stochastic gravitational waves. Since gravitational wave
production is gauge dependent, we
adopt the approach developed by Metaxas and Weinberg~\cite{Metaxas:1995ab} to obtain a gauge-independent effective bounce action,
as detailed in Appendix~\ref{App:GIEA}. Using this approach we compute the leading and the sub-leading actions which are both gauge
independent. Thus the leading terms in the total action $S_{\rm eff}$ are given by
\begin{align}
  S_{\rm eff} &= S_0 + S_1\,,\label{stotal}\\
  S_0 &= \beta\int {\rm d}^3 x\left[\frac{1}{2}(\partial_\mu\phi_b)^2 + V_{g_x^4}(\phi_b)\right]\,,\label{s0}\\
  S_1 &= \beta\int {\rm d}^3 x\left[\frac{1}{2}Z_{g_x^2}(\phi_b,\xi)(\partial_\mu\phi_b)^2 + V_{g_x^6}(\phi_b,\xi)\right]\,.
  \label{s1}
\end{align}
where the definitions of the quantities that enter in Eqs.~(\ref{s0}) and (\ref{s1})
are given in Appendix~\ref{App:GIEA}. Here $S_0$ is gauge
invariant by construction and $S_1$ is gauge invariant due to Nielsen Identity.
Details are given in Appendix~\ref{App:GIEA}.
For comparative analysis, we calculate the gauge-dependent effective bounce action employing the original effective potential given in Eq.~(\ref{eq:Veff}), with $Z(\phi) = 1$ and utilizing the tree level mass for $m_G$, such that the gauge dependent effective bounce action $S^{\rm G.D.}_{\rm eff}$ is defined so that
 \begin{align}
   S^{\rm G.D.}_{\rm eff} = \beta\int {\rm d} x^3\left[\frac{1}{2}(\partial_\mu\phi)^2 + V_{\rm eff}(\phi,\xi) \right]\,.
  \label{SG.D.}
\end{align}
A standard gauge choice is the Landau gauge ($\xi = 0$), which is adopted in the majority of studies in the literature.

\subsection{Numerical results using gauge invariant effective bounce action}
To implement this methodology, it is not necessary to solve for the complete profile of $\phi_b(x)$ with the renormalization factor $Z(\phi)$, as only the bounce solution for the leading-order term is required, where the leading order of $Z(\phi)$ is unity. Therefore, only a minor modification to the action $S_1$ calculation is needed. We thus use the \textbf{CosmoTransition} package with appropriate modifications for this analysis.
Once we determined this gauge invariant effective action, we follow the pipeline given in our previous paper~\cite{Li:2025nja} to finish the rest of the analysis on gravitational wave prediction. A brief summary is given in Appendix~\ref{App:Details}. One of the major difference here is that in this paper, we have the hidden sector in thermalization with the visible sector during the phase transition, i.e. $T_h = T$.

We first examine the validity of this method under low-temperature conditions, i.e. $T/m_{A_x} < 1$. We evaluate the associated uncertainties of this approach across different temperature regimes, particularly for cases where $T/m_{A_x} \ll 1$ and $T/m_{A_x} \sim 1$. In Table \ref{tab:benchmark1} we provide three benchmark model with different temperature to mass ratios. Also, we provide a benchmark model (*) which is an example of supercooled phase transition that stayed valid after all kind of constraints.
\begin{table}[h]
    \centering
    \begin{tabular}{c|ccccc}
        \hline
         Model& $g_x$ & $\lambda_x$      & $v_x$[GeV] & $T_p$[GeV] & $T_p/m_{A_x}$  \\
        \hline
          (*) & 0.58 & $2.21\times10^{-3}$& 0.239 & 0.01 & 0.074  \\
          (a) & 0.691  & $4.6\times10^{-3}$ & 1 & 0.093 & 0.13 \\
          (b) & 0.400  & $1\times10^{-3}$ & 1 & 0.128 & 0.32 \\
          (c) & 0.145  & $1\times10^{-4}$ & 1 & 0.135 & 0.93 \\
    \end{tabular}
    \caption{Benchmark models for test of gauge invariance action up to one loop order. Here $g_x$ is the $U(1)_x$ gauge coupling, $\lambda_x$ is the Higgs self-coupling, and $v_x$ is the vacuum expectation value
        which are free parameters. $T_p$ is the percolation temperature defined by Eq.~(\ref{eq:percolation}). $T_p/m_{A_x}$ gives the temperature-mass ratio. We note that the model labeled (*) originates from a
     supercooled phase transition which remains valid after all the relevant constraints are satisfied.  }
    \label{tab:benchmark1}
\end{table}
\begin{figure}[h]
    \centering
    \includegraphics[width=0.24\linewidth]{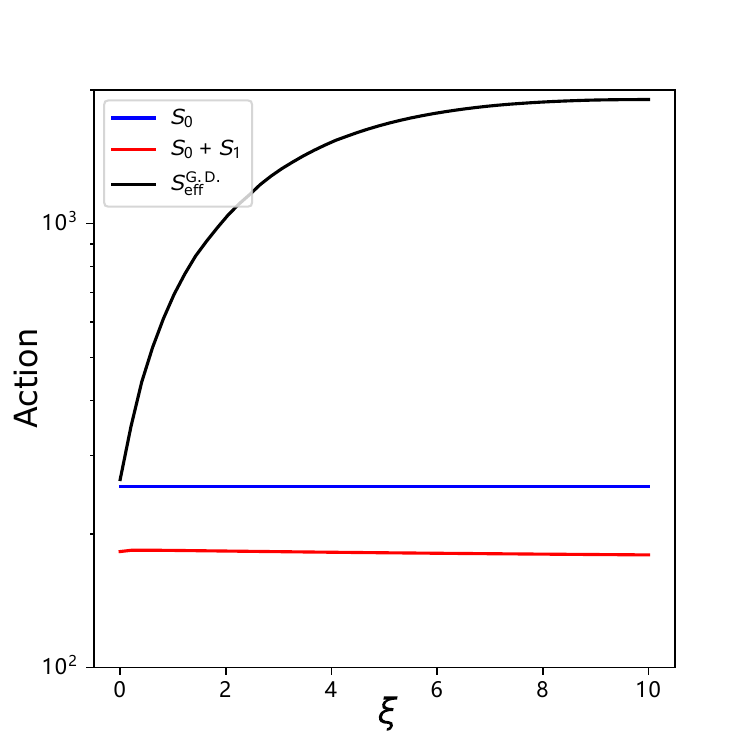}
    \includegraphics[width=0.24\linewidth]{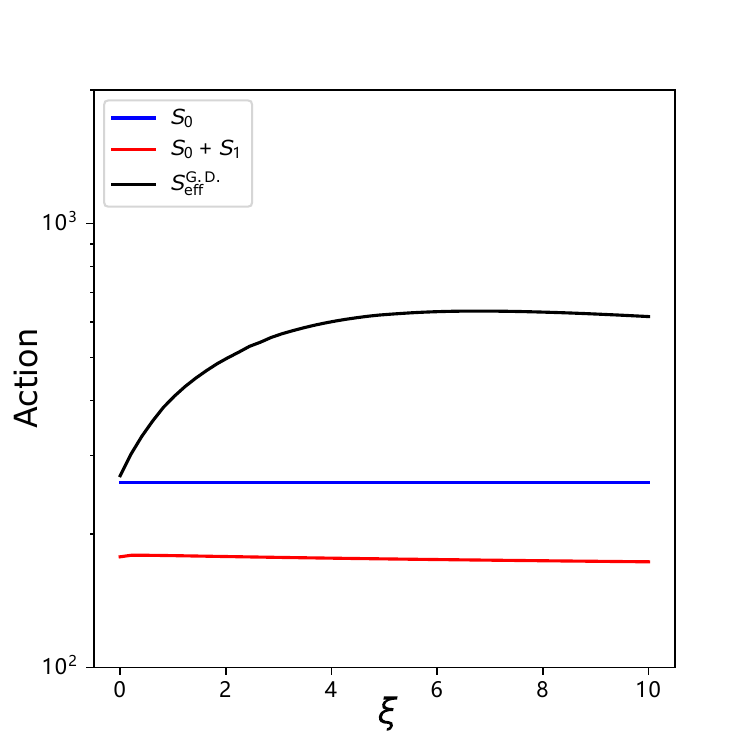}
    \includegraphics[width=0.24\linewidth]{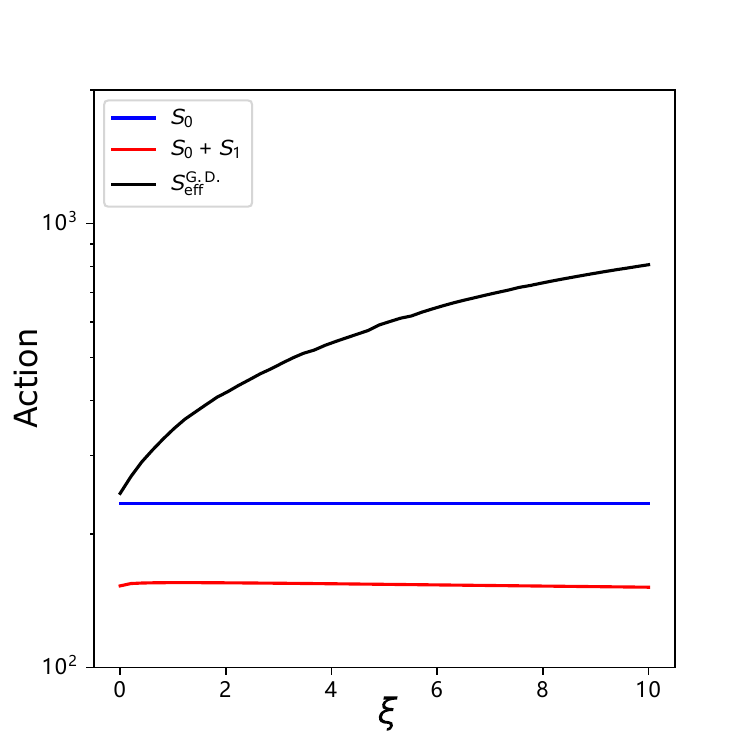}
    \includegraphics[width=0.24\linewidth]{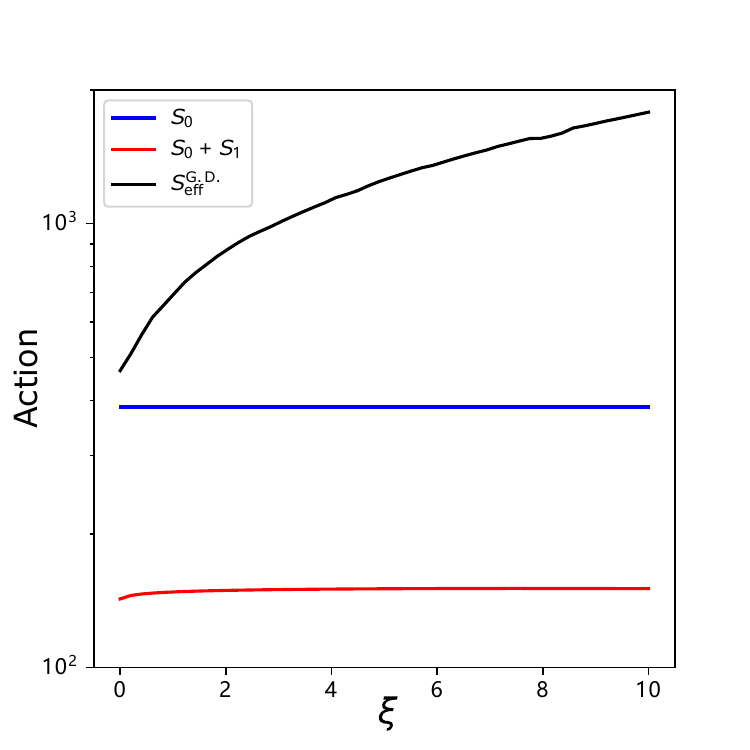}
    \caption{A display of effective bounce actions $S^{\rm G.D.}_{\rm eff}$ defined in Eq. (\ref{SG.D.})
     in Landau gauge ($\xi = 0$), along with $S_0$ and $S_0+S_1$ with different choices of $\xi$. From left to right, the models are (*),(a), (b) and (c) where the models are defined in  Table \ref{tab:benchmark1}.}
    \label{fig:Actions}
\end{figure}

Using the parameter values specified in Table~\ref{tab:benchmark1}, we present the computational results for $S_{\text{eff}}^{\rm G.D.}$ with Landau gauge ($\xi = 0$), $S_0$, and $S_0+S_1$ in Fig.~\ref{fig:Actions}. The temperature is set to the percolation temperature $T_p$, which represents a critical parameter in gravitational wave analysis since it determines the onset of efficient bubble nucleation and sets the scale for the phase transition dynamics. Our calculations show that the sum $S_{0} + S_1$ remains approximately constant across different values of the parameter $\xi$ for all three examined cases, with a maximum deviation of only about~5\%. This stability provides clear evidence for the successful derivation of a gauge-invariant effective action across all investigated temperature conditions where $T/m_{A_x} < 1$. By contrast, $S_{\rm eff}$ exhibits a pronounced and systematic dependence on $\xi$, with values that can vary by nearly an order of magnitude as $\xi$ increases from 0 to 10. This spread illustrates the potential pitfalls of relying on gauge-dependent quantities, since even modest changes in the gauge parameter can lead to large shifts in the predicted nucleation dynamics, and by extension in the resulting gravitational wave signals.

\subsection{Gravitational wave prediction sensitivity on gauge dependence and gauge-invariant result}

\begin{figure}[h]
    \centering
    \includegraphics[width=0.4\linewidth]{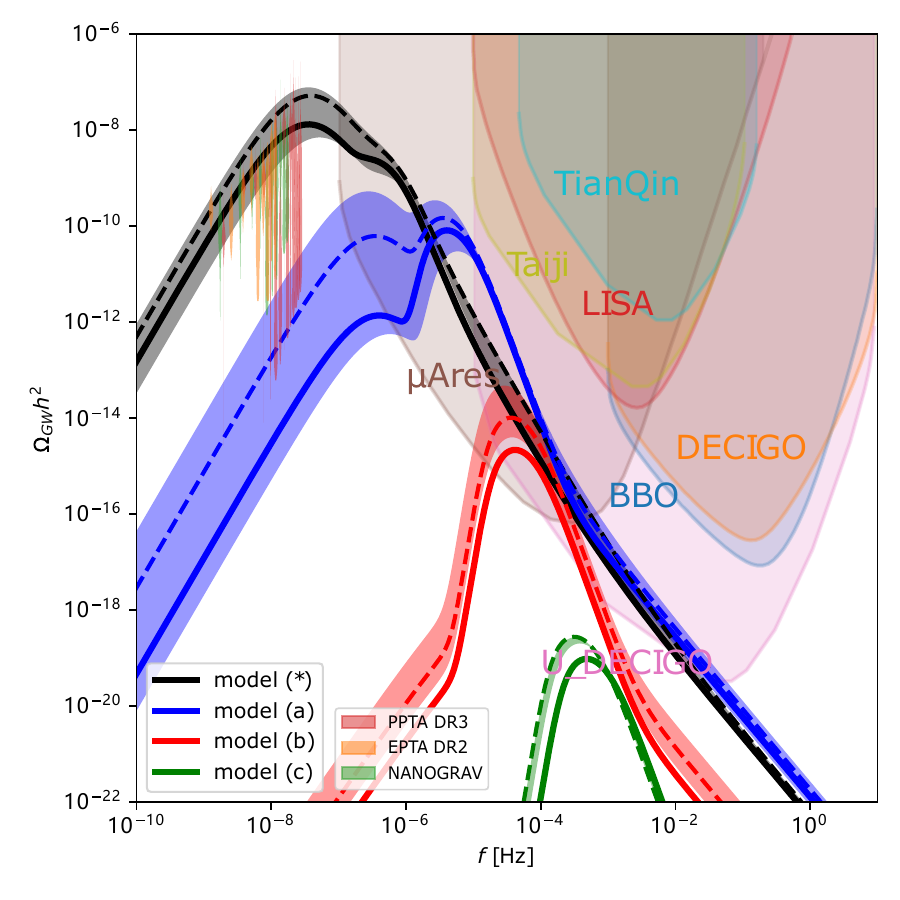}
    \caption{Analysis of gravitational wave power spectra for models (*), (a), (b), and (c) defined in
    Table  \ref{tab:benchmark1}
     which shows variations in the gravitational wave power spectrum $\Omega_{GW} h^2$  on the  gauge parameter $\xi$ across the range $(-1,1)$ exhibited by the color bands for each model. The solid line displays the gauge-invariant solution.The dashed line displays the Landau gauge solution. Shaded regions represent pulsar timing array signals from NANOGrav~\cite{NANOGrav:2023gor}, PPTA~\cite{Reardon:2023gzh}, and EPTA~\cite{EPTA:2023fyk}, while detection regions for planned space-based gravitational wave detectors are shown, including LISA \cite{LISA:2017pwj,Baker:2019nia,AmaroSeoane:2012km}, BBO\cite{Grojean:2006bp}, Decigo\cite{Kawamura:2006up}, Taiji\cite{Ruan:2018tsw}, TianQin\cite{TianQin:2015yph}, and $\mu$Ares\cite{Sesana:2019vho}.}
    \label{fig:GWpower_xivariation}
\end{figure}

We perform further analysis of bubble nucleation and gravitational wave production using benchmark models in Table~\ref{tab:benchmark1}, with the gauge fixing parameter $\xi$ chosen from the range $(-1,1)$. The results are shown in Fig.~\ref{fig:GWpower_xivariation}. Our analysis demonstrates that variations in $\xi$ across this parameter range produce power spectrum curves that differ by several orders of magnitude.
The Landau gauge selection typically employed in most of studies shows substantial differences from our gauge-invariant findings. These results emphasize the critical importance of implementing gauge-invariant approaches in such computations.
From Fig.~\ref{fig:GWpower_xivariation},
we see that the spread in predictions due to gauge choice is not uniform across the spectrum. The gauge dependence is particularly large in regions where the gravitational wave power is the  highest, i.e., where signals would be the most promising for detection by pulsar timing arrays or future space-based interferometers. This implies that the very signals most relevant for near-future experiments are also the most vulnerable to spurious gauge effects. Without gauge-invariant treatment, the predicted amplitude and peak frequency in these regions could be significantly
misestimated, leading to false expectations about their observability.
By contrast, the solid line in Fig.~\ref{fig:GWpower_xivariation} represents the gauge-invariant result, which eliminates the $\xi$ dependent effects and provides a robust prediction.

\bl{
The impact of gauge dependence on phase transition dynamics becomes particularly significant in the case of strongly supercooled transitions. At leading order, the partial cancellation between ghosts and Goldstone boson contributions leads to only modest gauge dependence at next-to-leading order. However, when the potential barrier is extremely small, as in supercooled cases, even small gauge-induced variations can dramatically affect the nucleation rate and the resulting gravitational wave signal. This behavior can be clearly seen in Fig.~\ref{fig:GWpower_xivariation}: for the non-supercooled benchmark models (b) and (c), the spread due to gauge choice is minimal, whereas in the supercooled benchmarks (models (*) and (a)), gauge dependence becomes large. Previous studies such as Ref.~\cite{Croon:2020cgk} have analyzed theoretical uncertainties including gauge dependence mainly in the context of general non-supercooled transitions where the effect is less pronounced. Our research highlights that in the supercooled regime, gauge dependence can lead to significant error. This underscores the critical necessity of adopting gauge-invariant methods to obtain reliable predictions, especially for supercooled phase transitions.
}

\subsection{Parameter scan for the \texorpdfstring{$U(1)_x$}{U(1)x} extended model}\label{sec:Parameterscan}
\begin{figure}[!h]
    \centering
    \includegraphics[width=0.4\linewidth]{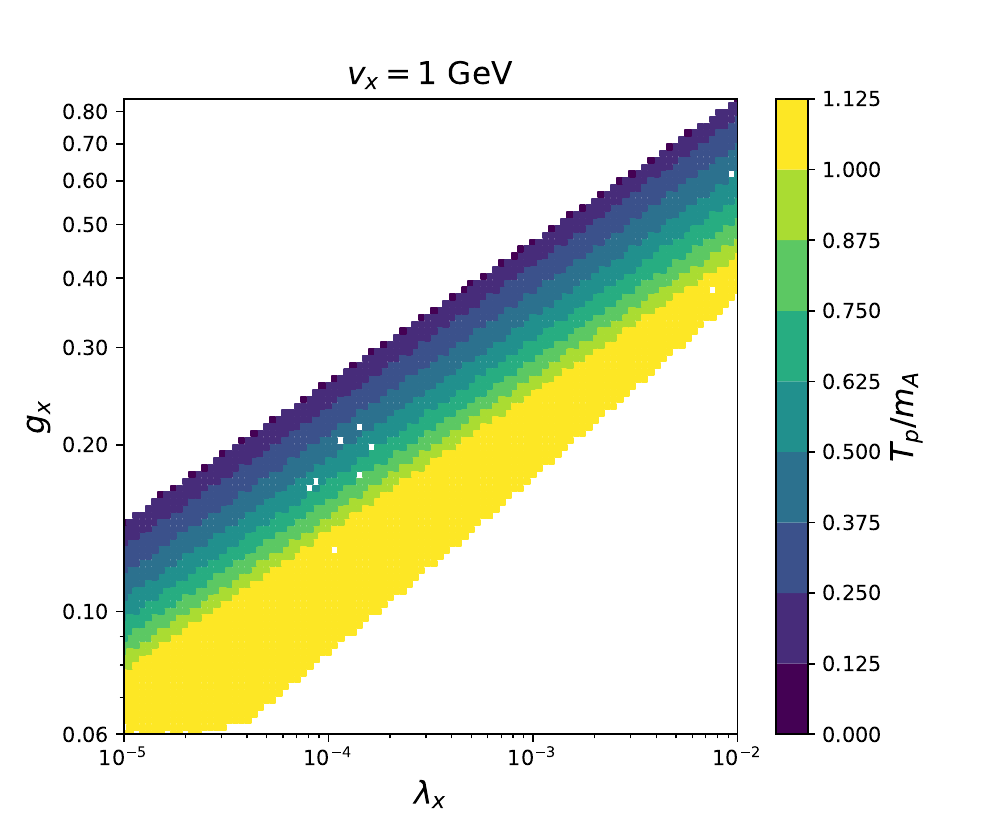}
    \includegraphics[width=0.4\linewidth]{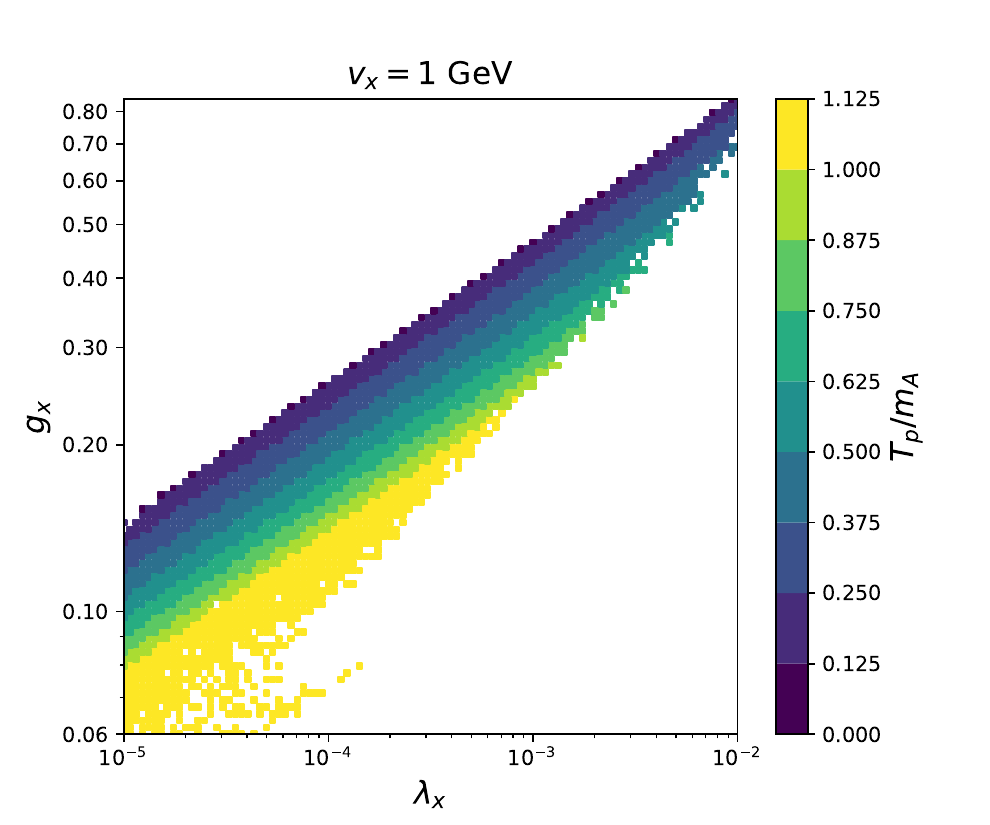}
    \includegraphics[width=0.4\linewidth]{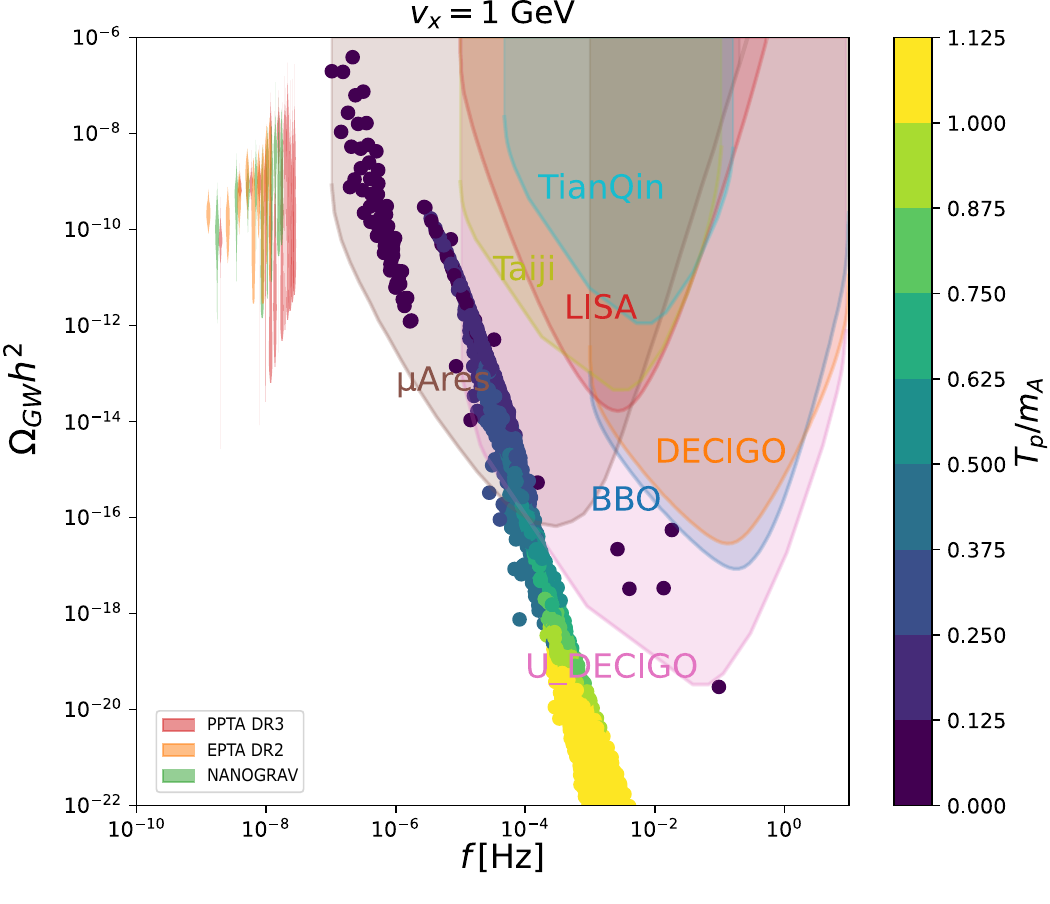}
    \includegraphics[width=0.4\linewidth]{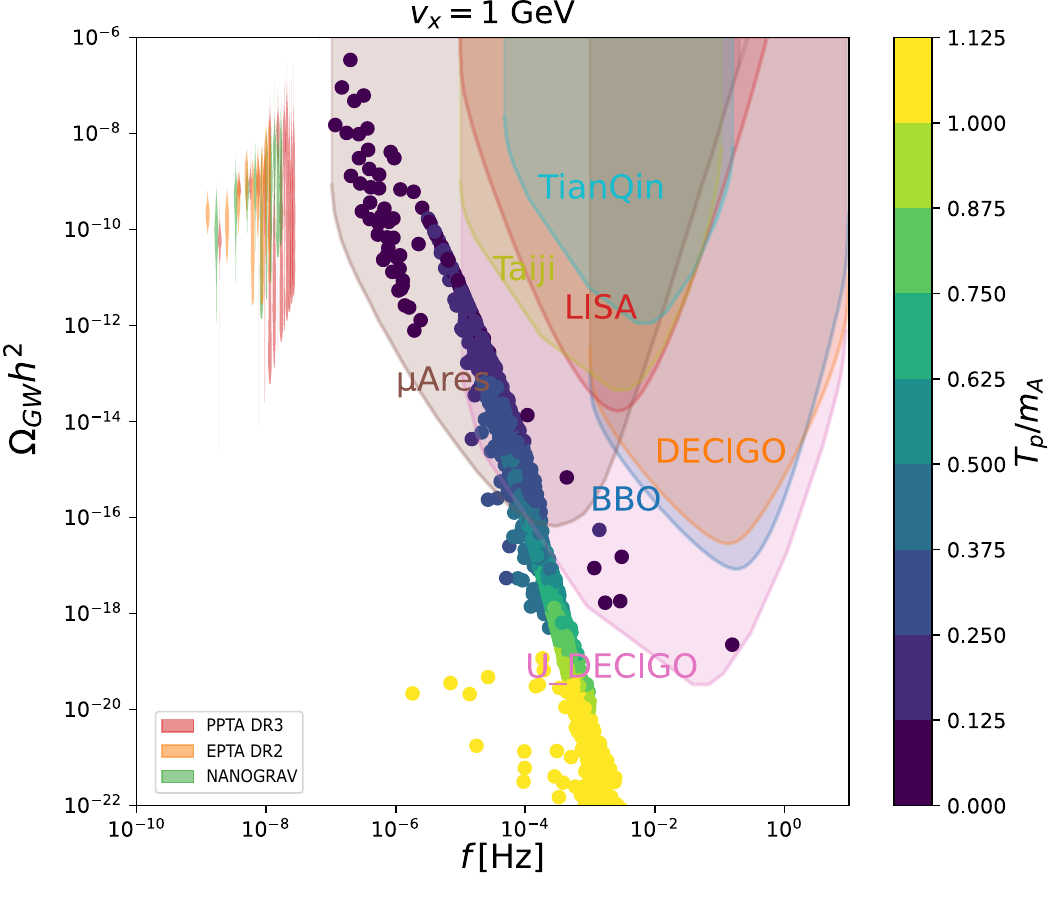}
    \caption{Upper left panel: Parameter scan within the $g_x - \lambda_x$ parameter plane, with $v_x$ fixed at 1 GeV. It presents results obtained using gauge-dependent effective action with Landau gauge selection ($\xi = 0$). Upper right panel: Parameter scan within the $g_x - \lambda_x$ parameter plane, with $v_x$ fixed at 1 GeV. It presents results obtained using gauge invariant analysis.
   The color coding along the left vertical axis indicates the ratio $T_p/m_{A_x}$ for both left and
   right panels.
       Lower panels: Theoretical predictions for gravitational wave power spectra $\Omega_{GW} h^2$
    as a function of the frequency corresponding to the parameter configurations displayed in the upper panels. Each point marks the peak of the power spectrum.}
    \label{fig:scan}
\end{figure}
We examine the cosmological phase transition in a minimal $U(1)_x$ extension of the Standard Model. The one-loop finite temperature effective potential depends on three parameters:
\begin{align}
   g_x\,,\qquad \lambda_x\,,\qquad v_x = \frac{\mu_x}{\sqrt{\lambda_x}}\,,
\end{align}
We conduct a parameter space scan over the following ranges:
$g_x\in(0.06,\,0.8),\,\lambda_x\in(10^{-5},\,10^{-2}),\,v_x = 1~\text{GeV}$.
For comparison purposes, we apply the scan using the gauge-dependent effective action and select the Landau gauge ($\xi = 0$), which is the gauge choice employed in most other similar studies. Then, we apply the scan with the gauge invariant analysis. The result is shown in Fig.~\ref{fig:scan}. The scans illustrate how gauge treatment reshapes both the allowed region in the $(g_x,\lambda_x)$ plane and the distribution of corresponding gravitational wave spectra. In top panels, colored points indicate parameters for which a first-order transition nucleates and percolates; notably, the envelope and internal texture of the viable band change when moving from Landau gauge ($\xi{=}0$) to the gauge-invariant analysis. In particular, part of the parameter space that appears allowed in the gauge-dependent case is removed once the gauge-invariant construction is imposed.
The bottom panels convert these identical parameter values into gravitational wave power spectra. A notable observation is that the color gradient additionally demonstrates that numerous phenomenologically significant points concentrate near $T_p/m_{A_x} \leq 1$, confirming that these transitions take place at comparatively low temperatures. This indicates that the frequently employed high-temperature expansion ($T/m\gg 1$) may not be applicable in this context, and a low-temperature, gauge-invariant approach should be implemented instead.

\begin{figure}[!h]
    \centering
    \includegraphics[width=0.24\linewidth]{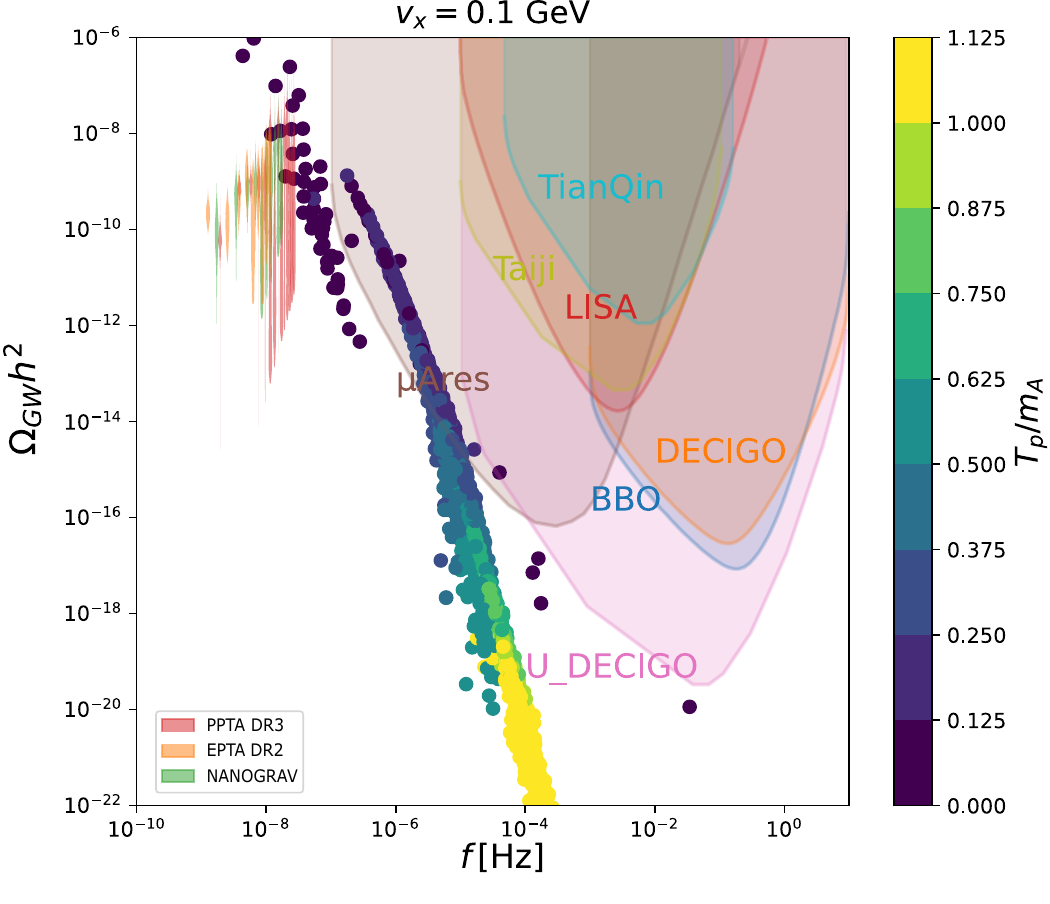}
    \includegraphics[width=0.24\linewidth]{Figs/GWv1Vgi.pdf}
    \includegraphics[width=0.24\linewidth]{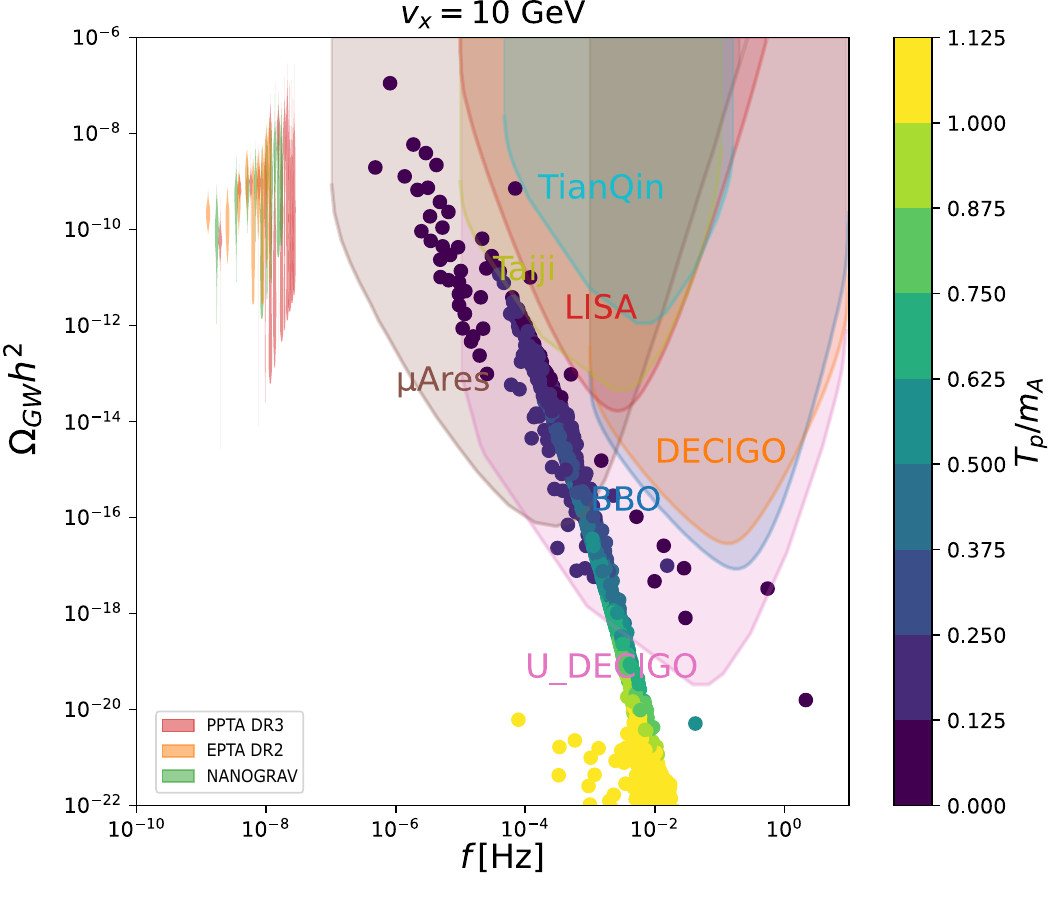}
    \includegraphics[width=0.24\linewidth]{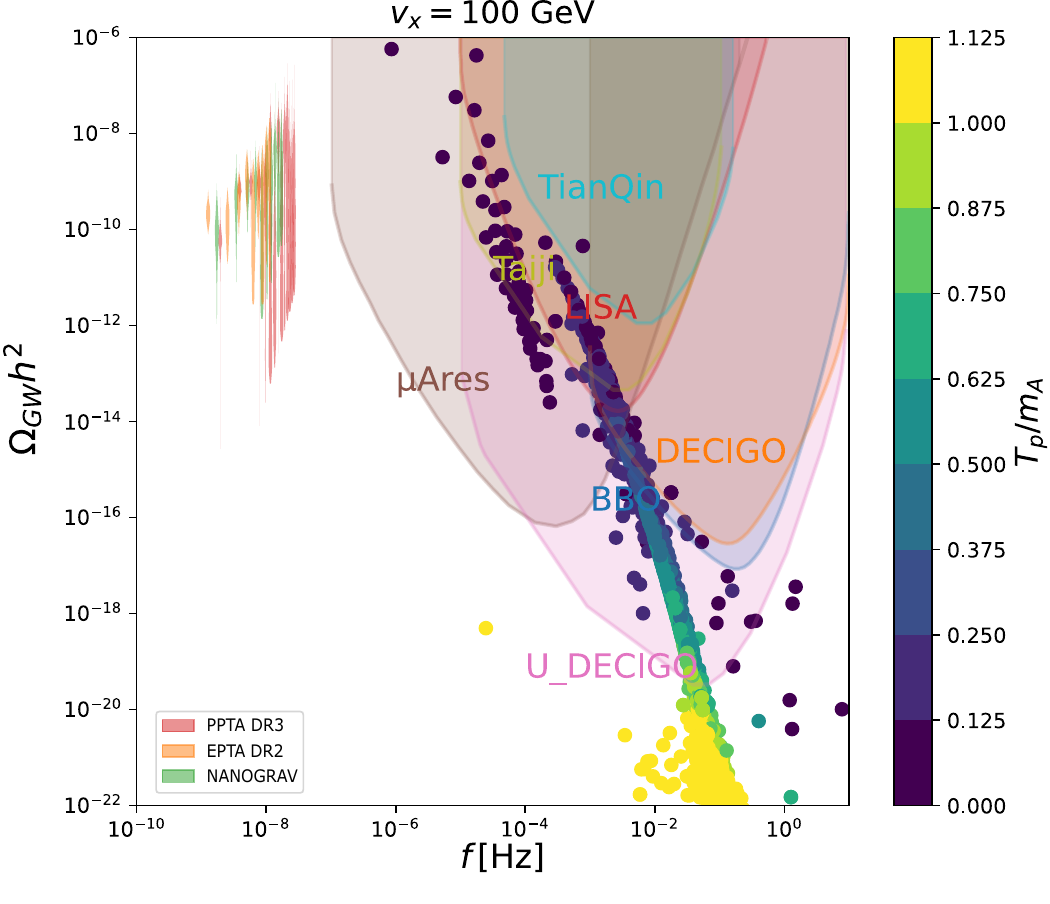}
    \caption{
     Exhibition of the dependence of the gravitational wave power spectrum for four choices for
     the hidden sector VEV $v_x$ where $v_x=(0.1, 1, 10, 100)$ GeV. Aside from using different values of
     $v_x$ in each for the four cases the same parameter scan range is chosen using gauge-invariant analysis. Each point marks the peak of the power spectrum. The color coding on the right vertical axis
      indicates the ratio $T_p/m_{A_x}$.}
    \label{fig:scan2}
\end{figure}

The four panels in Fig.~\ref{fig:scan2} show gauge-invariant gravitational wave spectra obtained with the same scan ranges in $(g_x,\lambda_x)$ but different symmetry breaking scales $v_x=\{0.1,1,10,100\}\,\mathrm{GeV}$. Across all choices of $v_x$, the colored points concentrate near $T_p/m_{A_x}\!\lesssim\!1$, indicating that the relevant transitions occur in a low-temperature regime. Varying $v_x$ primarily reorganizes the distribution of peak frequencies and amplitudes across PTA-to-mHz bands, which falls at different detection regions.

\subsection{Contributions to the gravitational wave power spectrum}

From Fig.~\ref{fig:scan} and Fig.~\ref{fig:scan2} in the previous section, one may notice that the peak points of the gravitational wave power spectra tend to separate into two distinct patterns: one group characterized by lower frequencies but higher amplitudes, and another with higher frequencies but lower amplitudes. This difference in patterns arises because the dominant gravitational wave sources differ between these two cases. In the first case, the signal is dominated by direct bubble collisions, while in the second case the sound wave contribution plays the leading role.
\begin{figure}[h!]
   \centering
   \includegraphics[width=0.4\linewidth]{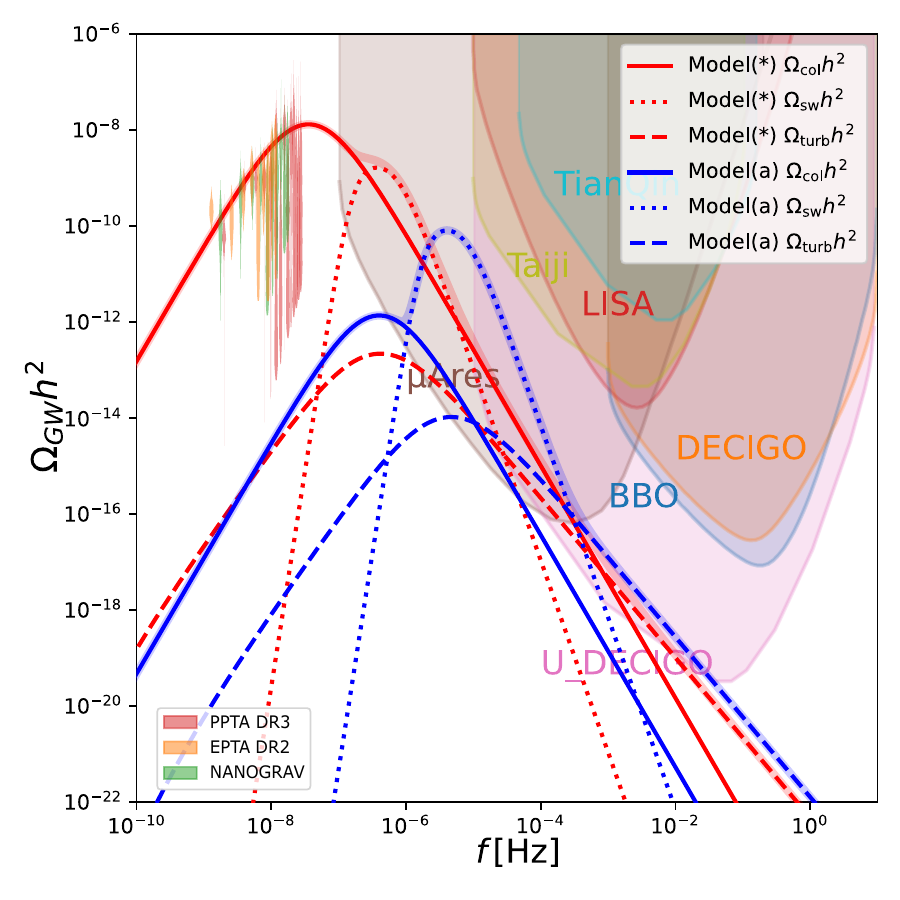}
   \caption{Individual gravitational wave source contributions for two
    benchmark models (*) and (a) defined in Table.~\ref{tab:benchmark1}. For these model points the  plots
     show the relative contribution of bubble collisions (``col''), sound waves (``sw''), and turbulence (``turb''). Model (*) exemplifies a supercooled phase transition dominated by bubble collisions, while model (a) illustrates a case where sound waves dominate.}
   \label{fig:GWcontributions}
\end{figure}

Fig.~\ref{fig:GWcontributions} illustrates this behavior explicitly by showing the relative contributions of different gravitational wave sources for benchmark model (*) and model (a) in Table.~\ref{tab:benchmark1}. For model~(*), the spectrum is largely shaped by the bubble collision component, while the sound wave and turbulence contributions remain subdominant. In contrast, for model~(a), the sound wave component provides the dominant contribution to the gravitational wave spectrum, yielding a higher-frequency peak but with reduced overall amplitude compared to model~(*).
This finding is consistent with our earlier work~\cite{Li:2025nja}, where it was shown that strongly supercooled phase transitions capable of generating signals within the sensitivity range of pulsar timing arrays are generally dominated by bubble collision induced gravitational waves. Meanwhile, scenarios with less supercooling and weaker latent heat tend to fall into the sound wave dominated regime, producing higher-frequency but comparatively weaker signals.

\subsection{Validation range of the perturbative gauge-invariant method}\label{sec:pertubativeValid}

As detailed in Appendix~\ref{App:GIEA}, the gauge-invariant framework adopted in this work is constructed as a perturbative expansion, whose accuracy is expected to improve in the weak-coupling limit. In particular, the expansion is formally justified when the gauge coupling $g_x$ is small, ensuring that higher-order corrections to the effective bounce action remain under control. However, parts of the parameter space relevant to our study extend into the regime of comparatively large couplings, $g_x \sim \mathcal{O}(1)$. This raises the question of whether the perturbative construction continues to provide a quantitatively reliable description in such regions. To address this, a pragmatic consistency check is to verify that the next-to-leading correction $S_1$ to the bounce action remains subdominant relative to the leading-order term $S_0$ at the percolation temperature.

\begin{figure}[!h]
    \centering
    \includegraphics[width=0.4\linewidth]{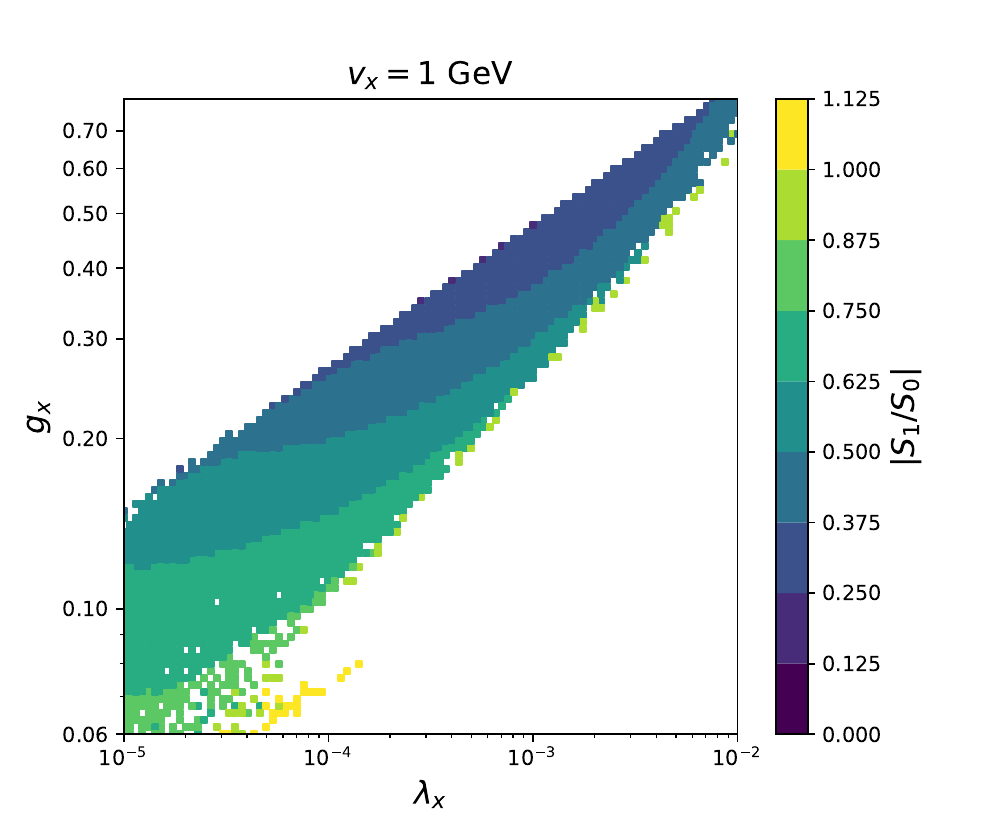}
    \includegraphics[width=0.35\linewidth]{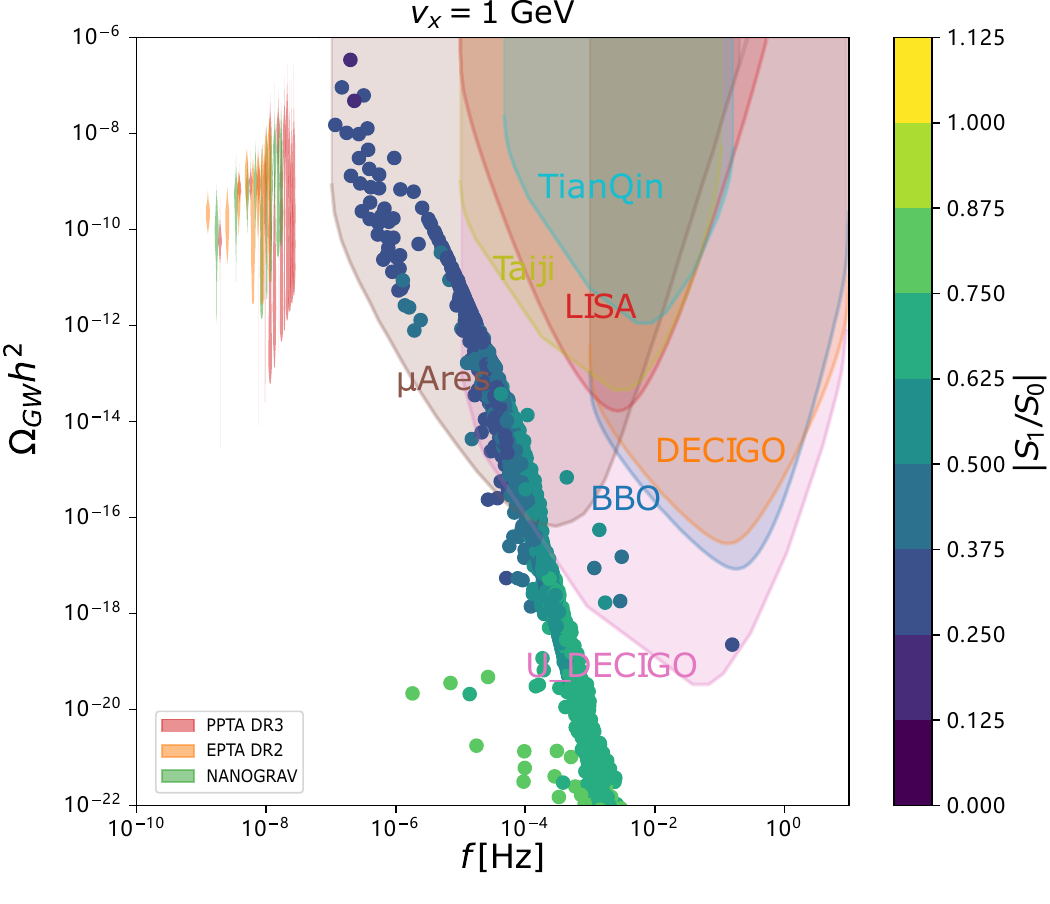}
    \caption{Left Panel:  Ratio \(|S_1/S_0|\) at $T_\mathrm{p}$ across $(g_x,\lambda_x)$ with $v_x=1\ \mathrm{GeV}$. The phase-transition-completed region satisfies \(|S_1/S_0|<1\), indicating that next-to-leading corrections remain subdominant to the leading bounce action.
    Right Panel:  Correlation between gravitational wave spectra and \(|S_1/S_0|\) at $v_x=1\ \mathrm{GeV}$. Models yielding the largest gravitational wave peak amplitudes preferentially exhibit smaller \(|S_1/S_0|\), showing that perturbative control is strongest in the phenomenologically most relevant regions.
    }
    \label{fig:B1B0_combined}
\end{figure}

As shown in Fig.~\ref{fig:B1B0_combined}, the perturbative expansion is well controlled across the region in which the phase transition completes, with $|S_1/S_0|<1$ broadly satisfied. Across the scanned range of $g_x$ we do not observe a monotonic decrease toward smaller $g_x$; rather, bands of reduced $|S_1/S_0|$ appear at multiple $g_x$ values (depending on $\lambda_x$), indicating that perturbative control is not determined by $g_x$ alone. Even toward moderately strong coupling, $g_x\sim\mathcal{O}(1)$, the ratio remains below unity in the phenomenologically relevant swath of parameter space, supporting the quantitative reliability of the gauge-invariant perturbative framework. Importantly, the right panel shows that the subset of models producing the largest gravitational wave power systematically resides in the lower $|S_1/S_0|$ region. In other words, the points of greatest phenomenological interest are also those for which the perturbative expansion is most reliable.
While the condition $|S_1/S_0|<1$ is not by itself a strict proof of convergence, it provides a practical and widely adopted check that higher-order terms remain subdominant. The combined evidence in Fig.~\ref{fig:B1B0_combined}, together with the stability of observables across the explored parameter space, indicates that the perturbative gauge-invariant framework remains under good control, particularly in the regions of greatest phenomenological relevance.

\subsection{Monte Carlo analysis and cogenesis constraint}

In the end, we run a Monte Carlo analysis with the same scan range of $g_x$ and $\lambda_x$ as in Section~\ref{sec:Parameterscan}, but with $v_x$ extended to the range $(0.01,100)\,\mathrm{GeV}$. The resulting parameter points are then subjected to the cogenesis constraint shown in Fig.~\ref{fig:gxbd}, which delineates the region of parameter space consistent with efficient secluded annihilation of the symmetric dark matter component.
\begin{figure}[!h]
    \centering
    \includegraphics[width=0.43\linewidth]{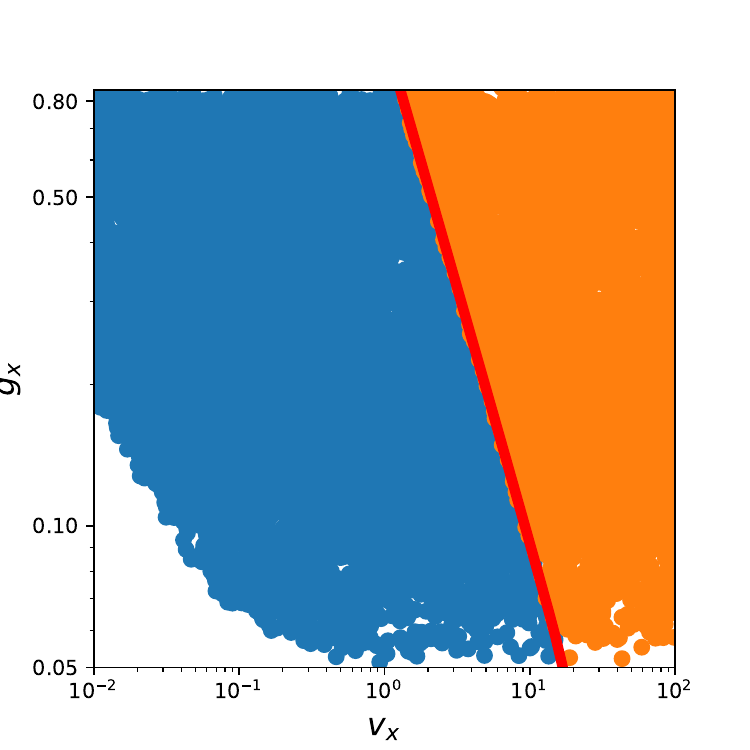}
    \includegraphics[width=0.4\linewidth]{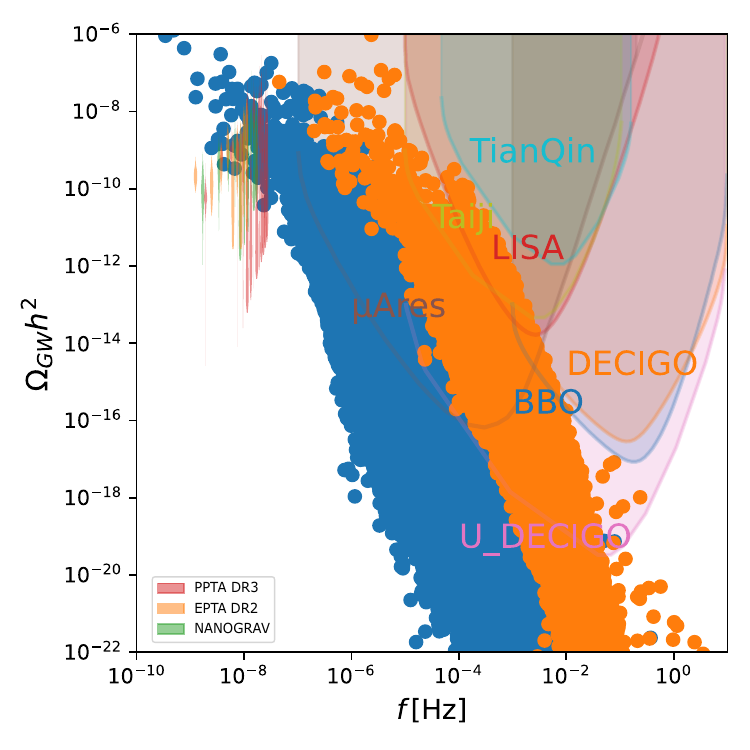}
    \caption{Left Panel: Exhibition of the parameter space in the ($g_x - v_x$ plane) allowed by the cogenesis constraint, ensuring adequate annihilation of the symmetric dark matter component.
    Blue points represent allowed models, while orange points are excluded.
     Right Panel: gravitational wave power spectrum  $\Omega_{GW} h^2$  corresponding to the left panel
    where the power spectrum is plotted vs the frequency $f(Hz)$ and where the experimental reach of several experiments are also shown. It is seen that the blue region has intercept with the
    NANOGrav, EPTA, PPTA data as well can probe future data in the higher frequency region.}
    \label{fig:cogenesisConstraint}
\end{figure}
As illustrated in Fig.~\ref{fig:cogenesisConstraint}, the cogenesis requirement significantly narrows the viable parameter space. In particular, models that would otherwise reach the detection range of space-based interferometers such as LISA are excluded once the annihilation condition is imposed. However, a subset of models in the lower-frequency, higher-power-region precisely those overlapping the pulsar-timing array (PTA) sensitivity bands remain viable.

\begin{figure}[!h]
    \centering
    \includegraphics[width=0.4\linewidth]{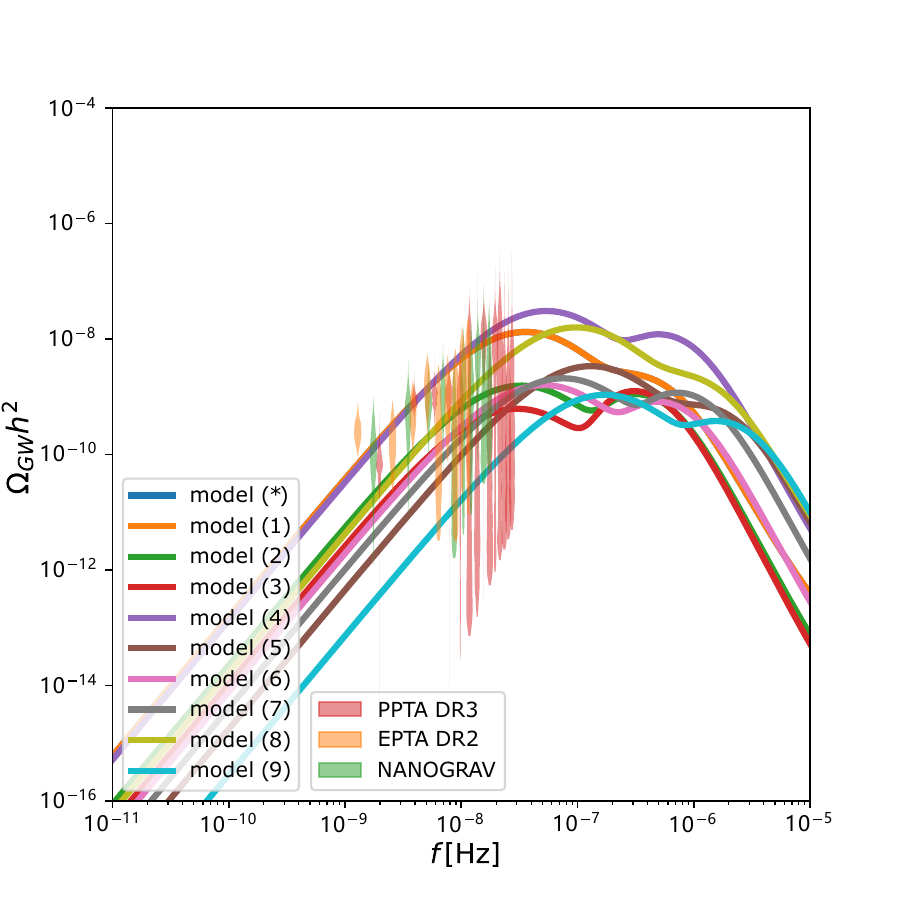}
    \caption{Plots of the gravitational wave power spectrum $\Omega_{GW} h^2$ as a function of the
    frequency f(Hz) for the 10 benchmark points in Table.~\ref{tab:benchmark2}. All the benchmark curves
    intercept the observed PTA signals.}
    \label{fig:GWpower_PTA}
\end{figure}

In Table.~\ref{tab:benchmark2}, we list a series of benchmark models that explains the observed PTA signals while all relevant experimental and cosmological constraints are simultaneously satisfied, including the relic density requirement, cogenesis, bounds from Big Bang nucleosynthesis (BBN), and collider searches. The gravitational wave power spectrum curves of these models are also shown in Fig.~\ref{fig:GWpower_PTA}.
\begin{table}[h!]
    \centering
    \begin{tabular}{c|ccccccccccc}
        \hline
Model & $g_x$ & $\lambda_x$ & $v_x$[GeV]  & $m_{A_x}$[GeV] & $T_p$[GeV] & $T_p/m_{A_x}$  \\
\hline
(*) & 0.58 & $2.21\times10^{-3}$&  0.239&  0.14 & 0.01 & 0.074  \\
(1) & 0.629 & $3.07\times10^{-3}$&  0.186&  0.12 & 0.011 & 0.094  \\
(2) & 0.684 & $4.24\times10^{-3}$&  0.183&  0.13 & 0.013 & 0.11  \\
(3) & 0.752 & $6.01\times10^{-3}$&  0.996&  0.75 & 0.064 & 0.086  \\
(4) & 0.529 & $1.54\times10^{-3}$&  0.565&  0.3 & 0.023 & 0.076  \\
(5) & 0.597 & $2.50\times10^{-3}$&  0.255&  0.15 & 0.014 & 0.09  \\
(6) & 0.638 & $3.24\times10^{-3}$&  0.454&  0.29 & 0.027 & 0.092  \\
(7) & 0.575 & $2.13\times10^{-3}$&  0.643&  0.37 & 0.026 & 0.071  \\
(8) & 0.541 & $1.70\times10^{-3}$&  0.653&  0.35 & 0.03 & 0.085  \\
(9) & 0.781 & $6.91\times10^{-3}$&  0.107&  0.084 & 0.01 & 0.12  \\
    \end{tabular}
    \caption{Benchmark models that accommodate pulsar timing array signals while satisfying other experimental and cosmological constraints.
    The relic densities of the symmetric components of $X$ and $X^\prime$ are computed to be $\Omega^X_{sym}h^2, \Omega^{X^\prime}_{sym}h^2 \ll 10^{-10}$.
    The dark matter abundance is thus overwhelmingly set by the asymmetric components. The BBN constraints are satisfied as all hidden sector species freeze out before $T = 0.1$ GeV. }
    \label{tab:benchmark2}
\end{table}
Thus the analysis demonstrates that while the LISA-accessible region is ruled out under the cogenesis condition, parameter points capable of producing PTA-scale signals survive. These surviving scenarios constitute a phenomenologically consistent and observationally motivated region of parameter space, offering a potential explanation for the stochastic gravitational wave background hinted at in recent PTA observations.

\section{Conclusion}\label{Sec:Con}

In this work we have presented an extension of the standard model to include a  dark sector with a $U(1)_x$ gauge symmetry which includes matter which is
$B-L$ conserving, which can account of several phenomena outside the realm of a possible standard model explanation.  These include the possibility of explaining baryon asymmetry and observable gravitational wave power spectrum via a first order phase transition in the dark sector. The baryon asymmetry is generated via cogenesis which also leads to a dark matter relic density consistent with current data.
Thus in the $U(1)_x$ extension, out-of-equilibrium decays of heavy Majorana fermions generate equal and opposite $B\!-L$ asymmetries in the visible and in the hidden sectors. Sphaleron interactions partially convert the visible sector lepton asymmetry into a baryon asymmetry, while the dark sector asymmetry is deposited in two Dirac fermions, $X$ and $X'$, yielding an asymmetric dark matter population.
This setup naturally accounts for the observed similarity between the cosmic abundances of baryons and dark matter, implying a characteristic dark matter mass $m_{\rm DM}\sim \mathcal{O}(\mathrm{GeV}$).
The symmetric component efficiently annihilates through secluded channels $\overline{X}X\to A_xA_x$ and $\overline{X^\prime}X^\prime\to A_xA_x$, with the dark photon subsequently decaying into Standard Model states, consistent with existing kinetic mixing constraints.
We further use  this framework for analysis of first-order phase transitions and production of
their stochastic gravitational wave backgrounds. Building on the Metaxas-Weinberg approach, we constructed a gauge-invariant effective action that controls bubble nucleation in the low-temperature regime relevant for observable signals. We demonstrate that gauge-invariant analysis is relevant for reliable predictions for the gravitational wave backgrounds since in general gravitational
wave power spectrum is sensitively dependent of the gauge choice.
Our parameter scans reveal broad regions in which successful cogenesis coincides with supercooled or near-supercooled transitions ($T_p/m_{A_x}\lesssim1$). In these regions, the gauge-invariant prediction yields robust gravitational wave signals, with amplitudes and peak frequencies that can fall within the reach of pulsar timing arrays experiments and future space-based interferometers.
Taken together, these results provide correlated targets for multi-messenger searches: gravitational wave observations of first-order phase transitions, laboratory probes of sub-GeV dark photons with small kinetic mixing,
direct detection of $\mathcal{O}(\mathrm{GeV}$) asymmetric dark matter,
and cosmological tests of neutrino properties and non-unitarity associated with the extended fermion sector. The framework thus  links otherwise disparate phenomena and their observability.

\noindent
\section{Acknowledgements}
The research of WZF and ZHY was supported in part by the National Natural Science Foundation of China under Grant No. 11935009,
and Tianjin University Self-Innovation Fund Extreme Basic Research Project Grant No. 2025XJ21-0007.
The research of PN and JL was supported in part by the NSF Grant PHY-220993.

\appendix

\section{The generation of baryon asymmetry including washout}\label{App:GenAsy}

In the early universe, the heavy $N_{i}$ decay generates an asymmetry
in $\psi$ and $\phi$, which is subsequently transferred to both
the visible and dark sectors. The corresponding Lagrangian is given
by
\begin{align}
\mathcal{L}_{{\rm asy}}=\lambda_{i}\overline{N}_{i}\psi\phi+\lambda_{i}^{*}\overline{\psi}N_{i}\phi^{*}\,,
\end{align}
which gives rise to the following decays
\begin{align}
N\to\psi+\phi\,,\qquad N\to\overline{\psi}+\overline{\phi}\,,
\end{align}
where we use $\psi,\phi$ to denote particles and $\overline{\psi},\overline{\phi}$
antiparticles. We further define
\begin{align}
|\text{\ensuremath{\mathcal{M}_{N}}}|^{2}\equiv|\text{\ensuremath{\ensuremath{\mathcal{M}_{N\rightarrow\psi\phi}}}}|^{2}+|\text{\ensuremath{\mathcal{M}_{N\rightarrow\overline{\psi}\overline{\phi}}}}|^{2}
\end{align}
and thus
\begin{align}
|\text{\ensuremath{\mathcal{M}_{N\rightarrow\psi\phi}}}|^{2} & =\frac{(1+\epsilon)}{2}|\text{\ensuremath{\mathcal{M}}}_{N}|^{2}\,,\\
|\text{\ensuremath{\mathcal{M}_{N\rightarrow\overline{\psi}\overline{\phi}}}}|^{2} & =\frac{(1-\epsilon)}{2}|\text{\ensuremath{\mathcal{M}}}_{N}|^{2}
\end{align}
The Boltzmann equation governing the evolution of $N$ reads
\begin{align}
\dot{n_{N}}+3Hn_{N} & =-\int{\rm d}\Pi_{p}{\rm d}\Pi_{k_{1}}{\rm d}\Pi_{k_{2}}(2\pi)^{4}|\text{\ensuremath{\mathcal{M}_{N\rightarrow\psi\phi}}}|^{2}\delta^{4}(p-k_{1}-k_{2})\big[f_{N}(p)-f_{\psi}(k_{1})f_{\phi}(k_{2})\big]\nonumber \\
 & \quad-\int{\rm d}\Pi_{p}{\rm d}\Pi_{k_{1}}{\rm d}\Pi_{k_{2}}(2\pi)^{4}|\text{\ensuremath{\mathcal{M}_{N\rightarrow\overline{\psi}\overline{\phi}}}}|^{2}\delta^{4}(p-k_{1}-k_{2})\big[f_{N}(p)-f_{\overline{\psi}}(k_{1})f_{\overline{\phi}}(k_{2})\big]\,,
\end{align}
where the distribution functions are given by
\begin{gather}
f_{\psi}={\rm e}^{-(E_{\psi}-\mu_{\psi})/T}\,,\qquad f_{\overline{\psi}}={\rm e}^{-(E_{\overline{\psi}}+\mu_{\psi})/T}\,,\\
f_{\phi}={\rm e}^{-(E_{\phi}-\mu_{\phi})/T}\,,\qquad f_{\overline{\phi}}={\rm e}^{-(E_{\overline{\phi}}+\mu_{\phi})/T}\,.
\end{gather}
At high temperatures, $\psi$ denotes the fermionic particle carrying
2 degrees of freedom while $\phi$ denotes the complex scalar particle
carrying 1 degrees of freedom. Their corresponding antiparticles $\overline{\psi}$
and $\overline{\phi}$ also carry 2 and 1 degree of freedom respectively.
Given that the decay of $N$ will generate equal amount of asymmetry
in both $\psi$ and $\phi$
\begin{align}
n_{\psi}-n_{\overline{\psi}}=n_{\phi}-n_{\overline{\phi}}\,,
\end{align}
one thus derives
\begin{align}
\mu_{\psi}=\mu_{\phi}\,.
\end{align}
The net $\psi$ and $\phi$ comoving number densities are expressed
by respectively\footnote{We adopt the standard notation for the number density, defined as
($+$ for fermion and $-$ for bosons)
\begin{align}
n_{i}^{{\rm EQ}} & =\frac{g_{i}}{2\pi^{2}}\int_{m}^{\infty}{\rm d}E\,\frac{E\sqrt{E^{2}-m^{2}}}{{\rm e}^{(E-\mu)/T}\pm1}=\frac{\zeta(3)}{\pi^{2}}g_{i}T^{3}\ ({\rm Bosons})\,,\quad\frac{3}{4}\frac{\zeta(3)}{\pi^{2}}g_{i}T^{3}\ ({\rm Fermions})\,,
\end{align}
where $g_{i}$ is the degrees of freedom of only the particle $i$
but not the antiparticle of $i$.  For high temperatures ($T\gg m$), the net particle number density is given by
\begin{align}
n_{i}-n_{\bar{i}} & =\frac{g_{i}}{2\pi^{2}}\int_{0}^{\infty}{\rm d}E\,E\sqrt{E^{2}-m_{i}^{2}}\left\{ \big[{\rm e}^{(E-\mu_{i})/T}\pm1\big]^{-1}-\big[{\rm e}^{(E+\mu_{i})/T}\pm1\big]^{-1}\right\} \approx\frac{g_{i}T^{3}c}{6}\frac{\mu_{i}}{T}\,,
\end{align}
where the bar denotes the antiparticle, and $c=2$ for bosons and
$c=1$ for fermions. }
\begin{align}
Y_{\Psi} & \equiv\frac{n_{\psi}-n_{\overline{\psi}}}{s}=Y_{\psi}-Y_{\overline{\psi}}=\frac{15g_{\psi}}{4\pi^{2}g_{*S}}\frac{\mu_{\psi}}{T}=\frac{15}{2\pi^{2}g_{*S}}\frac{\mu_{\psi}}{T}\,,\\
Y_{\Phi} & \equiv\frac{n_{\phi}-n_{\overline{\phi}}}{s}=Y_{\phi}-Y_{\overline{\phi}}=\frac{15g_{\phi}}{2\pi^{2}g_{*S}}\frac{\mu_{\psi}}{T}=\frac{15}{2\pi^{2}g_{*S}}\frac{\mu_{\phi}}{T}\,.
\end{align}
Thus the product $f_{\psi}f_{\phi}$ can be further expressed by
\begin{align}
f_{\psi}f_{\phi} & ={\rm e}^{-(E_{\psi}+E_{\phi})/T}{\rm e}^{(\mu_{\psi}+\mu_{\phi})/T}=f_{N}^{{\rm EQ}}{\rm e}^{2\mu_{\psi}/T}\approx f_{N}^{{\rm EQ}}\left[1+\frac{4\pi^{2}g_{*S}}{15}Y_{\Psi}\right]\,,\\
f_{\overline{\psi}}f_{\overline{\phi}} & ={\rm e}^{-(E_{\psi}+E_{\phi})/T}{\rm e}^{-(\mu_{\psi}+\mu_{\phi})/T}=f_{N}^{{\rm EQ}}{\rm e}^{-2\mu_{\psi}/T}\approx f_{N}^{{\rm EQ}}\left[1-\frac{4\pi^{2}g_{*S}}{15}Y_{\Psi}\right]\,.
\end{align}
The Boltzmann equation of $N$ then becomes
\begin{align}
\dot{n_{N}}+3Hn_{N} & =-\int{\rm d}\Pi_{p}{\rm d}\Pi_{k_{1}}{\rm d}\Pi_{k_{2}}(2\pi)^{4}|\text{\ensuremath{\mathcal{M}}}_{N}|^{2}\delta^{4}(p-k_{1}-k_{2})\times\nonumber \\
 & \qquad\qquad\qquad\left\{ \big[f_{N}(p)-f_{N}^{{\rm EQ}}(p)\big]-\epsilon f_{N}^{{\rm EQ}}(p)\times\frac{4\pi^{2}g_{*S}}{15}Y_{\Psi}\right\} \nonumber \\
 & =-\left\langle \Gamma_{N}\right\rangle (n_{N}-n_{N}^{{\rm EQ}})+\epsilon\left\langle \Gamma_{N}\right\rangle n_{N}^{{\rm EQ}}\times\frac{4\pi^{2}g_{*S}}{15}Y_{\Psi}\,.
\end{align}
We now rewrite the above Boltzmann equation in terms of the comoving
number density $Y_{N}$ as
\begin{align}
\frac{{\rm d}Y_{N}}{{\rm d}T}=\frac{1}{T\bar{H}}\left\{ \left\langle \Gamma_{N}\right\rangle (Y_{N}-Y_{N}^{{\rm EQ}})-\epsilon\left\langle \Gamma_{N}\right\rangle Y_{N}^{EQ}\times\frac{4\pi^{2}g_{*S}}{15}Y_{\Psi}\right\} \,.
\end{align}
The $2\to2$ scattering processes contributing to the change of $N$
number density are $NN\leftrightarrow\overline{\psi}\psi$ and $NN\leftrightarrow\overline{\phi}\phi$,
and their corresponding collision terms are derived as follows
{\footnotesize
\begin{align}
\mathrm{C}_{NN\leftrightarrow\overline{\psi}\psi} & =-2\int{\rm d}\Pi_{p_{1}}{\rm d}\Pi_{p_{2}}{\rm d}\Pi_{k_{1}}{\rm d}\Pi_{k_{2}}(2\pi)^{4}\frac{1}{2}\big|\text{\ensuremath{\mathcal{M}_{NN\rightarrow\overline{\psi}\psi}}}\big|^{2}\delta^{4}(k_{1}+k_{2}-p_{1}-p_{2})\big[f_{N}(p_{1})f_{N}(p_{2})-f_{\psi}(k_{1})f_{\overline{\psi}}(k_{2})\big]\nonumber \\
 & =-2\int{\rm d}\Pi_{p_{1}}{\rm d}\Pi_{p_{2}}{\rm d}\Pi_{k_{1}}{\rm d}\Pi_{k_{2}}(2\pi)^{4}\frac{1}{2}\big|\text{\ensuremath{\mathcal{M}_{NN\rightarrow\overline{\psi}\psi}}}\big|^{2}\delta^{4}(k_{1}+k_{2}-p_{1}-p_{2})\big[f_{N}(p_{1})f_{N}(p_{2})-f_{N}^{{\rm EQ}}(p_{1})f_{N}^{{\rm EQ}}(p_{2})\big]\nonumber \\
 & =-2\left\langle \sigma v\right\rangle _{NN\rightarrow\overline{\psi}\psi}\big(n_{N}n_{N}-n_{N}^{{\rm EQ}}n_{N}^{{\rm EQ}}\big)\,,
\end{align}
\begin{align}
\mathrm{C}_{NN\leftrightarrow\overline{\phi}\phi} & =-2\int{\rm d}\Pi_{p_{1}}{\rm d}\Pi_{p_{2}}{\rm d}\Pi_{k_{1}}{\rm d}\Pi_{k_{2}}(2\pi)^{4}\frac{1}{2}\big|\text{\ensuremath{\mathcal{M}_{NN\rightarrow\overline{\phi}\phi}}}\big|^{2}\delta^{4}(k_{1}+k_{2}-p_{1}-p_{2})\big[f_{N}(p_{1})f_{N}(p_{2})-f_{\phi}(k_{1})f_{\overline{\phi}}(k_{2})\big]\nonumber \\
 & =-2\int{\rm d}\Pi_{p_{1}}{\rm d}\Pi_{p_{2}}{\rm d}\Pi_{k_{1}}{\rm d}\Pi_{k_{2}}(2\pi)^{4}\frac{1}{2}\big|\text{\ensuremath{\mathcal{M}_{NN\rightarrow\overline{\phi}\phi}}}\big|^{2}\delta^{4}(k_{1}+k_{2}-p_{1}-p_{2})\big[f_{N}(p_{1})f_{N}(p_{2})-f_{N}^{{\rm EQ}}(p_{1})f_{N}^{{\rm EQ}}(p_{2})\big]\nonumber \\
 & =-2\left\langle \sigma v\right\rangle _{NN\rightarrow\overline{\phi}\phi}\big(n_{N}n_{N}-n_{N}^{{\rm EQ}}n_{N}^{{\rm EQ}}\big)\,.
\end{align}}
In the above equations, the coefficient $-2$ in the front indicates
that two units of $N$ are annihilated. The factor of $1/2$ in front
of the squared amplitude accounts for the avoidance of double counting
when integrating over the phase space of the two identical $N$ particles,
which is however absorbed into the definition of the cross-section $\sigma$.

The evolution of the generated net $\psi$ number can be obtained
by subtracting the two Boltzmann equations $\dot{n_{\psi}}+3Hn_{\psi}$
and $\dot{n}_{\overline{\psi}}+3Hn_{\overline{\psi}}$ and we obtain
\begin{align}
\dot{n}_{\Psi}+3Hn_{\Psi} & =\dot{n_{\psi}}+3Hn_{\psi}-\big(\dot{n}_{\overline{\psi}}+3Hn_{\overline{\psi}}\big) \nonumber\\
 & ={\rm C}_{\psi\phi\leftrightarrow N}+{\rm C}_{\overline{\psi}\overline{\phi}\leftrightarrow N}+{\rm C}_{\psi\psi\leftrightarrow\overline{\phi}\overline{\phi}}+{\rm C}_{\overline{\psi}\overline{\psi}\leftrightarrow\phi\phi}+{\rm C}_{\psi\phi\leftrightarrow\overline{\psi}\overline{\phi}}\,,
\end{align}
where the collision terms are
\begin{align}
{\rm C}_{\psi\phi\leftrightarrow N}+{\rm C}_{\overline{\psi}\overline{\phi}\leftrightarrow N} & =-\int{\rm d}\Pi_{p}{\rm d}\Pi_{k_{1}}{\rm d}\Pi_{k_{2}}(2\pi)^{4}|\text{\ensuremath{\mathcal{M}_{\psi\phi\leftrightarrow N}}}|^{2}\delta^{4}(p-k_{1}-k_{2})\big[f_{\psi}(k_{1})f_{\phi}(k_{2})-f_{N}(p)\big]\nonumber \\
 & \quad+\int{\rm d}\Pi_{p}{\rm d}\Pi_{k_{1}}{\rm d}\Pi_{k_{2}}(2\pi)^{4}|\text{\ensuremath{\mathcal{M}_{\overline{\psi}\overline{\phi}\leftrightarrow N}}}|^{2}\delta^{4}(p-k_{1}-k_{2})\big[f_{\overline{\psi}}(k_{1})f_{\overline{\phi}}(k_{2})-f_{N}(p)\big]\nonumber \\
 & =-\epsilon\left\langle \Gamma_{N}\right\rangle (n_{N}^{{\rm EQ}}-n_{N})-\left\langle \Gamma_{N}\right\rangle n_{N}^{{\rm EQ}}\times\frac{4\pi^{2}g_{*S}}{15}Y_{\Psi}\,,
\end{align}
and
{\footnotesize
\begin{align}
{\rm C}_{\psi\psi\leftrightarrow\overline{\phi}\overline{\phi}}+{\rm C}_{\overline{\psi}\overline{\psi}\leftrightarrow\phi\phi} & =-2\int{\rm d}\Pi_{p_{1}}{\rm d}\Pi_{p_{2}}{\rm d}\Pi_{k_{1}}{\rm d}\Pi_{k_{2}}(2\pi)^{4}\frac{1}{4}\big|\text{\ensuremath{\mathcal{M}_{\psi\psi\rightarrow\overline{\phi}\overline{\phi}}}}\big|^{2}\delta^{4}(k_{1}+k_{2}-p_{1}-p_{2})\big[f_{\psi}(p_{1})f_{\psi}(p_{2})-f_{\overline{\phi}}(k_{1})f_{\overline{\phi}}(k_{2})\big]\nonumber \\
 & \quad+2\int{\rm d}\Pi_{p_{1}}{\rm d}\Pi_{p_{2}}{\rm d}\Pi_{k_{1}}{\rm d}\Pi_{k_{2}}(2\pi)^{4}\frac{1}{4}\big|\text{\ensuremath{\mathcal{M}_{\overline{\psi}\overline{\psi}\leftrightarrow\phi\phi}}}\big|^{2}\delta^{4}(k_{1}+k_{2}-p_{1}-p_{2})\big[f_{\overline{\psi}}(p_{1})f_{\overline{\psi}}(p_{2})-f_{\phi}(k_{1})f_{\phi}(k_{2})\big]\nonumber \\
 & =-2\left\langle \sigma v\right\rangle _{\psi\psi\rightarrow\overline{\phi}\overline{\phi}}n_{\psi}^{{\rm EQ}}n_{\psi}^{{\rm EQ}}\times\frac{8\pi^{2}g_{*S}}{15}+2\left\langle \sigma v\right\rangle _{\overline{\psi}\overline{\psi}\leftrightarrow\phi\phi}n_{\psi}^{{\rm EQ}}n_{\psi}^{{\rm EQ}}\times\left(-\frac{8\pi^{2}g_{*S}}{15}\right)\nonumber \\
 & =-2\left\langle \sigma v\right\rangle _{\psi\psi\rightarrow\overline{\phi}\overline{\phi}}n_{\psi}^{{\rm EQ}}n_{\psi}^{{\rm EQ}}\times\frac{16\pi^{2}g_{*S}}{15}\,,
\end{align}
\begin{align}
{\rm C}_{\psi\phi\leftrightarrow\overline{\psi}\overline{\phi}} & =-\int{\rm d}\Pi_{p_{1}}{\rm d}\Pi_{p_{2}}{\rm d}\Pi_{k_{1}}{\rm d}\Pi_{k_{2}}(2\pi)^{4}\big|\text{\ensuremath{\mathcal{M}_{\psi\phi\rightarrow\overline{\psi}\overline{\phi}}}}\big|^{2}\delta^{4}(k_{1}+k_{2}-p_{1}-p_{2})\big[f_{\psi}(p_{1})f_{\phi}(p_{2})-f_{\overline{\psi}}(k_{1})f_{\overline{\phi}}(k_{2})\big]\nonumber \\
 & \quad+\int{\rm d}\Pi_{p_{1}}{\rm d}\Pi_{p_{2}}{\rm d}\Pi_{k_{1}}{\rm d}\Pi_{k_{2}}(2\pi)^{4}\big|\text{\ensuremath{\mathcal{M}_{\overline{\psi}\overline{\phi}\to\psi\phi}}}\big|^{2}\delta^{4}(k_{1}+k_{2}-p_{1}-p_{2})\big[f_{\overline{\psi}}(p_{1})f_{\overline{\phi}}(p_{2})-f_{\psi}(k_{1})f_{\phi}(k_{2})\big]\nonumber \\
 & =-\left\langle \sigma v\right\rangle _{\psi\phi\rightarrow\overline{\psi}\overline{\phi}}n_{\psi}^{{\rm EQ}}n_{\phi}^{{\rm EQ}}\times\frac{8\pi^{2}g_{*S}}{15}+\left\langle \sigma v\right\rangle _{\overline{\psi}\overline{\phi}\to\psi\phi}n_{\psi}^{{\rm EQ}}n_{\phi}^{{\rm EQ}}\times\left(-\frac{8\pi^{2}g_{*S}}{15}\right)\nonumber \\
 & =-\left\langle \sigma v\right\rangle _{\psi\phi\rightarrow\overline{\psi}\overline{\phi}}n_{\psi}^{{\rm EQ}}n_{\phi}^{{\rm EQ}}\times\frac{16\pi^{2}g_{*S}}{15}\,.
\end{align}}
Combining all above collision terms, the Boltzmann equation tracking
the generation of the asymmetries are written as
\begin{align}
\frac{{\rm d}Y_{N}}{{\rm d}T} & =\frac{1}{T\bar{H}}\left[\left\langle \Gamma_{N}\right\rangle (Y_{N}-Y_{N}^{{\rm EQ}})-\epsilon\left\langle \Gamma_{N}\right\rangle Y_{N}^{\rm EQ}\times\frac{4\pi^{2}g_{*S}}{15}Y_{\Psi}\right]\nonumber \\
 & +\frac{2s}{T\bar{H}}\Big(\left\langle \sigma v\right\rangle _{NN\rightarrow\phi\overline{\phi}}+\left\langle \sigma v\right\rangle _{NN\rightarrow\psi\overline{\psi}}\Big)\Big(Y_{N}Y_{N}-Y_{N}^{{\rm EQ}}Y_{N}^{{\rm EQ}}\Big)\,,\\
\frac{{\rm d}Y_{\Psi}}{{\rm d}T} & =\frac{{\rm d}\left(\frac{n_{\psi}-n_{\overline{\psi}}}{s}\right)}{{\rm d}T}=-\frac{1}{T\bar{H}}\epsilon\left\langle \Gamma_{N}\right\rangle (Y_{N}-Y_{N}^{{\rm EQ}})\nonumber \\
 & +\frac{s}{T\bar{H}}\frac{4\pi^{2}g_{*S}}{15}Y_{\Psi}\left(\frac{\left\langle \Gamma_{N}\right\rangle }{s}Y_{N}^{{\rm EQ}}+8\left\langle \sigma v\right\rangle _{\psi\psi\rightarrow\overline{\phi}\overline{\phi}}n_{\psi}^{{\rm EQ}}n_{\psi}^{{\rm EQ}}+4\left\langle \sigma v\right\rangle _{\psi\phi\rightarrow\overline{\psi}\overline{\phi}}n_{\psi}^{{\rm EQ}}n_{\phi}^{{\rm EQ}}\right)\,.
\end{align}
Since $N$ is much heavier than the combined masses of $\psi$ and
$\phi$, the thermally averaged cross-section $\left\langle \sigma v\right\rangle _{\psi\psi\rightarrow\overline{\phi}\overline{\phi}}$
and $\left\langle \sigma v\right\rangle _{\psi\phi\rightarrow\overline{\psi}\overline{\phi}}$
should be replaced by the Real Intermediate State (RIS)-subtracted
cross-sections, denoted with the superscript ``RIS'', to avoid the
double counting the contribution from the on-shell $N$ propagator,
as was shown in Eq.~(\ref{Eq:GenAsy}).

The Boltzmann equation governing the net $\phi$ number density $Y_{\Phi}$
takes the same form as that for $Y_{\Psi}$. As a result, the decay
of $N$ generates equal amounts of asymmetry in both $\psi$ and $\phi$,
including all washout processes.

\section{A set of parameters generating the neutrino mixing}\label{App:NuPara}

We provide a representative set of parameters that yields three light neutrinos in agreement with current experimental observations (including the observed Dirac phase),
where the neutrino mass matrix given by Eq.~(\ref{eq:nuMMS}) reads
{\footnotesize
\begin{align}
\left(\begin{array}{cccc}
2.056\times 10^{-11} & 0 & 0 & 3.198\times 10^{-5}\\
0 & 1.361\times 10^{-11} & 0 & 3.805\times 10^{-4} \\
0 & 0 & 2.560\times 10^{-11} & 3.314\times 10^{-4}\\
(-9.681-2.343{\rm i})\times 10^{-5} & (-1.084+3.700{\rm i})\times 10^{-4} & (-3.168+13.23{\rm i})\times 10^{-5} & 4.708\times 10^{3}
\end{array}\right)\,,
\end{align}}
with all elements expressed in units of GeV.

\section{$U(1)_x-$hypercharge kinetic mixing}\label{App:KM}
\begin{figure}[h]
\centering
\includegraphics[width=0.6\linewidth]{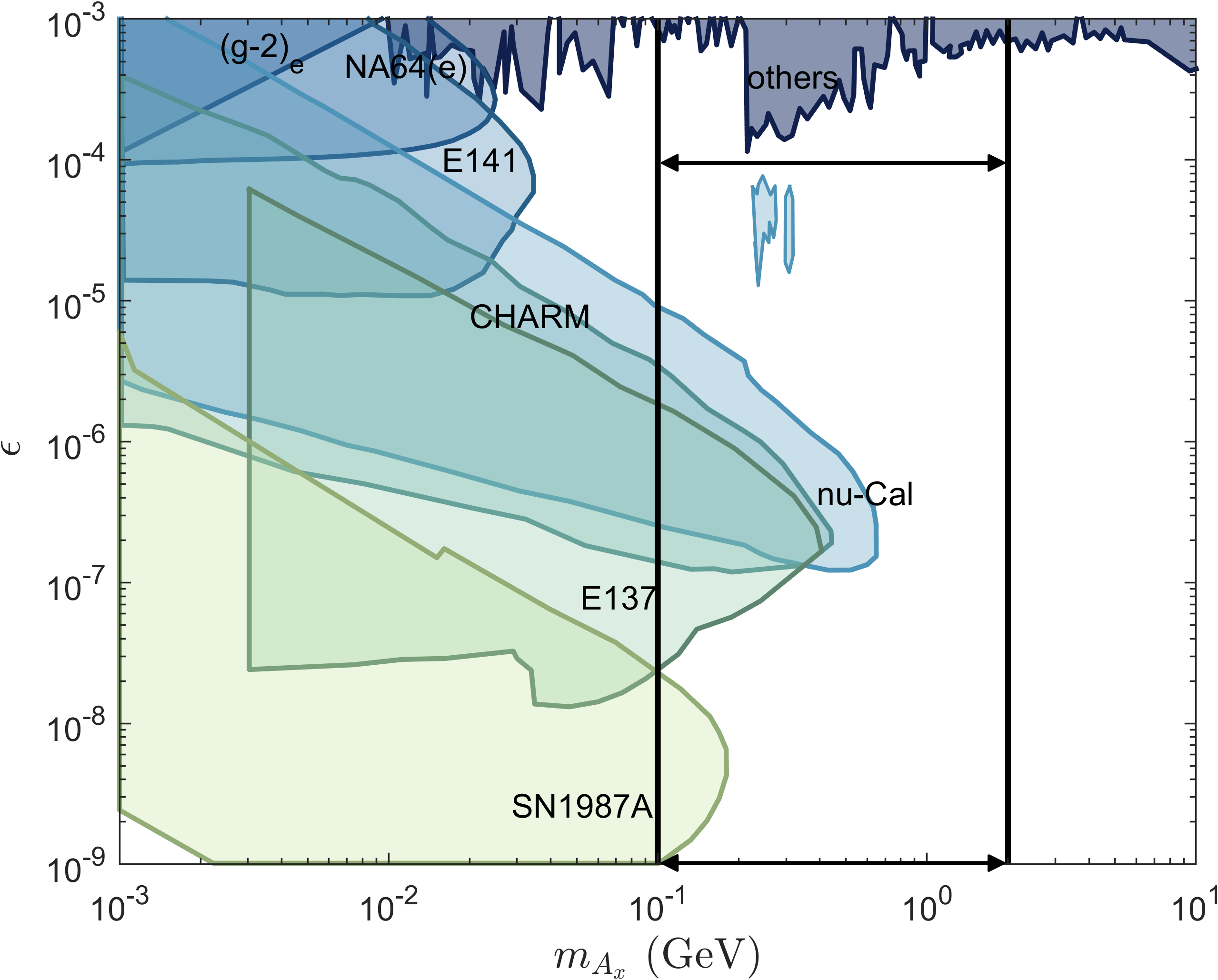}
\caption{Experimental constraints for the kinetic mixing parameter $\epsilon$ given in Eqs.~(\ref{eq:KM}) and (\ref{eq:realeps}),
for the mass of the dark $U(1)_x$ gauge boson $A_x$ ranging from 1~MeV to 10~GeV. 
For all benchmark points, 
the kinetic mixing parameters $\epsilon = 10^{-4}$ and $\epsilon = 10^{-9}$, shown as black lines,
are consistent with current experimental constraints.
Constraints on the kinetic mixing parameter $\epsilon$ are shown in the shaded regions, including bounds from 
SN1987A~\cite{Chang:2016ntp},
E137~\cite{Bjorken:1988as,Batell:2014mga,Marsicano:2018krp},
CHARM~\cite{Gninenko:2012eq},
$\nu$-Cal~\cite{Blumlein:2013cua,Blumlein:2011mv},
E141~\cite{Riordan:1987aw},
$(g-2)_e$~\cite{Pospelov:2008zw},
NA64(e)~\cite{NA64:2018iqr},
and others(A1~\cite{Merkel:2014avp}, BaBar~\cite{BaBar:2014zli},NA48/2~\cite{Piandani:2016iam}, KLOE~\cite{KLOE-2:2011hhj,KLOE-2:2012lii,KLOE-2:2014qxg,KLOE-2:2016ydq},
LHCb~\cite{LHCb:2019vmc},
CMS~\cite{CMS:2019kiy}).}
\label{fig:DPKM}
\end{figure}

\VF{
An additional $U(1)_x$ gauge symmetry can mix with hypercharge via gauge kinetic terms,
induced by, e.g., loop effects of heavy bifundamental fermions charged under both $U(1)_x$ and $U(1)_Y$~\cite{Holdom:1985ag}.
After integrating out the heavy states, in low energies one has
\begin{equation}
\mathcal{L}_{{\rm kin}}=-\frac{1}{4}F_{Y\,\mu\nu}F_{Y}^{\mu\nu}-\frac{1}{4}F_{x\,\mu\nu}F_{x}^{\mu\nu}-\frac{\epsilon}{2}F_{x\,\mu\nu}F_{Y}^{\mu\nu}\,,\label{eq:KM}
\end{equation}
where $\epsilon$ is the kinetic mixing parameter.
The kinetic mixing thus provides a portal that connects the $U(1)_x$ dark sector to the Standard Model.
After simultaneously diagonalizing the kinetic and mass mixing matrices,
in the mass eigenbasis the dark photon couples to the Standard Model fermions.

The neutral current couplings of $A_x$ and $Z$ to the SM fermions $f_i$ and to the dark matter fields $X$ and $X'$ can be written as~\cite{Feldman:2007wj}
\begin{align}
-\mathcal{L}_{A_{x},Z} & =\frac{1}{2}\bar{f}_{i}\gamma_{\mu}\big[\big(v_{i}^{\prime}-a_{i}^{\prime}\gamma^{5}\big)f_{i}A_{x}^{\mu}+\big(v_{i}-a_{i}\gamma^{5}\big)f_{i}Z^{\mu}\big]\nonumber \\
 & \,+g_{x}\left(R_{11}A_{x}^{\mu}+R_{13}Z^{\mu}\right)\big(\overline{X}\gamma_{\mu}X-\overline{X^{\prime}}\gamma_{\mu}X^{\prime}\big)\,.
\end{align}
where
\begin{align}
v_{i} & =\left(g_{2}R_{33}-g_{Y}R_{23}\right)T_{i}^{3}+2g_{Y}R_{23}Q_{i}\,,\\
a_{i} & =\left(g_{2}R_{33}-g_{Y}R_{23}\right)T_{i}^{3}\,,\\
v_{i}^{\prime} & =\left(g_{2}R_{31}-g_{Y}R_{21}\right)T_{i}^{3}+2g_{Y}R_{21}Q_{i}\,,\\
a_{i}^{\prime} & =\left(g_{2}R_{31}-g_{Y}R_{21}\right)T_{i}^{3}\,,
\end{align}
$g_Y,g_2,g_x$ are the hypercharge, weak and dark gauge couplings respectively,
and the rotation matrix is given by
\begin{equation}
R=\left(\begin{array}{ccc}
c_{\delta}\cos\psi & 0 & c_{\delta}\sin\psi\\
-s_{\delta}\cos\psi+\sin\psi\sin\theta_{W} & \cos\theta_{W} & -s_{\delta}\sin\psi-\cos\psi\sin\theta_{W}\\
-\sin\psi\cos\theta_{W} & \sin\theta_{W} & \cos\psi\cos\theta_{W}
\end{array}\right)\,.
\end{equation}
After mixing, the photon couplings to Standard Model fermions remain unchanged,
and the photon does not to couple to the dark matter fields.

In the literature, constraints on the kinetic mixing parameter $\boldsymbol{\epsilon}$
is obtained from a simplified model that the dark photon mixed directly with the photon field.
The experimental constraints on the $U(1)_x-$hypercharge kinetic mixing is obtained by the relation~\cite{Feng:2023ubl,Feng:2024nkh}
\begin{equation}
\epsilon=\frac{\sqrt{g_{2}^{2}+g_{Y}^{2}}}{g_{2}}\boldsymbol{\epsilon}\,, \label{eq:realeps}
\end{equation}
and are shown in Fig.~\ref{fig:DPKM}.
In this paper, we focus on $\mathcal{O}(1)$~GeV dark matter annihilating into dark photons
with masses in the sub-GeV to $\mathcal{O}(1)$~GeV range.
Accordingly, we consider two representative values for the kinetic mixing parameter,
$\epsilon = 10^{-4}$ and $\epsilon = 10^{-9}$,
which are shown as black lines within the allowed regions in Fig.~\ref{fig:DPKM}.}

\section{Gauge invariant effective action for first order phase transition}\label{App:GIEA}

\subsection{Generating functional and effective potential/action}
We start with the generating functional and connected Greens functions with $J$ to be the source, $\phi$ to be the scale field and $\mathcal{L}$ to be Eq.~(\ref{eq:Lagrangian}), then
\begin{align}
    Z[J] =\int \mathcal{D}\phi \exp{\left[{\rm i}\int {\rm d}^4x(\mathcal{L}+J\phi)\right]}\,,\quad W[J] = -{\rm i}\ln{Z[J]}\,.
\end{align}
The effective action/potential is obtained through the Legendre transform of $W[J]$ that
\begin{align}
    \Gamma[\phi_c] = W[J]-\int {\rm d}^4xJ(x)\phi_c(x)\,,\quad \phi_c(x) = \frac{\partial W[J]}{\partial J}\,.
\end{align}
Write it as the expansion in powers of scalar field
\begin{align}
    S_{\rm eff}(\phi_c) = \Gamma(\phi_c) = \sum^\infty_{n=0}\frac{1}{n!} \int {\rm d}^4x_1\cdots{\rm d}^4x_n\Gamma^{n}(x_1,\cdots,x_n){\phi_c}(x_1)\cdots{\phi_c}(x_n)\,,
\end{align}
where $\Gamma^{n}$ are n-point 1PI Green's function. Following \cite{Hua:2025fap,Sher:1988mj} we take Fourier transformation on the classic background field $\phi_c$ and Green's function
\begin{align}
    \phi_c(x) =\int \frac{{\rm d}^4 p}{(2\pi)^4}\bar{\phi}(p)e^{ipx}\,,\quad
    \Gamma^{n}(x_i)= \int\prod_{i=1}^n\frac{{\rm d}^4p_i}{(2\pi)^4}e^{ip_ix_i}(2\pi)^4\delta^4\left(\sum_ip_i\right)\Gamma^n(p_i)\,.
\end{align}
Then, our effective action will be
\begin{align}
    S_{\rm eff}(\bar\phi) =&\sum_{n=1}^\infty\frac{1}{n!}\int\prod_{i=1}^n\frac{{\rm d}^4p_i}{(2\pi)^4}\bar{\phi}(-p_i)(2\pi)^4\delta^4\left(\sum_ip_i\right)\Gamma^n(p_i)\non
    =&\sum_{n=1}^\infty\frac{1}{n!}\int\prod_{i=1}^n\frac{{\rm d}^4p_i}{(2\pi)^4}\bar{\phi}(-p_i)\Big|_{p_i=0}(2\pi)^4\delta^4\left(0\right)\Gamma^n(p_i)\Big|_{p_i=0}\non
    &+\frac{1}{2}\int\frac{{\rm d}^4p}{(2\pi)^4}\left.\frac{\partial \Gamma^2(p^2)}{\partial p^2}\right|_{p^2=0}p^2\bar{\phi}(-p)\bar{\phi}(p)+\cdots\,,
\end{align}
where the first term represents the zero-order derivative expansion and the second term represents the leading term of the first-order expansion. Let us define the quantity
\begin{align}
Z = -\left.\frac{\partial \Gamma^2(p^2)}{\partial p^2}\right|_{p^2=0}\,.
\end{align}
Upon applying the Fourier transformation again, we obtain
\begin{align}
    S_{\rm eff}(\phi_c)
    =&\int {\rm d}^4x\sum_{n=0}^\infty\frac{(2\pi)^4}{n!}\phi_c^n(x)\Gamma^n(p)\Big|_{p=0} + \frac{Z(\phi_c)}{2}\int {\rm d}^4x\partial_\mu\phi_c\partial^\mu\phi_c\non
    =&\int {\rm d}^4x \left[V_{\rm eff}(\phi_c)+\frac{Z(\phi_c)}{2}\partial_\mu\phi_c\partial^\mu\phi_c\right]\,.
\end{align}

\subsection{Gauge invariant effective action using the derivative expansion}
Here, we follow the method by Metaxas and Weinberg \cite{Metaxas:1995ab,Arunasalam:2021zrs} to get a gauge independent effective bounce action. According to Nielsen Idensity, which is directly from BRST invariance of effective action, we have
\begin{align}
    \xi\frac{\partial S_{\rm eff}}{\partial \xi} &= -\int {\rm d}^4x \, C(x)\frac{\delta S_{\rm eff}}{\delta \phi(x)}\,, \label{eq:Nielsen}\\
    C(x) &= -\frac{ie}{2}\int {\rm d}^4y \left<\bar{c}(x)G(x)c(y)\left[\partial_\mu A^\mu(y) - g_x\xi\bar{\phi}G(y)\right]\right>\,.
\end{align}
Performing gradient expansion on $S_{\rm eff}$ at finite temperature, $C(x)$ and $\delta S_{\rm eff}/\delta \phi$, we get
\begin{align}
    S_{\rm eff} &= \beta\int {\rm d}^3x\left[V_{\rm eff}(\phi) + \frac{1}{2} Z(\phi)(\partial_\mu\phi)^2+\cdots\right]\,,\\
    C(x) & = C_0(\phi) + D(\phi)(\partial_\mu \phi)^2 + \cdots\,,\\
    \frac{\delta S_{\rm eff}}{\delta \phi}& = \frac{\partial V_{\rm eff}}{\partial \phi}+ \frac{1}{2}\frac{\partial Z}{\partial \phi}(\partial_\mu \phi)^2 - \partial_\mu[Z(\phi)\partial_\mu\phi]+\cdots\,,
\end{align}
where $\beta = 1/T$. By expanding $V_{\rm eff}(\phi),Z(\phi)$ in power series of coupling constant $g_x$ that
\begin{align}
    V_{\rm eff} &= V_{g_x^4} + V_{g_x^6}+\cdots\,,\\
    Z& = 1+Z_{g_x^2}+\cdots\,.
\end{align}
Nielsen Identity for effective action Eq.~(\ref{eq:Nielsen}) turns into the following form for leading and sub-leading contributions that
\begin{align}
    \xi\frac{\partial V_{g_x^4}}{\partial\xi}& = 0\,, \label{eq:Nielsen1} \\
     \xi\frac{\partial V_{g_x^6}}{\partial\xi}& = -C_{g_x^2}\frac{\partial V_{g_x^4}}{\partial\phi}\,, \label{eq:Nielsen2}\\
      \xi\frac{\partial Z_{g_x^2}}{\partial\xi} &=-2\frac{\partial C_{g_x^2}}{\partial \phi}\,.
      \label{eq:Nielsen3}
\end{align}
Meanwhile, we have the leading and sub-leading order terms of effective action to be
\begin{align}
  S_{\rm eff} &= S_0 + S_1\,,\\
  S_0 &= \beta\int {\rm d}^3 x\left[\frac{1}{2}(\partial_\mu\phi_b)^2 + V_{g_x^4}(\phi_b)\right]\,,\\
  S_1 &= \beta\int {\rm d}^3 x\left[\frac{1}{2}Z_{g_x^2}(\phi_b,\xi)(\partial_\mu\phi_b)^2 + V_{g_x^6}(\phi_b,\xi)\right]\,.
\end{align}
where $\phi_b(x)$ is a bounce solution for the leading approximation to the effective action given by
\begin{align}
    \square\phi_b = \frac{\partial V_{g_x^4}}{\partial \phi}\,.
    \label{eq:bounce}
\end{align}
By Eq.~(\ref{eq:Nielsen1}) we directly have $S_0$ to be gauge independent. For $S_1$ we have
\begin{align}
    \xi\frac{\partial S_1}{\partial \xi} & =  \xi\frac{\partial }{\partial \xi} \beta \int {\rm d}^3x\left[\frac{1}{2}Z_{g_x^2}(\phi_b)(\partial_\mu\phi_b)^2 + V_{g_x^6}(\phi_b)\right] \non
    & = \beta\int {\rm d}^3x \left[C_{g_x^2}\frac{\partial V_{g_x^4}}{\partial \phi} + \frac{\partial C_{g_x^2}}{\partial \phi} (\partial_\mu\phi_b)^2\right]\non
    & = \beta\int {\rm d}^3x \left[C_{g_x^2}\frac{\partial V_{g_x^4}}{\partial \phi} + (\partial_\mu C_{g_x^2}) (\partial_\mu\phi_b)\right]\non
    & = \beta\int {\rm d}^3x C_{g_x^2} \left[\frac{\partial V_{g_x^4}}{\partial \phi} - \square\phi_b\right]\non
    & = 0\,.
\end{align}
Thus $S_1$ is also gauge independent.

\subsection{Gauge dependent effective potential}
Before discussion of how to generate a gauge invariant effective action, we introduce
the gauge dependent effective potential which provides the starting point for
further analysis.
We begin by exhibiting the gauge dependence of the effective potential in $R_\xi$ gauge fixing \cite{Patel:2011th}.
According to Eq.~(\ref{eq:Lagrangian}), the part involved in gravitational wave generation is
\begin{align} \Delta\mathcal{L} = & -\frac{1}{4} F_{\mu\nu} F^{\mu\nu} -  \frac{\delta}{2} F_{\mu\nu} B^{\mu\nu}-\Big|(\partial_\mu - {\rm i} g_x A_\mu) \Phi\Big|^2 - V_{0}^x(\Phi)\,, \label{bsm} \end{align}
 where
\begin{align}
 V_{0}^x&= -\mu_x^2\Phi\Phi^* +\lambda_x (\Phi^*\Phi)^2\,,\qquad
\Phi= \frac{1}{\sqrt 2} (\phi_c+ \phi+ {\rm i} G)\,.
\label{pot-hid}
\end{align}
By expanding the covariant derivative term we have that
\begin{align}
\Big|(\partial_\mu - i g_x A_\mu) \Phi\Big|^2  = \frac{1}{2}(\partial_\mu \phi)^2 + \frac{1}{2}(\partial_\mu G)^2 + \frac{1}{2}g_x^2\phi_c^2A_\mu A^\mu+ g_x\phi_cA^\mu\partial_\mu G+\mathcal{O}({\rm cubic})\,.\label{covariant}
\end{align}
The last term $g_x\phi_cA^\mu\partial_\mu G$ gives the mixing. In order to cancel this mixing and fix the gauge, we introduce $R_\xi$ gauge that
\begin{align}
    \mathcal{L}_{\rm gf} = -\frac{1}{2\xi}F^2
\end{align}
where,
\begin{align}
    F = \partial_\mu A^\mu - \xi g_A\phi_cG
\end{align}
Expanding it gives us
\begin{align}
    \mathcal{L}_{\rm gf} &= -\frac{1}{2\xi}(\partial_\mu A^\mu - \xi g_xvG)^2\nonumber\\
    & = -\frac{1}{2\xi}(\partial_\mu A^\mu)^2 + g_x\phi_cG\partial_\mu A^\mu-\frac{1}{2}\xi g_x^2\phi_c^2G^2\,.
\end{align}
Here $g_x\phi_cG\partial_\mu A^\mu$ cancel out $g_x\phi_cA^\mu \partial_\mu G $ in Eq.~(\ref{covariant}) with integral by parts. Beside this we also have ghost Lagrangian
\begin{align}
    \mathcal{L}_{\rm gh}= \bar{c}(-\partial^2-\xi g_x^2v^2)c + gf^{abc}(\partial^\mu c^{\dagger a})c^bA^c_\mu+ \cdots\,.
\end{align}
Add all Lagrangian together, we have expanded form to be
\begin{align}
    \Delta\mathcal{L} + \mathcal{L}_{\rm gf} + \mathcal{L}_{\rm gh} =& \frac{1}{2}A_\mu\Big[\big(\partial^2g^{\mu\nu}-(1-1/\xi)\partial^\mu\partial^\nu\big)+m_{A_x}^2g^{\mu\nu}\Big]A_\nu\nonumber\\
    &+\frac{1}{2}\phi\big(-\partial^2-m_\phi^2\big)\phi+\frac{1}{2}G\big(-\partial^2-m_G^2-\xi m_{A_x}^2\big)G\nonumber\\
    &+ \bar{c}\big(-\partial^2-\xi m_{A_x}^2\big)c + \cdots,
\end{align}
where we have field dependent masses to be
\begin{align}
    m_{A_x}^2 = g_x^2\phi_c^2\,,\qquad m_\phi^2 = -\mu_x^2+3\lambda_x\phi_c^2\,,\qquad m_G^2 = -\mu_x^2+\lambda_x\phi_c^2\,.\label{eq:fielddependentmasses}
\end{align}
At one loop, we have Coleman-Weinberg correction to be
\begin{align}
    V_{\rm CW}(\phi_c,\xi)=  &  \frac{1}{4(4\pi)^2}\Bigg\{(m_\phi^2)^2\left[\ln{\left(\frac{m_\phi^2}{\Lambda^2}\right)}-\frac{3}{2}\right]+3(m_{A_x}^2)^2\left[\ln{\left(\frac{m_{A_x}^2}{\Lambda^2}\right)}-\frac{5}{6}\right]\nonumber\\
    &+\frac{1}{4(4\pi)^2}(m_G^2+\xi m_{A_x}^2)^2\left[\ln{\left(\frac{m_G^2+\xi m_{A_x}^2}{\Lambda^2}\right)}-\frac{3}{2}\right]\non
    &-\frac{1}{4(4\pi)^2}(\xi m_{A_x}^2)^2\left[\ln{\left(\frac{\xi m_{A_x}^2}{\Lambda^2}\right)}-\frac{3}{2}\right]\Bigg\}\,. \label{eq:coleman}
\end{align}
In this paper, we take on-shell-like schemes, which means renormalization conditions are defined such that the single and double derivatives of the one-loop effective potential vanish, i.e.
\begin{align}
    \left.\frac{\partial}{\partial\phi}\left(V_{\rm CW}+V_{\rm CT}\right)\right|_{\phi=v_0} = 0\,,\\
    \left.\frac{\partial^2}{\partial\phi^2}\left(V_{\rm CW}+V_{\rm CT}\right)\right|_{\phi=v_0} = 0\,.
\end{align}
By resolving these two equations, we can then determine the counterterms
so that
$\delta \mu_x^2$ and $\delta\lambda_x$ that
\begin{align}
    V_{\rm CT}(\phi_c,\xi) = \frac{\delta\mu_x^2}{2}\phi_c^2+\frac{\delta\lambda_h}{4}\phi_c^4\,.
\end{align}
Also, we have finite temperature correction to be
\begin{align}
    V_{T}(\phi_c,T,\xi) &= \frac{T^4}{2\pi^2}\Bigg[J_B\left(\frac{m_\phi^2}{T^2}\right)+3J_B\left(\frac{m_{A_x}^2}{T^2}\right)+J_B\left(\frac{m_G^2+\xi m_{A_x}^2}{T^2}\right)-J_B\left(\frac{\xi m_{A_x}^2}{T^2}\right)\Bigg]\,.
    \label{eq:thermal}
\end{align}
The one-loop corrections given above are unambiguous and well established. However, for $V_{\rm daisy}$ corrections, a further detailed discussion is required.  Infrared divergences appear at high temperatures and break down finite-temperature perturbation theory. To resolve this problem, daisy resummation is applied. Using the Arnold-Espinosa approach, which employs a high-temperature approximation, we have
\begin{align}
    V_{\rm daisy}(\phi,T) = -\frac{T}{12}\sum_B\left[\left(m_B^2+\Pi_B\right)^{3/2}-\left(m_B^2\right)^{3/2}\right]\,, \label{eq:daisy}
\end{align}
where for the $U(1)_x$ model we have
\begin{align}
        \Pi_{A}(T) =  \frac{2}{3}g_x^2 T^2\,,\qquad
    \Pi_{\phi}(T) = \frac{1}{4}g_x^2 T^2 + \frac{1}{3}\lambda_xT^2\,,\qquad \Pi_{G}(T) =\frac{1}{4}\lambda_xT^2 . \label{eq:daisy2}
\end{align}
The significance of Daisy resummation emerges in the high-temperature regime where $T \gg m$, as it captures essential infrared effects within the finite-temperature field theory framework. In scenarios characterized by supercooled or near-supercooled conditions, where the temperature satisfies $T \leq m$, the contribution of Daisy resummation to the effective potential becomes negligible. Consequently, we adopt the approximation of neglecting this term throughout the present analysis for computational simplicity \cite{Bahl:2024ykv,Hua:2025fap}.
For a more rigorous treatment of finite-temperature mass resummation in the low-temperature regime, we refer the reader to the comprehensive methodologies of ``Full Dressing'' and ``Partial Dressing'' as detailed in Ref.~\cite{Bittar:2025lcr,Curtin:2016urg}.
All together we have
\begin{align}
    V_{\rm eff}(\phi_c,T,\xi) = V_0(\phi_c) + V_{\rm CW}(\phi_c,\xi) + V_{\rm CT}(\phi_c,\xi)  + V_{T}(\phi_c,T,\xi)\,.\label{eq:Veff}
\end{align}

\subsection{Gauge invariant effective action for the $U(1)_x$ sector at low temperature}\label{sec:3.2}

We proceed to implement the gauge invariance analysis within the framework of our $U(1)$ extension model. This approach follows methodologies analogous to those employed in previous investigations of gauge-invariant formulations for $U(1)$ extension models, wherein the effective potential incorporates contributions from both tree-level diagrams and Coleman-Weinberg radiative corrections~\cite{Arunasalam:2021zrs}. To extend this theoretical framework to the present investigation, we must incorporate temperature-dependent corrections into our analysis. Previous investigations have established the requisite formalism for the high-temperature approximation regime, characterized by the condition $m/T \ll 1$ \cite{Lofgren:2021ogg, Hirvonen:2021zej}. However, the supercooled and near-supercooled phase transitions that may account for gravitational wave signals detectable by next generation instruments operate in a regime where the high-temperature approximation fails (see Section~\ref{sec:Parameterscan}). Specifically, these phenomena require consideration of the low-temperature regime where $m/T > 1$. As demonstrated in our preceding analysis, daisy summation corrections become negligible in the low-temperature limit and may therefore be omitted from the gauge-invariant treatment. This represents a fundamental departure from the methodologies employed in Refs.~\cite{Lofgren:2021ogg, Hirvonen:2021zej}, wherein such corrections constitute the dominant contribution to the effective potential. Consequently, a novel analytical approach tailored to the low-temperature regime is required.

To ensure the validity of our gauge invariance analysis, we first clarify the definitions of the $\xi$-independent term $V_{g_x^4}$ and the $\xi$-dependent term $V_{g_x^6}^\xi$. We impose the conditions that $\lambda_x + \delta\lambda_x \sim g_x^4$ and $\mu_x^2 + \delta\mu_x^2 \sim g_x^4\langle\phi_c\rangle^2$, where $\langle\phi_c\rangle$ is the vacuum expectation value. Since we are working in the on-shell scheme, we have $\langle\phi_c\rangle \sim v_0$. It is important to note that for certain choices of $g_x$ and $\lambda_x$, the relation $\lambda_x \sim g_x^4$ may not strictly hold; however, the inclusion of counterterms ensures that the power counting remains consistent.
Under these conditions, the tree-level potential with counterterms maintains order $g_x^4$ and exhibits $\xi$-independence. From Eq.~(\ref{eq:coleman}), we observe that the correction attributable to the dark photon scales as $m_{A_x}^4\sim g_x^4\langle\phi_c\rangle^4$, while the correction arising from the dark Higgs scales as $m_\phi^4\sim \lambda_x^2 \sim g_x^8\langle\phi_c\rangle^4$ which is thus negligible. Regarding the remaining two $\xi$-dependent correction terms, we note that
\begin{align}
    V_{\rm cw}^\xi = X(m_G^2+\xi m_{A_x}^2) - X(\xi m_{A_x}^2)\,,\quad X(m^2) = \frac{1}{4(4\pi)^2}[m^2]^2\left[\ln{\left(\frac{m^2}{\Lambda^2}\right)}-\frac{3}{2}\right]\,.
\end{align}
In the regime where $\xi m_{A_x}^2 \ll m_G^2$, we obtain $V_{\rm cw}^\xi \sim X(m_G^2)\sim m_G^2\sim g_x^8$, where $m_G^2 \sim g_x^4\phi^2$. This contribution is therefore negligible within our leading and sub-leading order analysis. Conversely, in the alternative regime, we have
\begin{align}
    V_{\rm cw}^\xi = X(m_G^2+\xi m_{A_x}^2) - X(\xi m_{A_x}^2) \simeq m_G^2X'(\xi m_{A_x}^2)\propto m_G^2m_{A_x}^2 \sim \mathcal{O}(g_x^6)\,.
\end{align}
Thus, this $\xi$-dependent term exhibits the anticipated scaling of order $g_x^6$.
In this study, we examine the low-temperature regime characterized by $m/T > 1$. We begin by analyzing two distinct cases based on the scaling of temperature relative to the scalar field expectation value $\phi$ and the gauge coupling $g_x$.

\textbf{Case 1: $T \sim g_x\phi$}. This corresponds to the limiting condition $m/T \sim 1$. Under this condition, the $\xi$-independent thermal correction term, which includes contributions from both the dark Higgs $\phi$ and dark photon $A$, scales as
\begin{align}
   V_T \sim T^4 \sim g_x^4\phi^4\,,
\end{align}
which is of order $\mathcal{O}(g_x^4)$ and contributes to the leading-order potential. The gauge-dependent component can be written as
\begin{align}
   V_T^{\xi}  &=  \frac{T^4}{2\pi^2}\left[J_B\left(\frac{m_G^2+\xi m_{A_x}^2}{T^2}\right)-J_B\left(\frac{\xi m_{A_x}^2}{T^2}\right)\right]
  \nonumber\\ &\simeq \frac{T^4}{2\pi^2}\frac{m_G^2}{T^2}\left[J_B'\left(\frac{\xi m_{A_x}^2}{T^2}\right)\right]\sim T^2m_G^2 \sim (g_x\phi)^2(g_x^4\phi^2) \sim \mathcal{O}(g_x^6)\,,
\end{align}
contributing to the sub-leading potential as expected. Combining these results, we define our leading and sub-leading potentials as
\begin{align}
   V_{g_x^4}(\phi,T) =& -\frac{\mu_x^2+\delta\mu_x^2}{2}\phi^2 + \frac{\lambda+\delta\lambda}{4}\phi^4 + \frac{3m_{A_x}(\phi)^4}{64\pi^2}\left[\ln{\left(\frac{m_{A_x}^2}{\Lambda^2}\right)}-\frac{5}{6}\right]\nonumber\\
   &+ \frac{T^4}{2\pi^2}\left[J_B\left(\frac{m_\phi^2}{T^2}\right)+3J_B\left(\frac{m_{A_x}^2}{T^2}\right)\right]\,, \label{eq:Ve4}\\
   V_{g_x^6}(\phi,\xi,T)=& \frac{(m_G^2+\xi m_{A_x}^2)^2}{64\pi^2}\left[\ln{\left(\frac{m_G^2+\xi m_{A_x}^2}{\Lambda^2}\right)}-\frac{3}{2}\right] - \frac{\xi^2 m_{A_x}^4}{64\pi^2}\left[\ln{\left(\frac{\xi m_{A_x}^2}{\Lambda^2}\right)}-\frac{3}{2}\right]\nonumber\\
   &+ \frac{T^4}{2\pi^2}\left[J_B\left(\frac{m_G^2+\xi m_{A_x}^2}{T^2}\right)-J_B\left(\frac{\xi m_{A_x}^2}{T^2}\right)\right]\,.
   \label{eq:Ve6}
\end{align}

\textbf{Case 2: $T \ll g_x\phi$}. This corresponds to the very low-temperature or strongly supercooled regime, where $m/T \gg 1$. In this limit, all temperature-dependent corrections, including the thermal potential and daisy resummation, are exponentially suppressed by Boltzmann factors of the form $e^{-m/T}$. These contributions are therefore negligible to all relevant orders of perturbation theory. The effective potential reduces to the zero-temperature Coleman-Weinberg potential, and our analysis simplifies to the treatment presented in~\cite{Arunasalam:2021zrs}.

As in~\cite{Arunasalam:2021zrs}, three key modifications are required to ensure the gauge invariance of the effective action. In the low-temperature regime, these modifications are slightly different, and we will now discuss each of them in detail.
\begin{itemize}
\item First, we use the bounce solution derived from the leading-order potential for the calculation of both $S_0$ and $S_1$. In the supercooled regime (Case 2), the bounce solution is highly sensitive to thermal corrections, even though they are suppressed and do not contribute to the leading-order potential ($V_{g_x^4}$). Therefore, we always include the leading temperature correction term when solving for the bounce solution with Eq.~(\ref{eq:bounce}), i.e. Eq.~(\ref{eq:Ve4}) will be used for both cases. The leading temperature correction is $\xi$-independent and thus make sure Eq.~(\ref{eq:Nielsen1}) is satisfied.
\item Second, we use the dressed Goldstone boson mass, which is defined as
\begin{align}
m_G^2 = \frac{1}{\phi}\frac{\partial V_{g_x^4}}{\partial\phi}\,,
\end{align}
instead of the expression in Eq.~(\ref{eq:fielddependentmasses}). This ensures that the Nielsen identity (Eq.~(\ref{eq:Nielsen2})) is satisfied, preserving gauge invariance. This was verified at zero temperature in~\cite{Arunasalam:2021zrs}. The generalization to finite temperature is straightforward, requiring the replacement of the zero-temperature four-momentum integral with the Matsubara sum integral:
\begin{align}
  \int\frac{d^4p}{(2\pi)^4}\rightarrow T\sum_{n\in\mathbb Z}\int\frac{d^3p}{(2\pi)^3}\,,
  \qquad \omega_n\equiv 2\pi n T\,.
\end{align}
Define $E_{G'} = \sqrt{p^2+m_{G'}^2}$ and $E_{\rm FP}=\sqrt{p^2+m_{\rm FP}^2}$ with the finite-temperature ghost and dressed Goldstone masses
\begin{equation}
  m_{\rm FP}^2=2\xi g_x^2\phi^2,\qquad
  m_{G'}^2=\frac{1}{\phi}\frac{\partial V_{g_x^4}(\phi,T)}{\partial\phi} + 2\xi g_x^2\phi^2.
  \label{eq:thermal-masses}
\end{equation}
The one-loop ghost-Goldstone sector then gives
\begin{align}
  C^T_{g_x^2}(\phi,T)= -g_x^2\xi\phi^2 T\sum_{n\in\mathbb Z}\int\frac{d^3p}{(2\pi)^3}
  \frac{1}{\big(\omega_n^2+E_{G'}^2\big)\big(\omega_n^2+E_{\rm FP}^2\big)}\,.
  \label{eq:Cee-sum}
\end{align}
The $\xi$-dependent part of the finite temperature correction in $V_{g_x^6}$ is the difference between the Goldstone and ghost determinants,
\begin{align}
  V^{\xi,T}_{g_x^6}(\phi,T)
  =\frac{T}{2}\sum_{n\in\mathbb Z}\int\!\frac{d^3p}{(2\pi)^3}\,
  \Big[\ln\big(\omega_n^2+E_{G'}^2\big)-\ln\big(\omega_n^2+E_{\rm FP}^2\big)\Big]\,.
  \label{eq:Vxi-thermal}
\end{align}
Differentiating Eq.~\eqref{eq:Vxi-thermal} with respect to $\xi$ and using $\partial_\xi m_{G'}^2=\partial_\xi m_{\rm FP}^2=2g_x^2\phi^2$ gives
\begin{align}
  \xi\frac{\partial V^{\xi,T}_{g_x^6}(\phi,T)}{\partial\xi}
  = -\xi g_x^2\phi^2\, T\sum_{n}\int\frac{d^3p}{(2\pi)^3}\!
  \bigg(\frac{1}{\omega_n^2+E_{\rm FP}^2}-\frac{1}{\omega_n^2+E_{G'}^2}\bigg)\,.
\end{align}
Using the relation $\tfrac{1}{A}-\tfrac{1}{B}=\tfrac{B-A}{AB}$ with
\begin{align}
  B-A=m_{G'}^2-m_{\rm FP}^2=\frac{1}{\phi}\frac{\partial V_{g_x^4}(\phi,T)}{\partial \phi}\,,
\end{align}
and identifying the sum-integral defining $C^T_{g_x^2}(\phi,T)$ leads to
\begin{align}
  \xi\frac{\partial V^{\xi,T}_{g_x^6}(\phi,T)}{\partial\xi}= -\,C^T_{g_x^2}(\phi,T)\frac{\partial V_{g_x^4}(\phi,T)}{\partial \phi}\,,
  \label{eq:thermal-identity-V}
\end{align}
which is the thermal analogue of the zero-temperature identity.
\item Third, we calculate the normalization factor $Z(\phi,\xi)$ and its perturbative expansion. This quantity is determined through the relation
\begin{align}
Z = -\left.\frac{\partial \Gamma^2(p^2)}{\partial p^2}\right|_{p^2=0}\,,
\end{align}
where $\Gamma$ denotes the one-particle irreducible (1PI) Green's function, which is evaluated from the sum of one-particle irreducible diagrams. We should expect that
\begin{align}
    Z(\phi) &= 1 + Z_{g_x^2}(\phi,T=0) +  \delta Z_{g_x^2}(\phi,T)  +\mathcal{O}(g_x^4) \,,
\end{align}
where the zero-temperature factor $Z_{g_x^2}(\phi,T=0)$ has been provided in \cite{Arunasalam:2021zrs} that,
\begin{align}
    Z_{g_x^2}(\phi,T=0)&= \frac{g_x^2}{16\pi^2}\left[\xi\ln{\left(\frac{g_x^2\phi^2\xi}{\Lambda^2}\right)}+3\ln{\left(\frac{g_x^2\phi^2}{\Lambda^2}\right)+\xi}\right]\,.
\end{align}
For the leading thermal correction $\delta Z_{g_x^2}(\phi,T) $, one must again re-evaluate these loop diagrams using the Matsubara formalism following \cite{Garny:2012cg}. In the low-temperature regime considered here, the thermal correction these corrections take the general form of a sum over the particle species, with each term proportional to a Boltzmann suppression factor $e^{-m/T}$, and therefore does not contribute to the sub-leading order term $Z_{g_x^2}$.
\end{itemize}

\section{First order phase transitions, hydrodynamics and gravitational waves Details}\label{App:Details}

\subsection{Percolation temperature}

As established in the literature \cite{Athron:2022mmm,Athron:2023mer,Caprini:2019egz,Wang:2020jrd}, the percolation temperature $T_{p}$ serves as the characteristic temperature scale relevant for gravitational wave generation during phase transitions. The quantity $T_{p}$ represents the hidden sector temperature at which 71\% of the universe volume remains in the false vacuum state, and is determined by the condition
\begin{align}
P_f(T_{p}) &= 0.71\,, \\
        P_f(T)&=\exp\left[-\frac{4\pi}{3}v_w^3\int_{T}^{T_{c}}{\rm d}T'\frac{\Gamma(T')}{T'^4H(T')}\left(\int_{T}^{T'}{\rm d}T''\frac{1}{H(T'')}\right)^3\right]\,,\label{eq:percolation} \\
        \Gamma(T) &\simeq T^4 \left[\frac{S_{\rm eff}(T)}{2\pi T} \right]^{3/2}e^{-S_{\rm eff}(T)/T}\,,
        \label{pf}
\end{align}
where $S_{\rm eff}(T)$ denotes the three-dimensional gauge invariant effective action calculated in the previous section, $T_{c}$ is the critical temperature, $v_w$ represents the bubble wall velocity, and $H$ is the Hubble parameter.

\subsection{Transition strength}

Two primary types of transition strength parameters are defined, as described in \cite{Bringmann:2023opz,Bai:2021ibt}:
\begin{align}
\label{alphatot}
    \alpha_{\rm tot} &= \frac{\Delta \Bar{\theta}(T_{h,p})}{\rho_{\rm rad}^v(T_{p}) + \rho_{\rm rad}^h(T_{p})}\,, \\
    \alpha_h &= \frac{\Delta \Bar{\theta}(T_{p})}{\rho_{\rm rad}^h(T_{p})}\,. \label{Eq.alphastar}
\end{align}
The quantity $\Delta \Bar{\theta}$ represents the vacuum energy released during the phase transition \cite{Giese:2020znk,Giese:2020rtr}, and is expressed as:
\begin{align}
    \Delta \Bar{\theta}(T_h)& = \Bar{\theta}_f(T_h) - \Bar{\theta}_t(T_h)\,,\\
     \Bar{\theta}_i(T_h) &= \frac{1}{4}\left[\rho^h_i - \frac{p^h_i(T_h)}{c_{s,t}^2(T_h)}\right]\,,
\end{align}
where
\begin{align}
    p^h_i(\phi_i,T_h) &=  \frac{\pi^2}{90}g^h_{\rm eff}(T_h)T_h^4 - V^h_{\rm eff}(\phi_i,T_h)\,,\label{eq:phi}\\
    \rho^h_i(\phi_i,T_h) &= T_h\frac{\partial p^h_i}{\partial T_h} - p^h_i\,,\label{eq:rhohi}\\
    c^2_{s,t}(T_h) &= \frac{\partial p^h_t}{\partial \rho^h_t} = \frac{\partial p^h_t/\partial T_h}{\partial \rho^h_t/\partial T_h}\,,
\end{align}
where $g_{\rm eff}^h$ is the effective degrees of freedom of the hidden sector. The subscripts ``$f$" and ``$t$" denote the false vacuum state and the true vacuum state, respectively. The parameter $\alpha_{tot}$ is utilized in gravitational wave power spectrum calculations, while $\alpha_h$ is employed in fluid dynamics calculations for the hidden sector to determine the efficiency parameter $\kappa$ and the bubble wall velocity $v_w$.

\subsection{Transition rate}

The mean bubble separation distance $R_*$ is used to characterize the transition rate instead of action derivative $\left.{\rm d}S_{\rm eff}/{\rm d}\log{T}\right|_{T_p} $ which does not provide a good approximation for supercooled phase transitions~\cite{Li:2025nja}. The value of $R_*$ is calculated from the bubble number density according to
\begin{align}
    R_*(T_h) = \big[n_B(T_h)\big]^{-\frac{1}{3}} = \left[T_h^3\int_{T_h}^{T_{h,c}}dT_h'\frac{\Gamma(T_h')P_f(T_h')}{{T_h'}^4H(T_h')}\right]^{-\frac{1}{3}}\,.\label{eq:Rstar}
\end{align}
The transition rate parameter is then defined in relation to this quantity as
\begin{align}
    \frac{\beta}{H_*} = \frac{(8\pi)^{1/3}v_w}{R_*H_*}\,.
\end{align}

\bl{
\subsection{Reheating temperature}

After bubble percolation, the vacuum energy released during the phase transition is converted into radiation, reheating the plasma. The reheating occurs primarily in the hidden sector where the phase transition takes place. Following~\cite{Li:2025nja}, assuming reheating completes instantaneously around the time of percolation, energy conservation in the hidden sector gives
\begin{align}
\rho_{\rm rad}^h(T_{h,{\rm reh}}) \simeq \rho_{\rm rad}^h(T_{h,p}) + \rho_{\rm vac}(T_{h,p})\,,
\end{align}
where $\rho_{\rm rad}^h = (\pi^2/30)g_{\rm eff}^h T_h^4$ is the hidden sector radiation energy density and $\rho_{\rm vac}$ is the vacuum energy density released during the transition. Using the definition of the hidden sector transition strength $\alpha_h$ (Eq.~\ref{Eq.alphastar}), the hidden sector reheating temperature is
\begin{align}
T_{h,{\rm reh}} = (1 + \alpha_h)^{1/4}\left(\frac{g_{\rm eff}^h(T_{h,p})}{g_{\rm eff}^h(T_{h,{\rm reh}})}\right)^{1/4} T_{h,p}\,.
\end{align}
For strong phase transitions with $\alpha_h \gg 1$, reheating significantly increases the hidden sector temperature: $T_{h,{\rm reh}} \gg T_{h,p}$.

Although the hidden sector is in thermal equilibrium with the visible sector before the phase transition ($T_h = T_v$), the phase transition completes on a timescale much shorter than the heat transfer timescale between sectors. Consequently, the vacuum energy is initially deposited entirely in the hidden sector, and the visible sector temperature remains essentially unchanged during reheating: $T_{v,{\rm reh}} \simeq T_{v,p}$. 

\subsection{Redshift factors for gravitational wave spectrum}

The gravitational wave spectrum produced at the phase transition must be redshifted to the present day. Following~\cite{Li:2025nja}, the frequency and amplitude redshift factors are defined as
\begin{align}
\mathcal{R}_f \equiv \frac{a_*}{a_0} = \left(\frac{h_{\rm eff}^0}{h_{\rm eff}(T_{h,{\rm reh}})}\right)^{1/3}\frac{T^0}{T_{v,{\rm reh}}}\,,
\end{align}
\begin{align}
\mathcal{R}_\Omega \equiv \left(\frac{a_*}{a_0}\right)^4\left(\frac{H_*}{H_0}\right)^2 \simeq 2.473\times 10^{-5}h^{-2}\left(\frac{h_{\rm eff}^0}{h_{\rm eff}(T_{h,{\rm reh}})}\right)^{4/3}\frac{g_{\rm eff}(T_{h,{\rm reh}})}{2}\,,
\end{align}
where $T^0 = 2.35\times 10^{-13}$~GeV is the present CMB temperature and $h_{\rm eff}^0 \simeq 3.91$ is the current effective entropy degrees of freedom. The effective degrees of freedom account for contributions from both sectors, weighted by the temperature ratio $\xi_{\rm reh} \equiv T_{h,{\rm reh}}/T_{v,{\rm reh}}$:
\begin{align}
g_{\rm eff}(T_{h,{\rm reh}}) &= g_{\rm eff}^h(T_{h,{\rm reh}}) + \xi_{\rm reh}^{-4}\,g_{\rm eff}^v(T_{v,{\rm reh}})\,,\\
h_{\rm eff}(T_{h,{\rm reh}}) &= h_{\rm eff}^h(T_{h,{\rm reh}}) + \xi_{\rm reh}^{-3}\,h_{\rm eff}^v(T_{v,{\rm reh}})\,,
\end{align}
where the superscripts $v$ and $h$ denote the visible and hidden sector contributions respectively.

\subsection{Hydrodynamics: efficiency factor and bubble wall velocity}

The efficiency factor $\kappa$ quantifies the fraction of vacuum energy converted into bulk kinetic energy of the plasma~\cite{Bodeker:2017cim,Espinosa:2010hh,Giese:2020rtr,Giese:2020znk,Ai:2021kak,Ai:2024uyw,Ellis:2019oqb}. In this paper, we follow the calculation in~\cite{Giese:2020rtr,Giese:2020znk} to get $\kappa(c_{s,f}^2,c_{s,t}^2,\alpha,v_w)$. Following~\cite{Li:2025nja}, we use the hidden sector transition strength $\alpha_h$ (Eq.~\ref{Eq.alphastar}) to determine the hydrodynamic properties. The bubble expansion proceeds in two regimes distinguished by $\alpha_{h,\infty}$, the critical parameter characterizing maximum plasma friction on the bubble wall:

\textbf{Non-runaway regime} ($\alpha_h < \alpha_{h,\infty}$): The bubble wall reaches a terminal velocity $v_w < 1$. The efficiency factor are:
\begin{align}
    \kappa_{\rm col} \simeq 0\,,\qquad
    \kappa_f \simeq \kappa(c_{s,f}^2,c_{s,t}^2,\alpha_h,v_J)\,,
\end{align}
, where the Chapman-Jouguet velocity is~\cite{Steinhardt:1981ct,Espinosa:2010hh}
\begin{align}
v_J = \frac{c_{s,f}\left(1 + \sqrt{3\alpha_h(1+c_{s,f}^2\alpha_h)}\right)}{1 + 3c_{s,f}^2\alpha_h}\,,
\end{align}
with $c_{s,f}/c_{s,t}$ the speed of sound in the false/true vacuum.

\textbf{Runaway regime} ($\alpha_h > \alpha_{h,\infty}$): The bubble wall accelerates toward $v_w \to 1$, with most vacuum energy going into the wall. The efficiency factors are~\cite{Bodeker:2017cim,Ellis:2019oqb}
\begin{align}
\kappa_{\rm col} \simeq 1 - \frac{\alpha_{h,\infty}}{\alpha_h}\,,\qquad
\kappa_f \simeq \frac{\alpha_{h,\infty}}{\alpha_h}\kappa(c_{s,f}^2,c_{s,t}^2,\alpha_{h,\inf},1)\,,
\end{align}
for bubble collisions and plasma motion respectively.

The bubble collision contribution becomes the dominant source of gravitational waves when $\alpha_h \gg \alpha_{h,\infty}$, which leads to $\kappa_{\rm col} \gg \kappa_f$. The critical parameter $\alpha_{h,\infty}$ characterizing maximum plasma friction is determined by the mass spectrum of particles coupling to the bubble wall~\cite{Bodeker:2017cim}:
\begin{align}
\alpha_{h,\infty} = \frac{T_p^2}{\rho_{\rm rad}(T_p)} \left(\sum_{i=\rm boson} n_i \frac{\Delta m_i^2}{24} + \sum_{i=\rm fermion} n_i \frac{\Delta m_i^2}{48}\right)\,,
\end{align}
where $n_i$ counts the degrees of freedom and $\Delta m_i^2$ is the change in the squared mass across the phase transition. In our hidden sector model, only two bosons participate in the phase transition: the dark photon and the dark scalar. Since both have small, sub-GeV masses, the value of $\alpha_{h,\infty}$ remains relatively small. Consequently, the collision-dominated condition $\alpha_h \gg \alpha_{h,\infty}$ is readily satisfied with moderate values of $\alpha_h \sim 10^1$--$10^3$, without requiring extreme supercooling. This contrasts with the classically conformal $U(1)_{B-L}$ models studied in Refs.~\cite{Ellis:2019oqb,Ellis:2020nnr}, where the phase transition involves Standard Model particles and additional TeV-scale bosons.

\subsection{Details of the gravitational wave power spectrum}

The total gravitational wave power spectrum produced by a first-order phase transition receives contributions from three distinct sources: bubble wall collisions, sound waves in the plasma, and turbulence~\cite{Caprini:2015zlo,Caprini:2019egz,Athron:2023mer,Hindmarsh:2016lnk,Hindmarsh:2019phv,Hindmarsh:2017gnf,Guo:2020grp}:
\begin{align}
\Omega_{\rm GW}h^2 = \Omega_{\rm col}h^2 + \Omega_{\rm sw}h^2 + \Omega_{\rm turb}h^2\,.
\end{align}
Following~\cite{Athron:2023mer}, the redshifted spectra are expressed in terms of the mean bubble separation $R_*$ and the kinetic energy fraction $K = \kappa_f\alpha_{\rm tot}/(1+\alpha_{\rm tot})$.

\textbf{Bubble collision contribution.}
For runaway bubble walls, the contribution from bubble collisions is~\cite{Huber:2008hg,Athron:2023mer}
\begin{align}
\Omega_{\rm col}(f) = \mathcal{R}_\Omega A \left(\frac{H_* R_*}{(8\pi)^{1/3}v_w}\right)^2 \kappa_{\rm col}^2 \left(\frac{\alpha_{\rm tot}}{1+\alpha_{\rm tot}}\right)^2 S_{\rm col}(f)\,,
\end{align}
where $A = 0.049$ and the spectral shape is a broken power law
\begin{align}
S_{\rm col}(f) = \frac{(a+b)^c}{\left[b(f/f_{\rm col})^{-a/c} + a(f/f_{\rm col})^{b/c}\right]^c}\,,
\end{align}
with spectral indices $(a,b,c) = (2.4, 2.3, 2.0)$ and peak frequency $f_{\rm col} = \mathcal{R}_f \cdot 0.77(8\pi)^{1/3}v_w/(2\pi R_*)$.

\textbf{Sound wave contribution.}
The sound shell model gives~\cite{Hindmarsh:2017gnf,Athron:2023mer}
\begin{align}
h^2\Omega_{\rm sw}(f) = 3\mathcal{R}_\Omega (H_* R_*) K^2 \frac{M(s,r_b,b)}{\mu_f(r_b)} \Upsilon(\tau_{\rm sw}) \tilde{\Omega}_{\rm gw}\,,
\end{align}
where $\tilde{\Omega}_{\rm gw} \approx 0.01$, $M(s,r_b,b)$ is the spectral shape function with $s = 2\pi f R_*/\mathcal{R}_f$ being the dimensionless frequency. The suppression factor accounting for finite sound wave lifetime is~\cite{Guo:2020grp}
\begin{align}
\Upsilon(\tau_{\rm sw}) = 1 - \frac{1}{\sqrt{1 + 2H_*\tau_{\rm sw}}}\,,
\end{align}
where $\tau_{\rm sw} \sim R_*/\bar{U}_f$ is the sound wave lifetime with $\bar{U}_f$ being the root-mean-square fluid velocity.

\textbf{Turbulence contribution.}
The MHD turbulence contribution is~\cite{Caprini:2009yp,Athron:2023mer}
\begin{align}
\Omega_{\rm turb}(f) = 9.0\,\mathcal{R}_\Omega (H_* R_*) (\kappa_{\rm turb} K)^{3/2} S_{\rm turb}(f)\,,
\end{align}
where $\kappa_{\rm turb} \simeq 0.05$ is the turbulence efficiency factor, and the spectral shape is
\begin{align}
S_{\rm turb}(f) = \frac{(f/f_{\rm turb})^3}{[1+(f/f_{\rm turb})]^{11/3}(1+8\pi f/(\mathcal{R}_f H_*))}\,,
\end{align}
with peak frequency $f_{\rm turb} = 3.5\mathcal{R}_f/R_*$.

}


\end{document}